\documentclass[10pt,twocolumn]{article}
\usepackage[letterpaper,margin=0.8in]{geometry}
\usepackage[utf8]{inputenc}
\usepackage{textcomp}
\usepackage{hyperref}
\usepackage{multirow}
\usepackage{varwidth}
\usepackage{booktabs}
\usepackage{threeparttable}
\usepackage{natbib}
\usepackage{astrojournals}
\usepackage{graphicx}
\usepackage{mathtools}
\usepackage{amsmath}
\usepackage{abstract}
\usepackage{lipsum}
\usepackage[charter]{mathdesign}
\usepackage{microtype}
\usepackage{xspace}
\usepackage{xcolor}
\usepackage{fixltx2e}
\usepackage{ragged2e}
\usepackage{paralist}

\RequirePackage{etoolbox}
\makeatletter

\patchcmd{\NAT@citex}
  {\@citea\NAT@hyper@{\NAT@nmfmt{\NAT@nm}\NAT@date}}
  {\@citea\NAT@nmfmt{\NAT@nm}\NAT@hyper@{\NAT@date}}
  {}% Do nothing if patch works
  {}% Do nothing if patch fails

\patchcmd{\NAT@citex}
  {\@citea\NAT@hyper@{%
     \NAT@nmfmt{\NAT@nm}%
     \hyper@natlinkbreak{\NAT@aysep\NAT@spacechar}{\@citeb\@extra@b@citeb}%
     \NAT@date}}
  {\@citea\NAT@nmfmt{\NAT@nm}%
   \NAT@aysep\NAT@spacechar%
   \NAT@hyper@{\NAT@date}}
  {}% Do nothing if patch works
  {}% Do nothing if patch fails

\patchcmd{\NAT@citex}
  {\@citea\NAT@hyper@{%
     \NAT@nmfmt{\NAT@nm}%
     \hyper@natlinkbreak{\NAT@spacechar\NAT@@open\if*#1*\else#1\NAT@spacechar\fi}%
       {\@citeb\@extra@b@citeb}%
     \NAT@date}}
  {\@citea\NAT@nmfmt{\NAT@nm}%
   \NAT@spacechar\NAT@@open\if*#1*\else#1\NAT@spacechar\fi%
   \NAT@hyper@{\NAT@date}}
  {}% Do nothing if patch works
  {}% Do nothing if patch fails

\makeatother
\usepackage{array}
\newcolumntype{L}{>{$}l<{$}}
\newcolumntype{C}{>{$}c<{$}}
\newcolumntype{R}{>{$}r<{$}}

\usepackage{caption}
\usepackage{sectsty}
\allsectionsfont{\raggedright\bfseries\large}
\subsectionfont{\raggedright\normalfont\itshape\normalsize}
\subsubsectionfont{\centering\normalfont\itshape\normalsize}

\newcommand{\eg}{e.\,g.\xspace}
\newcommand{\ie}{i.\,e.\xspace}

\newcommand{\cstcool}{\ensuremath{c_{\text{s}}{}t_{\text{cool}}}\xspace}
\newcommand{\rcloudlet}{\ensuremath{\ell_{\text{cloudlet}}}\xspace}

\renewcommand*{\vec}[1]{\boldsymbol{#1}}

\let\oldhat\hat
\renewcommand*{\hat}[1]{\vec{\oldhat{#1}}}

\hypersetup{%
 pdftitle={Cold Clumps in Hot Halos},
 pdfauthor={\textcopyright\ authors},
 bookmarksopen=true,
 colorlinks=true,
 linkcolor=black,
 citecolor=black,
 urlcolor=black}

\begin{document}
\fontsize{10}{14}\selectfont
\raggedbottom

%\title{Cold Clumps in Hot Halos}
%\title{Shattering Tiny, Cold Gas Clouds in Galaxy Halos}
\title{A Characteristic Scale for Cold Gas}

\author{Michael McCourt,%
\!\thanks{UC Santa Barbara, Santa Barbara, CA}%
\,\,\thanks{Hubble Fellow}
\, S.\ Peng Oh,\!\footnotemark[2]%
\,\,\, Ryan M.\ O'Leary,%
\thanks{JILA, University of Colorado, Boulder, CO}
\, \& Ann-Marie Madigan%
\footnotemark[4]}
\date{\today}

\twocolumn[
\addtocounter{footnote}{1}      % why?????
\maketitle
\begin{onecolabstract}
We find that clouds of optically-thin, pressure-confined gas are prone
to fragmentation as they cool below $\sim10^6$\,K.  This fragmentation
follows the lengthscale $\sim\cstcool$, ultimately reaching very small
scales ($\sim{0.1}\,\text{pc}/n$) as they reach the temperature
$\sim10^4$\,K at which hydrogen recombines.  While this lengthscale
depends on the ambient pressure confining the clouds, we find that the
column density through an individual fragment
$N_{\text{cloudlet}}\sim10^{17}\,\text{cm}^{-3}$ is essentially
independent of environment; this column density represents a
characteristic scale for atomic gas at $10^4$\,K.  We therefore
suggest that ``clouds'' of cold, atomic gas may in fact have the
structure of a mist or a fog, composed of tiny fragments dispersed
throughout the ambient medium.  We show that this scale emerges in
hydrodynamic simulations, and that the corresponding increase in the
surface area may imply rapid entrainment of cold gas.  We also apply
it to a number of observational puzzles, including the large covering
fraction of diffuse gas in galaxy halos, the broad line widths seen in
quasar and AGN spectra, and the entrainment of cold gas in galactic
winds.  While our simulations make a number of assumptions and thus
have associated uncertainties, we show that this characteristic scale
is consistent with a number of observations, across a wide range of
astrophysical environments.  We discuss future steps for testing,
improving, and extending our model.
\end{onecolabstract}
\vspace*{2\baselineskip}
]
\saythanks

\section{Introduction}
\label{sec:intro}
One of the unique features of astrophysical plasmas relative to
terrestrial fluids is the presence of \textit{multiphase gas}, which
spans a wide range of density and temperature.  Whereas local density
fluctuations of $\lesssim{1}\%$ might be typical for the air in our
atmosphere, cooling in astrophysics can lead to gas spanning orders of
magnitude in density.  Multiphase gas occurs everywhere from
cosmological scales such as the circumgalactic medium permeating
galaxy halos on $\sim100$\,kpc scales, to comparatively tiny scales
such as the interstellar medium in galaxy disks on $\sim10$\,pc
scales, and all the way down to compact objects such as the
broad-absorption line regions around accreting massive black holes on
scales of $\sim(0.01$\,--\,$0.1)$\,pc.  The dynamics of multiphase gas
thus enters into the evolution of a wide range of cosmic structures
and its formation and evolution are important topics of study.

This paper focuses on optically-thin multiphase gas, especially in the
temperature range between $10^4$\,K and $10^6$\,K, where radiation by
atomic lines can rapidly cool the gas.  We focus our attention on
long-lived clouds, which are confined by external pressure or by
magnetic fields.  We do not consider transient, unconfined clouds
(such as fluctuations due to supersonic turbulence), or
self-gravitating gas (\eg, in the intergalactic medium or in giant
molecular clouds and star-forming regions).  Though our assumptions
may sound restrictive, we find such conditions are frequently met.  We
discuss several applications of our results in
sections~\ref{sec:results} and~\ref{sec:discussion}.

When modeling the cold component of multiphase gas, it is often
assumed the cold gas is composed of monolithic, contiguous ``clouds.''
This assumption is not always motivated, however, and it may lead to a
number of the persistent inconsistencies related to the cold gas in
galaxies.  In particular:
\begin{compactenum}
\item Absorption line studies of the circumgalactic medium (CGM) on
  large scales in galaxy halos often find a relatively small amount of
  cold gas, with a high density and therefore small total volume, yet
  also with a very high covering fraction (\eg \citealt{Hennawi2015};
  see section~\ref{subsec:results-galaxies} for more references).  In
  other words, a tiny total amount of cold gas somehow manages to span
  the entire galaxy; it's difficult to understand how so little gas
  can be present everywhere.

\item Strongly supersonic turbulence is sometimes needed to explain
  the broad line widths seen in cold gas (see
  section~\ref{subsec:results-lines} for references and examples).
  But such motions are difficult to understand theoretically because
  supersonic turbulence should shock, rapidly either dissipating the
  motions, or heating the gas above line-emitting temperatures.
  Outside of dense star-forming regions, with rapid cooling and
  continual stirring, strongly suprathermal line widths are hard to
  explain.

\item Galactic winds often contain entrained cold gas moving at high
  velocity.  This is not a priori expected, however, because the
  timescale for a wind to destroy cold gas via shear instabilities is
  typically much shorter than the timescale to accelerate the cold gas
  via drag forces.  Though observations of such co-moving multiphase
  gas are quite common, they remain unexplained (this is the
  ``entrainment in trouble'' problem; \citealt{Zhang2015}).

\item High-resolution observations of atomic clouds in our own galaxy
  halo, such as high-velocity clouds (HVCs), indicate that the clouds
  are made up of many small fragments or filaments with scales less
  than 0.1\,pc (and possibly much smaller).  At least in regions which
  are not self-gravitating, no explanation for this small lengthscale
  currently exists.

\item The broad-line (BLR) and broad-absorption line (BAL) regions
  around quasars exhibit elements of \textit{all} of the above
  problems.  These regions are made of gas in close proximity
  ($\sim0.01$\,--\,$0.1$\,pc) to the quasar: this gas orbits with
  relativistic velocities ($\gtrsim0.1{}c$) and experiences an intense
  field of ionizing radiation.  Even under such extreme conditions,
  some fraction of the gas remains cold and neutral, creating the
  low-energy atomic lines for which these regions are named.  Several
  lines of evidence suggest these lines come from a large number of
  dense, small cloudlets with a tiny overall volume-filling fraction
  (see \S~\ref{subsec:results-bals} for references and details).  To
  our knowledge, no theoretical explanation currently exists for this
  tiny size, or for the enormous number of cloudlets required to match
  the measured line profiles.
\end{compactenum}
We discuss each of these problems in more detail in
section~\ref{sec:results}, where we suggest that all of them stem at
least in part from an assumption that cold gas comes in the form of
monolithic, contiguous clouds.

In this paper, we show that gas clouds larger than a characteristic
scale $\rcloudlet\sim\cstcool\sim(0.1\,\text{pc})/n$ are prone to
fragmentation, where $c_{\text{s}}$ is the sound speed for cold gas,
$t_{\text{cool}}$ is the cooling time, and $n$ is its volume density
in $\text{cm}^{-3}$ (see section~\ref{sec:theory} for details).  For
large clouds with an initial size $R_0\gg{}\rcloudlet$, this
fragmentation is so rapid that we refer to it as \textit{shattering}.

\begin{figure}
  \centering
  \includegraphics[width=\linewidth]{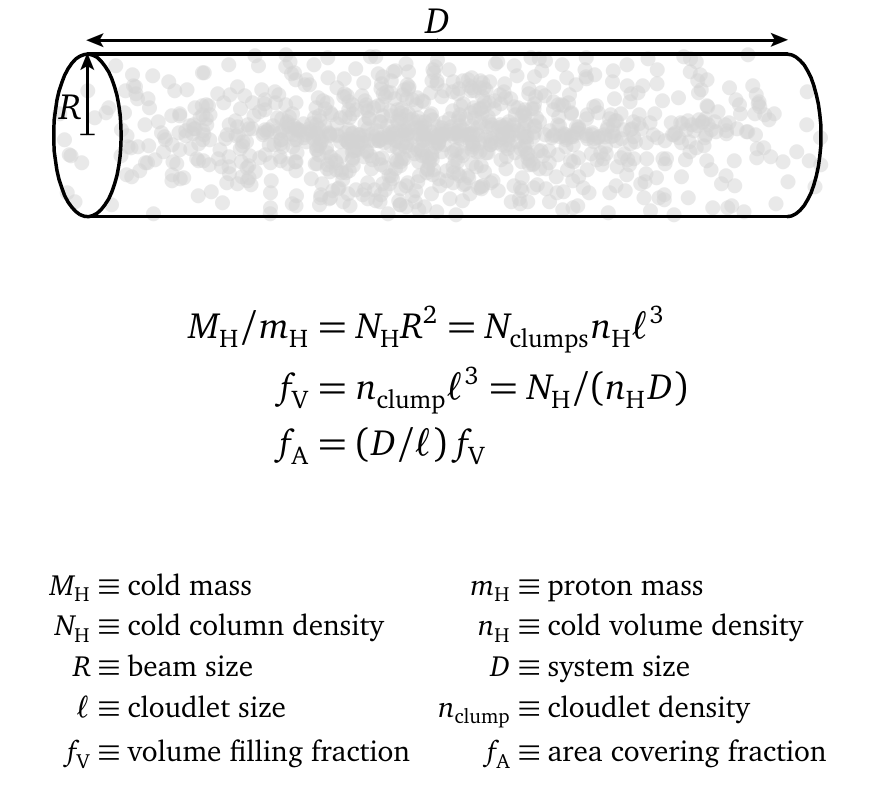}
  \caption{Sketch of a line of sight through a galaxy with virial
    radius $D$ and a pixel size $R$.  Shows the relations among
    observable quantities.\label{fig:diagram}}
\end{figure}
In all of the examples mentioned above -- CGM, HVCs, BALs, and
galactic winds -- we find that the scale \rcloudlet is orders of
magnitudes smaller than the size of the system.  Hence, we propose
that `clouds' of cold gas in these systems should be modeled in a
manner analogous to meteorological \textit{clouds}: they are composed
of tiny droplets (or `cloudlets') of dense gas distributed sparsely
throughout space, with a low overall volume-filling fraction.
Physically, dividing gas into many tiny cloudlets decouples its
area-covering fraction from the volume-filling fraction (or total
mass) of the gas: the ratio of the area-covering fraction
$f_{\text{A}}$ to the volume-filling fraction $f_{\text{V}}$
becomes:\footnote{We define the area-covering fraction $f_{\text{A}}$
  somewhat crudely as the mean number of cloudlets intercepted along a
  line of sight through the multiphase gas.  Thus, $f_{\text{A}}$ can
  be $>1$, and often $f_{\text{A}} \gg 1$.  This is related to, but
  not identical to, observable quantities such as the statistics of
  sightlines covered as a function of column density.}
\begin{align*}
  \frac{f_{\text{A}}}{f_{\text{V}}} \sim \frac{D}{\rcloudlet}
\end{align*}
where \rcloudlet is the size of an individual cloudlet and $D$ is the
size of the system of cloudlets (we sketch this geometry in
figure~\ref{fig:diagram}).  We show in section~\ref{sec:results} that
this ratio $D/\rcloudlet$ may be very large in astrophysical systems.

This paper is organized as follows: in section~\ref{sec:theory}, we
discuss the physics of shattering and our interpretation of the scale
\rcloudlet.  We also show simulations demonstrating how this scale
emerges in numerical simulations.  In section~\ref{sec:results}, we
discuss some observational evidence for shattering: we show how a
large area to volume ratio $f_{\text{A}}/f_{\text{V}}\gg{}1$ is
precisely what quasar-absorption studies of the circumgalactic medium
in galaxy halos imply (\S~\ref{subsec:results-galaxies}).  We also
find evidence for shattering in the broad, turbulent line-widths
sometimes found in the interstellar and circumgalactic media
(\S~\ref{subsec:results-lines}), as well as in the broad-line regions
around quasars, some aspects of which seem to imply very small clouds
(\S~\ref{subsec:results-bals}), and in high-resolution observations of
HVCs and other milky way clouds (\S~\ref{subsec:results-hvcs}).  We
also suggest that a large ratio $f_{\text{A}}/f_{\text{V}}\gg{}1$ may
explain the observed entrainment of cold gas in galaxy winds
(\S~\ref{subsec:results-winds}).  We summarize our model in
section~\ref{sec:discussion} and discuss more speculative
applications, such as the rapid evolution of the escape fraction
during the epoch of reionization and the observation that the milky
way has a $\sim10\times$ lower CGM mass than other comparable galaxies
(\S~\ref{subsec:disc-future}).  We discuss our assumptions and
theoretical uncertainties in this section
(\S~\ref{subsec:disc-uncertain}), as well as directions for further
work (\S~\ref{subsec:disc-future}).

\section{The Physics of Shattering}
\label{sec:theory}
In this section, we discuss physical reasons why we might expect cold
gas in astrophysics to sometimes come in the form of tiny cloudlets
dispersed throughout a much larger volume.  We begin with a
qualitative discussion in section~\ref{subsec:hand-wavy}, then
demonstrate shattering using preliminary numerical simulations in
section~\ref{subsec:sims}.  While we note that our numerical results
are somewhat uncertain, we compare our results with a range of
observations in section~\ref{sec:results} and with recent laboratory
experiments in section~\ref{sec:discussion}.

\subsection{Intuitive Discussion}
\label{subsec:hand-wavy}
\begin{figure}
  \centering
  \includegraphics[width=\columnwidth]{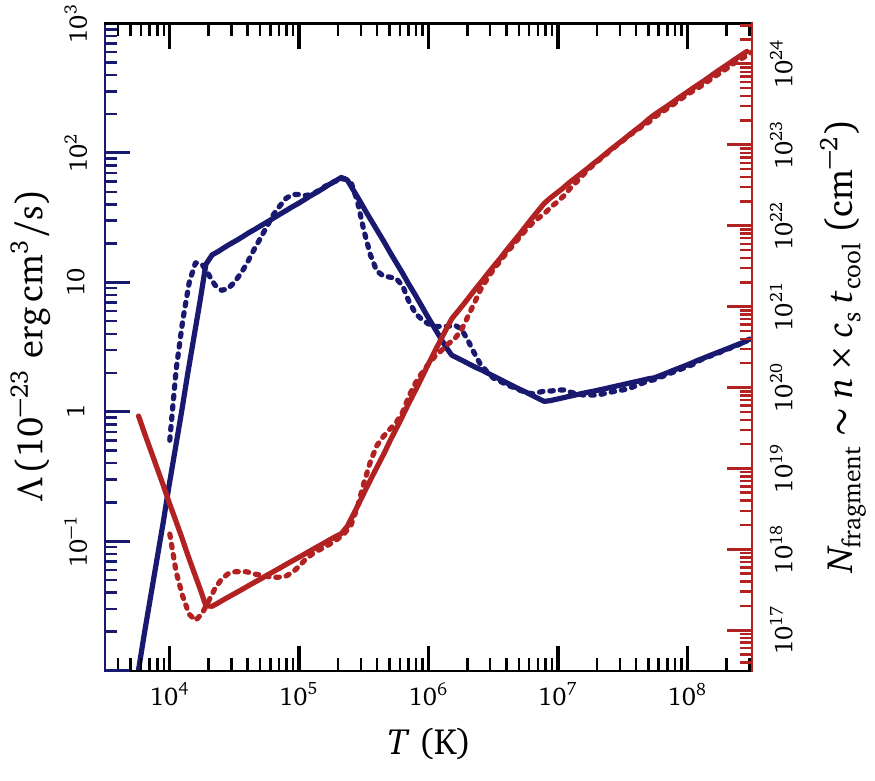}
  \caption{(\textit{blue}): Cooling curve for collisionally-ionized,
    solar-metallicity gas.  The dotted curve shows the calculation
    from \citet{Sutherland1993}, and the solid curve shows a
    simplified piecewise power-law fit used in our simulations
    (\S~\ref{subsec:sims}).  Gas below $\sim10^6$\,K cools rapidly
    down to $\sim10^4$\,K.  Such rapid cooling fragments the gas,
    implying a small fragment size, measured here by the column
    density through an individual fragment,
    $N_{\text{cloudlet}}\sim{n}\times\cstcool$, shown in the
    \textit{red} curve (see equation~\ref{eq:rcloud} in
    section~\ref{subsec:hand-wavy}).  As before, the dotted line is
    calculated from \citet{Sutherland1993} and the solid line is
    calculated from the simplified piecewise power-law fit used in our
    simulations.}\label{fig:cooling-curve}
\end{figure}
\begin{figure*}
  \centering
  \includegraphics[width=\textwidth]{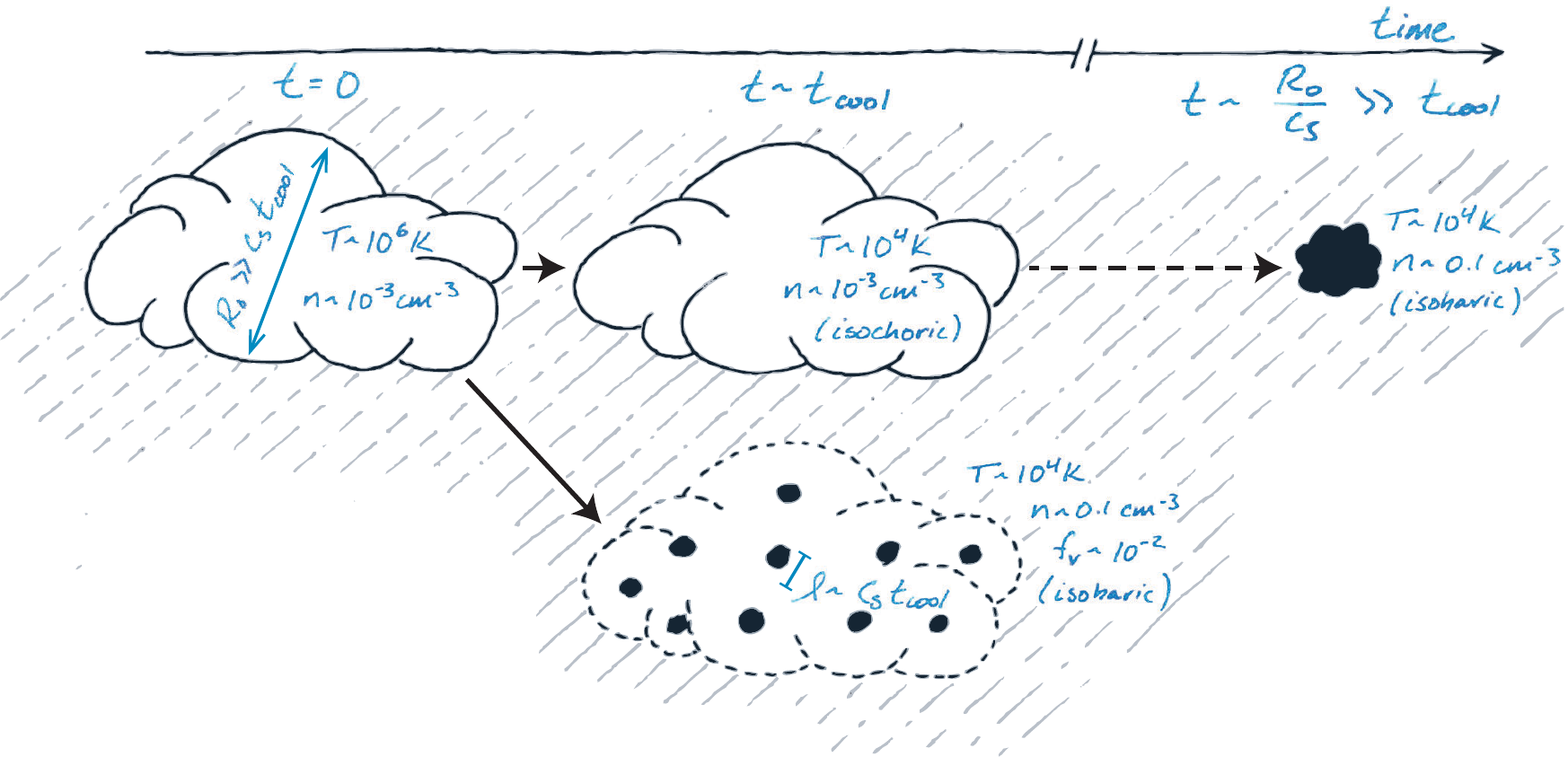}
  \caption{The fastest path to equilibrium is to shatter.  We imagine
    a cooling perturbation with an initial size $R_0\gg\cstcool$
    embedded within an ambient medium, represented by gray hatching.
    (\textit{top route}): It is commonly assumed
    \citep[\eg][]{Burkert2000,Field1965} that such a perturbation will
    cool isochorically, reaching $10^4$\,K after one cooling time,
    with little change in the cloud size or density.  The resulting
    cloud is severely under-pressured and out of equilibrium; it
    contracts on the (much larger) sound-crossing timescale, only
    afterwards reaching pressure equilibrium.  (\textit{bottom
      route}): A much more direct path to pressure equilibrium is for
    the cloud to fragment into smaller pieces, each with a scale
    $\sim\cstcool$; such fragments can cool isobarically, never
    leaving equilibrium.  (Not shown:) As these fragments cool, the
    lengthscale $\cstcool$ shrinks; when this happens the clumps
    fragment to yet smaller scales, never deviating dramatically from
    pressure equilibrium.}\label{fig:shattering-schematic}
\end{figure*}
The processes creating cold gas likely vary with application: the cool
CGM in galaxy halos, for example, may have an entirely different
physical origin than the atomic gas in the broad-line regions around a
quasar.  We expect our results apply to the dynamics of cold gas in
general, regardless of its exact (and often unknown) origin.  For
concreteness, however, in this section we imagine gas cooling out of a
hot, ambient medium, as might be produced by thermal instability
\citep[\eg][]{Field1965}.  We focus on the final non-linear outcome of
thermal instability, which is not directly predicted from the linear
analysis in Field's classic work.  Figure~\ref{fig:cooling-curve}
shows the cooling curve (\textit{blue}) for
collisionally-ionized,\footnote{We discuss the possible impact of
  photoionization in \S\ref{subsec:disc-uncertain}.}
solar-metallicity gas computed by \citet{Sutherland1993}.  When gas
cools below $\sim10^6$\,K, line emission can rapidly cool the gas down
to $\sim10^4$\,K.  Such rapid cooling between $10^6$\,K and $10^4$\,K
naturally suggests a two-phase medium, consisting of a hot phase with
$T\gtrsim10^6$\,K alongside a cold phase with $T\lesssim10^4$\,K.
This is especially true if some heating process is present to offset
the (slow) cooling in the hot phase and maintain it indefinitely.

Figure~\ref{fig:shattering-schematic} illustrates some gas cooling out
of a hot, pressure-supported background.  If the cooling perturbation
has an initial size $R_0\gg\cstcool$, where $c_{\text{s}}$ is the
sound speed (at which pressure forces are communicated throughout the
cloud), and $t_{\text{cool}}$ is the cooling time (the timescale for
the cloud to radiate away its energy), then the perturbation cannot
contract as quickly as it loses its pressure support.  It is commonly
assumed \citep[\eg][]{Burkert2000,Field1965} that such perturbations
cool \textit{isochorically}, i.\,e.\ at constant density, with little
change to the size or shape of the cloud.  The cloud is then severely
out of pressure balance, but contracts only on longer timescales,
reaching pressure equilibrium after the (far longer) sound-crossing
time.  We sketch this process in the top path of
figure~\ref{fig:shattering-schematic}.  However, such a configuration
is vastly out of force balance and is therefore extremely unstable.
The stable state is for cooling gas to reach $\sim10^4$\,K, and to be
in (at least approximate) pressure balance with its surroundings; we
emphasize that for initially large clouds, the isochoric route
represents an extremely slow path to equilibrium.

The fastest route to equilibrium is for the gas to split into smaller
pieces, each of which contracts more rapidly.\footnote{Though we have
  not specified the dynamics leading to this splitting, it would
  ultimately stem from compression by the ambient hot medium.  We
  expect the unbalanced pressure serves as a source of free energy to
  drive fluid motions or instabilities on a wide range of scales, even
  if the detailed processes are not yet identified.}  If the gas
breaks into many separate pieces, each with a scale $\sim\cstcool$,
such pieces are small enough to cool isobarically, essentially never
leaving pressure balance with their surroundings.  These fragments
contract as they radiate away their energy, but since they are out of
sonic contact with one another, each piece cools independently and
they pull away from each other as they contract.  The result is
something analogous to a mist or a fog, in which dense, small
cloudlets are spread throughout the volume and a hot, interstitial
medium permeates the space between the cloudlets.  We sketch this
process in the bottom path of figure~\ref{fig:shattering-schematic}:
here, the cloud reaches equilibrium after only a cooling time,
\textit{far} faster than implied by the isochoric route discussed
above.

For gas above $10^4$\,K, we note that the lengthscale \cstcool
plummets $\sim{T^3}$ as the gas cools (see the red curve in
figure~\ref{fig:cooling-curve}).  Each of the fragments in the bottom
picture of figure~\ref{fig:shattering-schematic} may therefore
fragment again to yet smaller scales, and again, repeatedly, until the
gas reaches the stable temperature near $10^4$\,K.  This splitting
therefore pushes to smaller scales as the gas cools, eventually
reaching the scale where \cstcool is minimized.  Putting in a
characteristic density, we find:
\begin{align}
  \rcloudlet \sim \text{min}\left(\cstcool\right) \sim
  (0.1\,\text{pc})\left(\frac{n}{\text{cm}^{-3}}\right)^{-1},\label{eq:rcloud}
\end{align}
where we have used the \citet{Sutherland1993} cooling curve, as shown
in figure~\ref{fig:cooling-curve}.  Since the cloudlet scale
$\rcloudlet\propto{1/n}$, the column density through an individual
cloudlet
\begin{equation}
N_{\text{cloudlet}}=n\,\rcloudlet\sim10^{17}\,\text{cm}^{-2}
\end{equation}
is essentially independent of the surrounding environment; we will
apply this useful result later in section~\ref{sec:results}.  We note
that this discussion is very similar to that in \citet{Voit1990}, who
arrive at an identical minimum column density $N_{\text{cloudlet}}$.

While we evaluate equation~\ref{eq:rcloud} at the temperature where
\cstcool is minimized, we note that one or more secondary processes
may also influence the cloudlet size (such as shock-heating; see
figure~\ref{fig:fragmentation-1e6} in Appendix~\ref{sec:sim-method}).
Despite a factor of $\lesssim10$ uncertainty in the shattering scale
for cold gas, it is orders of magnitude smaller than the typical sizes
for systems of cold gas, and below the scales usually probed in
simulations or in observations.  This is all that is needed for our
basic conclusions.

The fragmentation we discuss here is closely analogous to the Jeans
instability, which describes the collapse of pressure-supported gas
due to self-gravity \citep{Jeans1901,Low1976}.  In this case, clouds
smaller than a characteristic lengthscale:
\begin{align*}
  \lambda_{\text{J}} \sim c_{\text{s}} t_{\text{dyn}}
  \sim \frac{c_{\text{s}}}{\sqrt{G\rho}}
\end{align*}
can communicate pressure forces on a timescale faster than the
free-fall time; such clouds therefore establish hydrostatic
equilibrium, in which forces due to pressure balance those due to
gravity.  Small clouds resist collapse in this way, instead undergoing
stable oscillations about an equilibrium configuration.  Clouds larger
than this scale cannot communicate pressure forces rapidly enough to
establish hydrostatic equilibrium, however; with no way to balance
gravity, large clouds are therefore susceptible to collapse.  Since
regions separated by distances $\gtrsim\lambda_{\text{J}}$ are out of
causal contact as they collapse, each contracts independently,
fragmenting the cloud into distinct chunks of mass
$M_{\text{J}}\sim\rho\lambda_{\text{J}}^3$.  The ``shattering''
process we sketch in figure~\ref{fig:shattering-schematic} is similar
to the Jeans instability, with contraction driven by cooling and
external pressure, rather than self-gravity; the fragmentation driven
by these two processes is otherwise substantially identical.

This analogy with the Jeans instability extends further: in isothermal
gas, relevant for dense star-forming regions, the Jeans mass
$M_{\text{J}}\sim\rho\lambda_{\text{J}}^3\propto\rho^{-1/2}$ decreases
as the gas contracts to higher density.  Gas therefore continually
fragments to smaller scales as it collapses.  The final core size is
determined by the scale at which the gas becomes optically thick, such
that its contraction is no longer isothermal; above this density, the
Jeans mass increases with density and the gas becomes stable to
further fragmentation.  Physically, this transition represents the
smallest possible scale for $\lambda_{\text{J}}(\rho, T[\rho])$, and
results in core masses
$\sim{}M_{\text{J}}^{\text{(min)}}\ll{}1\,M_{\odot}$, rather than the
initial value $\sim{}M_{\text{J}}^{\text{(init)}}(1\,\text{cm}^{-3},
100\,\text{K}){}\sim{}10^5\,M_{\odot}$ characteristic of the ISM.  The
fragmentation cascade we describe in this paper is qualitatively
similar to Jeans' fragmentation; in our picture, pressure-confined gas
follows the scale $\cstcool(T)\sim{T^3}$ as it decreases along the
cooling curve, ultimately reaching a stable, minimum scale near
$10^4$\,K.  We therefore expect the relevant scale for cold gas is the
\textit{minimum} value of \cstcool, not simply its initial value.

This point highlights an essential difference between our scenario and
the theories presented by \citet{Burkert2000} and by
\citet{Hennebelle1999}, who also identify a scale $\sim\cstcool$ in
multiphase gas.  These theories are conceptually somewhat different,
however, and focus on the early development of thermal instability
producing the cold gas.  These theories therefore evaluate \cstcool in
the hot medium, \textit{before} the gas cools, and assume that large
perturbations cool isochorically, before being compressed
monolithically by their surroundings, such that
$l_{\text{cold}}\sim(\cstcool)^{\text{(hot)}}\times\delta^{-1/3}$,
where $\delta\equiv\rho_{\text{cool}}/\rho_{\text{hot}}$ is the volume
contraction ratio.  We instead suggest that cooling drives continued
fragmentation, and therefore that \cstcool should be evaluated in the
cold phase, near the temperature where this scale is minimized.  By
analogy with Jeans' instability, if the gas cools down to
$\sim10^4$\,K, we expect it will fragment to this minimum scale.
Thus, the scale \rcloudlet can be significantly smaller than the
scales identified by \citet{Burkert2000} and \citet{Hennebelle1999}.
In the following section, we show how this small physical lengthscale
appears in high-resolution numerical simulations with rapid cooling.
We furthermore show that the scale \rcloudlet does not depend on the
initial temperature of the cooling gas; this is consistent with the
repeated fragmentation discussed above, but not with the scales
predicted by \citet{Burkert2000} and by \citet{Hennebelle1999}.

\subsection{Simulation Results}
\label{subsec:sims}
Two common methods for numerically studying multiphase gas in
astrophysics are to simulate thermal instability, which exponentially
amplifies small temperature fluctuations and thus \textit{produces}
multiphase gas, or ``cloud-crushing,'' which studies the interaction
between hot and cold gas phases and therefore probes the detailed
evolution of multiphase gas after it forms.  It is well known that the
equations of adiabatic gas dynamics are scale-free; as a result,
simulations of, \eg, the Kelvin-Helmholtz instability may be equally
applied to weather systems on Earth or to gas stripping in galaxy
mergers.  Cooling breaks this self-similarity, however, and imprints a
scale: evolution on timescales shorter than the cooling time may be
radically different from evolution on longer timescales.
Equivalently, the evolution of small objects with a scale
$R\ll{\cstcool}$ differs from the evolution of larger objects.
Astrophysics simulations typically have somewhat limited resolution,
spanning a dynamic range of $\lesssim100$ between the minimum resolved
scale ($\gtrsim10$ resolution elements) and the size of the
computational domain.\footnote{Deeply nested AMR simulations such as
  those used to study star formation and cosmological structure
  formation can cover a fantastic range of scales, $\gtrsim10^6$.
  However, this technique does not lend itself to the fog-like
  distribution of cold gas discussed here, with rapidly-moving
  cloudlets widely distributed throughout the domain.} The domain is
usually scaled to match the overall physical region of interest,
limiting our ability to study small-scale evolution.  In light of the
fragmentation discussed above, however, here we revisit simulations of
thermal instability and cloud-crushing, now with an eye toward
small-scale evolution.  In particular, we design our simulations to
capture the tiny scale $\sim\cstcool\sim(0.1\,\text{pc})/n$ as
indicated in equation~\ref{eq:rcloud}.

We note that cloud crushing and the dynamics of multiphase gas are
thorny problems to study numerically, and that we have not controlled
for several numeric errors which may be present in our simulations
(see Appendix~\ref{sec:sim-problems} and
section~\ref{subsec:disc-uncertain} for a discussion).  Our
simulations presented here are only meant to be suggestive, and must
be followed with a more detailed study.

\subsubsection{Thermal Instability and Forming Multiphase Gas}
\label{subsubsec:ti-sims}
\begin{figure}
  \centering
  \includegraphics[width=\linewidth]{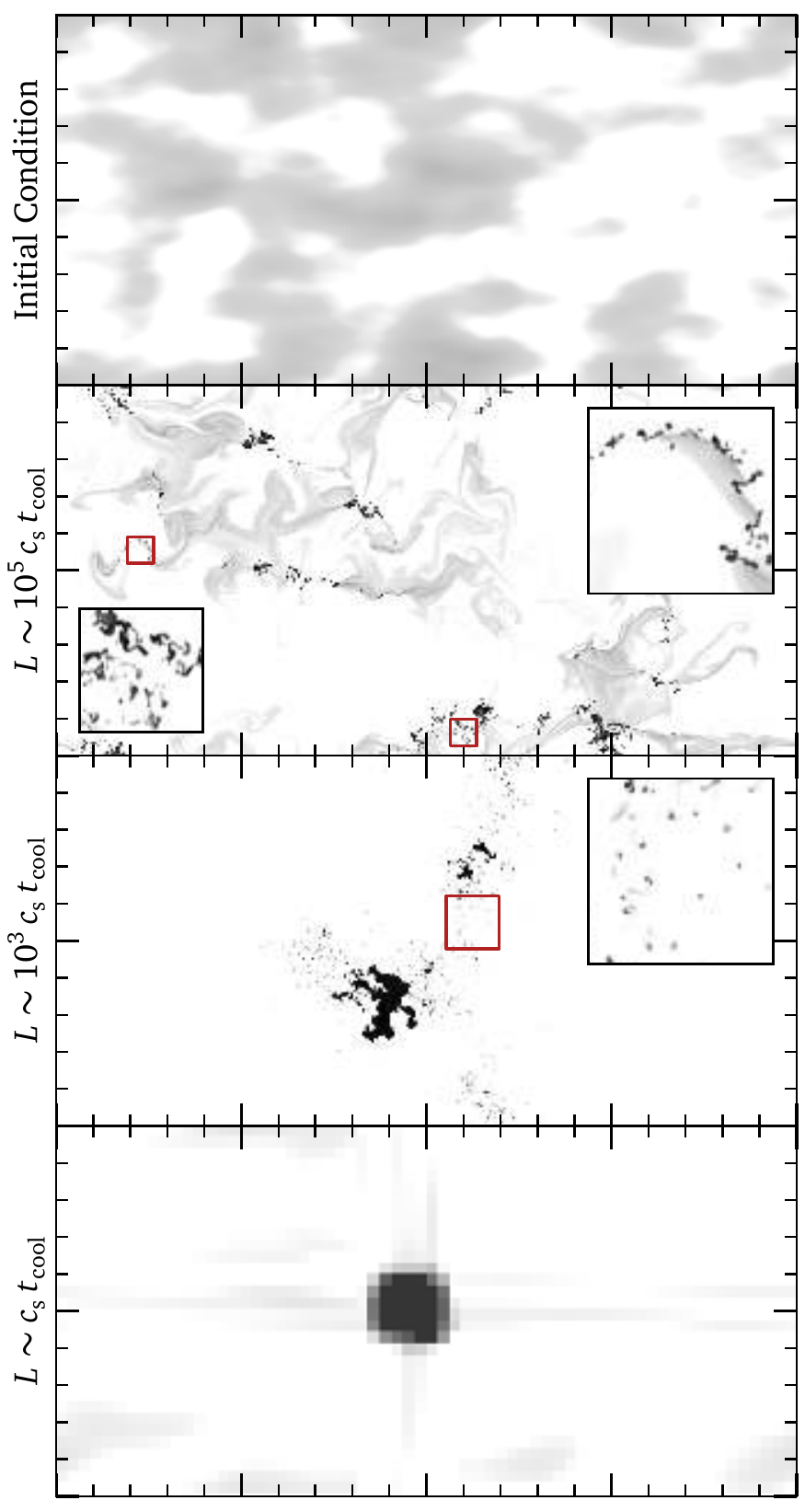}
  \caption{Cold clouds likely form small in the first place.  These
    simulations show an initial perturbation (\textit{top panel})
    cooling down to $10^4$\,K.  When the domain size is large
    ($\sim$\,kpc; \textit{second panel}), the perturbation shatters
    into tiny fragments with a characteristic scale
    $c_{\text{s}}{}t_{\text{cool}}$, as discussed in
    section~\ref{subsec:hand-wavy}.  When the domain size is small, of
    order $\sim{}c_{\text{s}}{}t_{\text{cool}}$ (\textit{bottom
      panel}), we see no such fragmentation.  Instead, the cloud
    assembles into a monolithic blob.  As argued in the text, these
    simulations suggest that \cstcool is a characteristic lengthscale
    for cold gas: perturbations much larger than this scale fragment
    into much smaller pieces, while perturbations smaller than this
    lengthscale evolve coherently as effectively single objects.  The
    third panel shows an intermediate case in which the cloud shatters
    and then re-forms as the cloudlets coagulate.  This result may be
    unrealistic, however; we discuss it further in
    section~\ref{subsubsec:ti-sims}.}\label{fig:cloud-plot}
  \vspace*{\fill}
\end{figure}
Figure~\ref{fig:cloud-plot} shows a perturbation cooling out of a hot
medium, analogous to the scenario sketched in
figure~\ref{fig:shattering-schematic}.  The top panel shows our
initial condition.  White shows an ambient, $\sim10^{7}$\,K
background, and gray fluff shows an isobaric perturbation at
$\sim10^{6}$\,K which will cool down to $\sim10^4$\,K.\footnote{We
  start with a large density perturbation so as to get a distinct
  cloud which will cool without needing thermal instability to grow it
  from small amplitude; the physics of thermal instability is
  application-dependent and can be sensitive to assumptions about,
  \eg, stratification and background heating (see, \eg
  \citealt{Field1969} for a theory relevant to the interstellar
  medium, or \citealt{McCourt2012} for a model relevant to halo gas in
  massive galaxies and galaxy clusters).  While important, the
  question of where cold gas comes from is beyond the scope of this
  paper.}  These are 2D hydro simulations evolved using the
second-order, Godunov-type code \textit{Athena}
\citep{Stone2008,Gardiner2008}.  We use the \citet{Sutherland1993}
cooling curve, implemented via the \citet{Townsend2009} algorithm, and
we run the simulations long enough for the $\sim10^6$\,K gas to cool,
but not long enough for the $\sim10^7$\,K gas to cool appreciably.
With this choice of parameters, we have a well-defined hot phase which
remains hot, and we avoid any need for ``feedback'' heating to stave
off a cooling catastrophe.  We detail our computational setup in
Appendix~\ref{sec:sim-method}.

When we make the domain size large compared to \cstcool, as shown in
the second panel of figure~\ref{fig:cloud-plot}, the perturbation
shatters into tiny fragments.  Here, the simulation domain is full of
tiny dots of cold gas.  In order to emphasize this point, the red
squares mark apparently diffuse regions of the domain.  Zooming in on
these regions, shown in the insets, reveals that they are full of
cloudlets with a low volume filling fraction and a tiny characteristic
size which is unresolved in this simulation.  If these cloudlets
eventually disperse (\eg, due to turbulence), the result would be
qualitatively very similar to the ``fog'' of cloudlets discussed in
the previous section.

When we make the domain size much smaller, comparable to the expected
cloudlet size $\rcloudlet\sim\cstcool$, as shown in the bottom panel,
we see no such fragmentation.  In this case, the entire perturbation
contracts as one coherent piece, and the gas assembles into a single,
monolithic cold mass.  The simulations in figure~\ref{fig:cloud-plot}
therefore suggest that cold gas \textit{forms} as a collection of
distributed small chunks, each with a characteristic scale
$\sim\cstcool$: perturbations much larger than this lengthscale
shatter into smaller pieces, while perturbations smaller than this
typical lengthscale maintain their integrity and behave as single
objects as they cool (see also the discussion of
figure~\ref{fig:fragmentation}).

We show an intermediate case in the third panel of
figure~\ref{fig:cloud-plot}.  Here, the perturbation initially cools
and shatters into tiny fragments (some of which are shown in the
inset).  But mixing between the hot and cold phases cools the
interstitial hot gas, lowering its pressure and driving the cloudlets
together.  In this simulation, the cold cloudlets will all eventually
collect into a single blob.  We note that this behavior depends on
mixing between the hot and cold phases, however, which is not resolved
in this simulation.  Moreover, this coagulation might easily be
suppressed by turbulent motions, which are expected in a realistic
astrophysical environment, but neglected in our simulations.  Hence,
we suspect this coagulation is as artifact of our simplified setup; we
defer a more detailed consideration of this intermediate case to a
future study.

\subsubsection{Cloud ``Crushing'' and Evolution of Multiphase Gas}
\label{subsubsec:crushing-sims}
\begin{figure*}
  \centering
  \includegraphics[width=\linewidth]{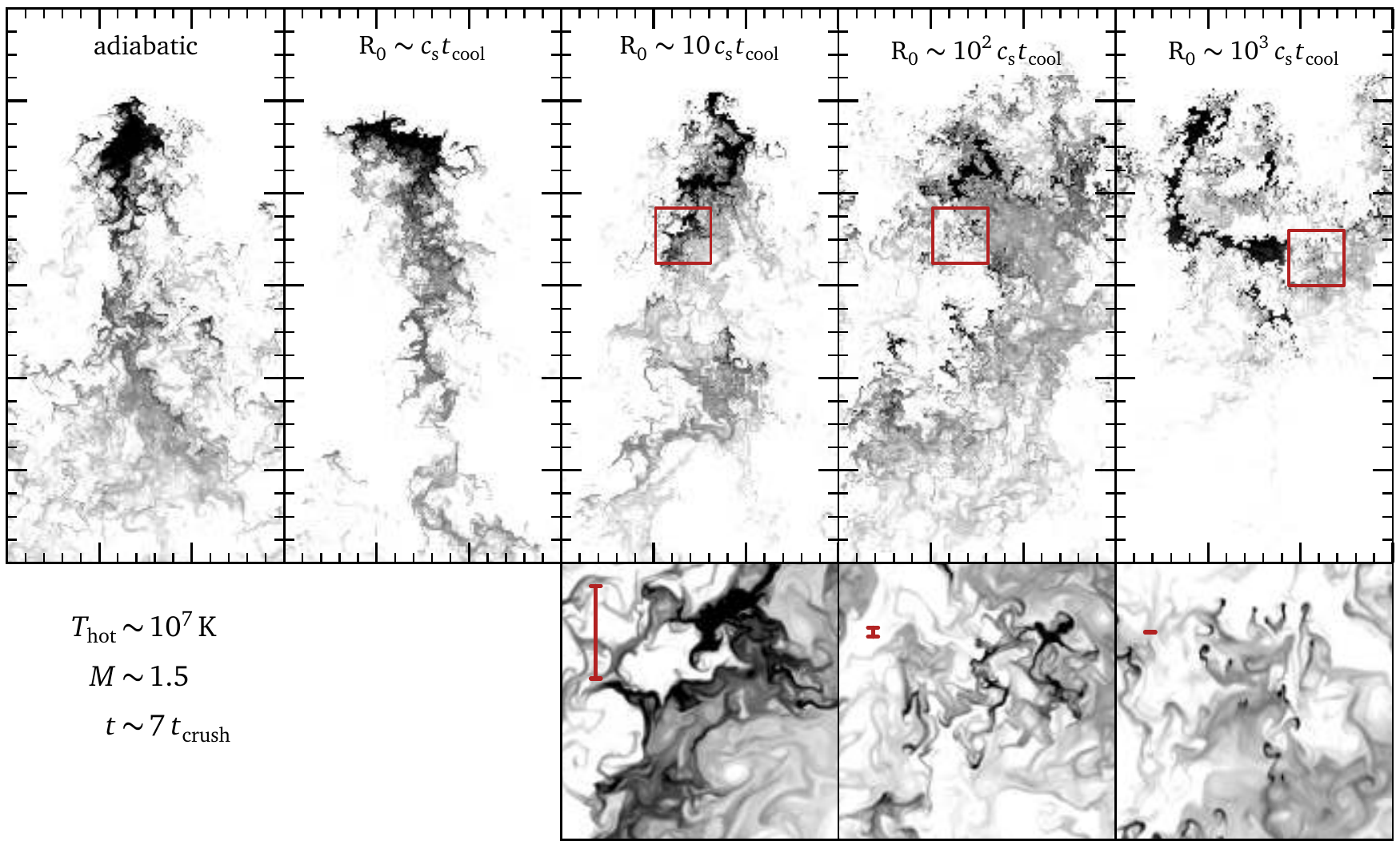}

  \caption{Clouds with initial sizes $R_0$ much greater than the
    characteristic size \cstcool easily fragment into smaller pieces.
    Each panel shows a 2D hydro simulation of an initially cold
    ($\sim10^{4}\,K$) cloud moving through an ambient medium at
    $\sim10^{7}$\,K.  We resolve the initial cloud radius with
    300\,cells; see figure~\ref{fig:res-1e2} for a resolution study.
    The initial size of the cloud decreases by a factor of 10 in each
    panel moving to the left.  The rightmost panel shows a cloud with
    an initial size $10^{3}\times{}\cstcool$; this would be a
    $\lesssim$\,kpc-sized cloud near the disk of the milky way, or out
    in the halo of a quasar host at redshift $z\sim{2}$.  The bottom
    panel, when shown, is a zoom-in of the red box in the top panel.
    Red error bars make the size \cstcool; when they are resolved by
    the simulation, the cloudlets roughly track this lengthscale.  We
    detail the simulation setup in Appendix~\ref{sec:sim-method}.
    These simulations are 2D and neglect both magnetic fields and
    viscosity; they cannot adequately model the disruption of the
    clouds.  Nonetheless, they do show that cooling breaks
    self-similarity in the cloud-crushing process and introduces a
    characteristic scale for the cold gas.
    \label{fig:fragmentation}}
\end{figure*}
\begin{figure*}
  \includegraphics[width=\textwidth]{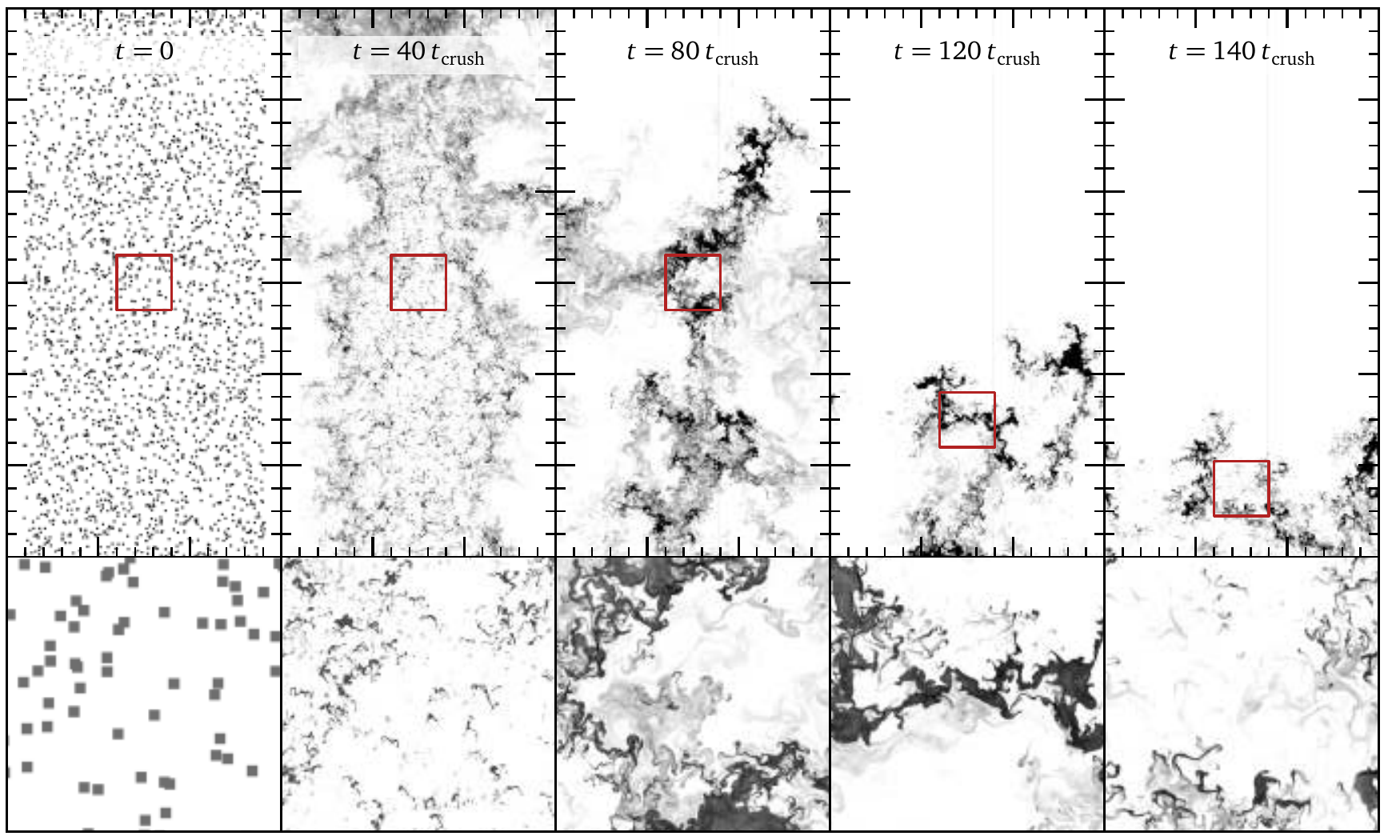}

  \vspace*{2\baselineskip}

  %% nb. barfs on the _opt version.  something weird with ghostscript?
  \includegraphics[width=\textwidth]{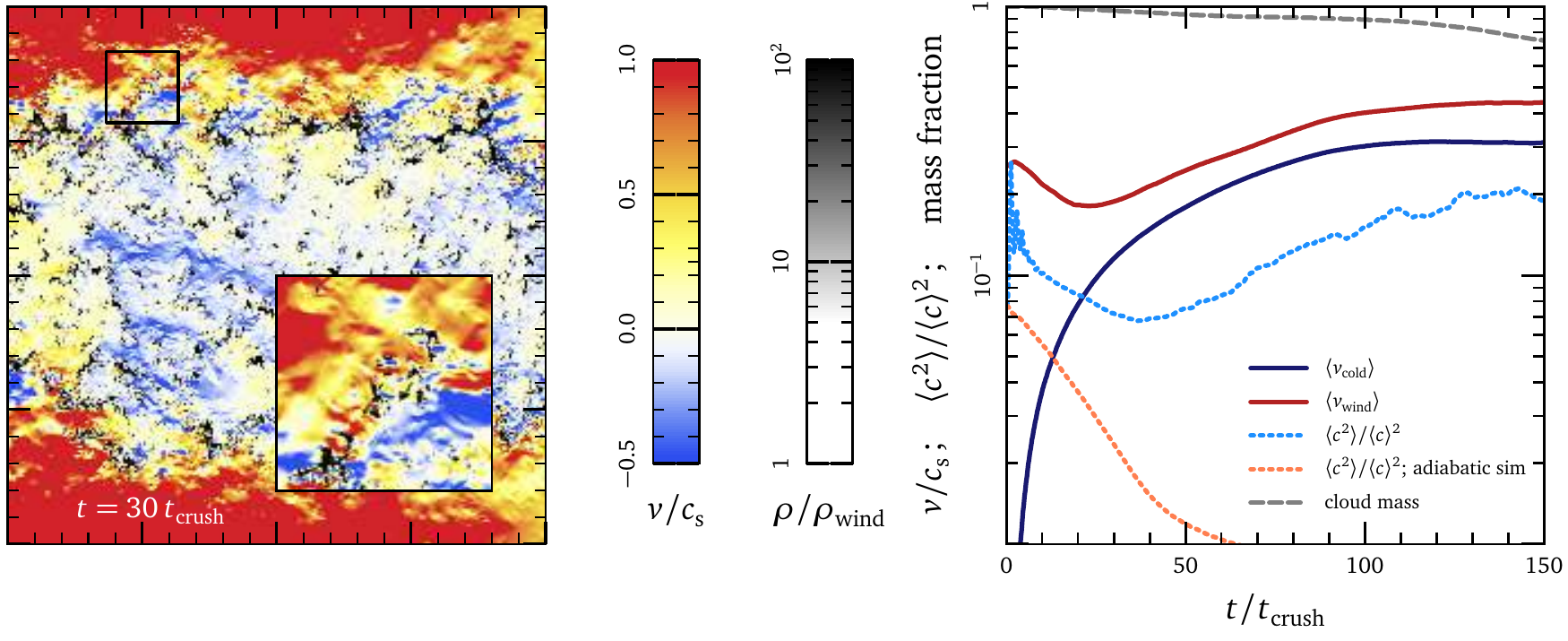}
  \caption{Entrainment of a system of cloudlets.  (\textit{top plot}):
    This figure shows a 2D simulation of a large number of cloudlets
    being hit with a wind coming from the top of the page.  The
    simulation has an aspect ratio of 4:1; here we show the middle
    half.  We resolve each (initially square) cloudlet with $(64)^2$
    cells; this is sufficient for the cloudlets to disrupt due to
    cloud crushing or shear instabilities.  While the cloudlets are
    disturbed by the incoming flow, and many partially disrupt, we
    find that the cold gas does not rapidly mix into the ambient hot
    wind.  The cold gas moves around, coalesces, and breaks apart
    again; however the clumping factor remains relatively high, and
    the cold gas effectively entrains in the ambient wind.
    (\textit{bottom right plot}): Quantitative evolution of the
    simulation.  We demonstrate entrainment by showing the mean
    velocity of cloud (\textit{blue}) and wind (\textit{red}) gas.
    Note that the dynamics are not well-represented by an average
    velocity, however; see the color plot on the left.  We also show
    clumping factors for the above simulation (\textit{blue, dashed};
    note that the figure above shows only half of the simulation
    domain), and for an analogous simulation without cooling
    (\textit{orange, dashed}) computed using passively advected
    scalars.  Clumping factors become inaccurate metrics of cloud
    disruption once an appreciable fraction of cloud material leaves
    the domain (\textit{gray, dashed}). (\textit{bottom left plot}):
    Velocity distribution for wind gas, showing the central 1/4 of the
    domain.  Red shows movement in the mean wind direction; blue shows
    a back-flow which develops within the bulk of the cloudlets.  Most
    of the interstitial flow between cloudlets is sub-sonic.  Cold gas
    may re-assemble due to streaming instabilities, which collect gas
    in converging flows in the wakes of dense knots.  This figure is
    only meant to be suggestive; in particular, it remains to be shown
    that this entrainment works in 3D.  We will investigate these
    dynamics further in a future study.}
  \label{fig:entrainment}
\end{figure*}
The simulations shown in figure~\ref{fig:cloud-plot} suggest that cold
gas initially forms as small fragments within a much larger group,
like the water droplets in a fog.  Even if we somehow start with a
large, monolithic cloud, however, such a cloud may \textit{still}
shatter into much smaller fragments.  We show examples of this
fragmentation in figure~\ref{fig:fragmentation}.  In these 2D
simulations, a sphere of dense gas at $10^4$\,K interacts
trans-sonically with a hot, $10^7$\,K wind with a relative mach number
of $v/c_{\text{s}}^{\text{(hot)}}=1.5$.  (This setup is roughly
appropriate for our intended applications; we show in
Appendix~\ref{sec:sim-method} that we obtain similar results for other
choices of the parameters).  Running these simulations in 2D enables
us to use resolutions of up to $\sim600$\,cells per cloud radius,
significantly higher than the resolutions $\sim$32\,--\,64 cells per
cloud radius typical of state of the art cloud-crushing simulations in
3D \citep[\eg][]{Scannapieco2015}.  We discuss in
Appendix~\ref{sec:sim-problems} why such high resolution is essential
to study shattering.  While the hydrodynamic instabilities which
disrupt the cloud may behave differently in 2D and 3D, we note that
nothing about the shattering discussed in
section~\ref{subsec:hand-wavy} seems fundamentally three-dimensional.
We therefore expect our results to be meaningful, despite being 2D.
We will study shattering using 3D simulations in O'Leary et al.\ in
prep., and we compare our results with published 2D and 3D simulations
in Appendix~\ref{sec:sim-problems}.  As before, we detail the rest of
our computational setup in Appendix~\ref{sec:sim-method}.

When the clouds in figure~\ref{fig:fragmentation} interact with their
surroundings, they disrupt due to a combination of cloud ``crushing''
and shear instabilities \citep[\eg][]{Klein1994}.  However, in
simulations with very high resolution and with strong cooling, we find
that when the cloud disrupts, the cold material doesn't simply mix
into the wind and disappear.  Instead, at least in the setup we
tested, the cloud disrupts by fragmenting into tiny, dense, clumps.
We will show these clumps follow the same characteristic scale
\rcloudlet discussed above.

The rightmost column in figure~\ref{fig:fragmentation} shows a cloud
with an initial size $\sim{}10^{3}\,\cstcool$; this would be a
$\sim$\,kpc-sized cloud near the milky way disk, or out near the
virial radius of a quasar host at redshift $z\sim{}2$.  We chose this
scale because it is comparable to inferred sizes for large HVCs
\citep{Wakker1997}, or to the resolution limit in cosmological zoom-in
simulations (\eg, \citealt{Liang2016}; note that cosmological
simulations have comparatively poor resolution in low-density halo
gas).  When this cloud disrupts, material strips off its surface and
forms a turbulent wake behind the cloud.  We find this wake is filled
with little, unresolved chunks of cold gas: as in
figure~\ref{fig:cloud-plot}, the inset zooms in on the region marked
with the red square, showing that this apparently-diffuse region is
indeed filled with tiny dots of cold gas.  For comparison, the red
error bar in the inset shows the scale \cstcool, which is invisible at
the scale of this figure.

Resolving the cloudlets in this simulation would be extremely
expensive with our computational setup; instead, we study the
small-scale evolution by decreasing the initial cloud size.  This
reduces the dynamic range between the initial cloud and the final
shattering fragments, making it easier to resolve the whole shattering
process.  Moving to the left, the second panel of
figure~\ref{fig:fragmentation} shows a simulation of a $10\times$
smaller cloud.  The shattering fragments are now marginally resolved
(see the convergence study in figure~\ref{fig:res-1e2}), and the
evolution is qualitatively very similar to that shown in the first
panel.  The cloudlets now appear to track the lengthscale
$\sim\cstcool$ marked by the red error bar.  The third panel shows a
cloud smaller by yet another factor of 10.  In this case, the
evolution appears qualitatively different: here, the cloud does not
shatter into tiny pieces.  The cloud still fragments, however, and
features in its wake seem to track the lengthscale \cstcool; we note
that this interpretation is uncertain because there are only a small
number of fragments and because this simulation may lack the dynamic
range needed to capture the shattering discussed in
section~\ref{subsec:hand-wavy}.

Continuing to the left in figure~\ref{fig:fragmentation}, we show a
cloud with an initial size $R_0\lesssim\cstcool$, comparable to the
expected size of the shattering fragments.  Here, we see no
fragmentation into smaller cloudlets.  While the cloud disrupts due to
shear instabilities, its wake does not contain dense knots of gas as
seen in previous panels.  The overall evolution is very similar to the
adiabatic simulation shown in the final (leftmost) panel of
figure~\ref{fig:fragmentation}.  As in figure~\ref{fig:cloud-plot},
this again suggests that the lengthscale $\sim\cstcool$ is a
characteristic scale for cold gas in these simulations: clouds
significantly larger than this scale fragment into much smaller
pieces, while clouds comparable in size to this scale do not.

We note that the simulations shown in the leftmost panels of
figure~\ref{fig:fragmentation} are essentially identical to ``cloud
crushing'' simulations with weak or no cooling
\citep[\eg][]{Klein1994}.  Therefore, the cloud will eventually erode
due to hydro instabilities on a timescale
$\sim(\rho_{\text{cloud}}/\rho_{\text{wind}})^{1/2}{\,}\rcloudlet/v_{\text{wind}}$,
which is extremely short due to the tiny size of the cloudlet.  One
might therefore conclude that shattering does little to influence the
disruption of cold clouds: even if the cold gas passes through a phase
consisting of tiny cloudlets, it will nonetheless rapidly mix into the
ambient surroundings once these cloudlets disrupt.  However, it is
important to realize that the single-cloud simulation shown in the
leftmost panel of figure~\ref{fig:fragmentation} is not likely
representative of a full system of cloudlets.  In a full simulation of
a large cloud, as shown in the rightmost panels of
figure~\ref{fig:fragmentation}, the cloudlets rapidly become co-moving
with the ambient hot gas, and also shield one another from the full
force of the background flow.  These effects substantially reduce the
velocity shear which shreds the isolated cloudlet in the leftmost
panel of figure~\ref{fig:fragmentation}.

We illustrate this shielding in figure~\ref{fig:entrainment}, which
shows a simulation of a large number of cloudlets distributed with a
volume-filling factor $f_{\text{V}}\sim{2\times10^{-2}}$.  Each
(initially square) cloudlet is resolved by $(64)^2$ cells to ensure
they are not artificially stabilized by numerical viscosity and have
sufficient resolution to disrupt.  We report time in units of this
single-cloud crushing timescale.  When we drive a wind into the upper
edge of the domain, the cloudlets disrupt, but do not immediately mix
into their surroundings.  Instead, the dynamics of this population of
cloudlets is more complex, and the cold gas ends up largely co-moving
with the hot phase (\textit{bottom-right plot}).  This likely results
from both cloudlets shielding one another, and from mixed material
which re-cools in the cloudlet wakes, perhaps aided by streaming
instabilities.  We will delve into this process in more detail in
future work.  We note that, for cloudlets with a lengthscale
$R_{0}\sim\cstcool$, the ratio of the cooling time to the
cloud-crushing time is:
\begin{align}
  \frac{t_{\text{cool}}}{t_{\text{crush}}} \sim \frac{v}{c_{\text{s}}};
\end{align}
therefore, we might expect cloudlets to preferentially survive in
subsonic regions of the flow.  The lower-left plot in
figure~\ref{fig:entrainment} indeed suggests this is the case.  The
simulation in figure~\ref{fig:entrainment} is only meant to be
suggestive, and we will perform a more detailed study in the future.
In particular, it remains to be shown that this result holds in three
dimensions.

We close this section by noting that
figure~\ref{fig:fragmentation-1e6} in the appendix shows simulations
similar to those in figure~\ref{fig:fragmentation}, but with a density
ratio of 100 and a background temperature of $10^6$\,K.  Though the
cloud disruption proceeds differently in these simulations due to the
different density ratio, the cloudlets ultimately fragment to a
similar lengthscale.  This independence of the initial temperature is
consistent with the fragmentation cascade discussed in
\citet{Voit1990} and in section~\ref{subsec:hand-wavy}, but not with
models based on linear theory such as those discussed in
\citet{Field1969}, \citet{Hennebelle1999}, \citet{Burkert2000}.

\section{Observational Evidence for Shattering}
\label{sec:results}
In this section, we outline several applications of shattering to
unexplained observations of cold gas.  We discuss galaxy halo gas,
ranging from the cold circumgalactic medium (CGM) probed by absorption
studies of galaxies (\S~\ref{subsec:results-galaxies}), as well as
HVCs and other resolved clouds in the Milky Way
(\S~\ref{subsec:results-hvcs}), to the broad line profiles seen in the
circumgalactic and interstellar media (\S~\ref{subsec:results-lines}),
to the atomic gas near quasars (\S~\ref{subsec:results-bals}), and to
multi-phase galactic winds, in which gas components with a wide range
of densities and temperatures are frequently found to be co-moving
(\S~\ref{subsec:results-winds}).

Dividing our discussion in this way is somewhat artificial since these
applications likely share physical connections: galaxy winds (some
driven by quasar outflows) may populate much of the CGM.  And HVCs may
essentially represent a high-density component of CGM gas close to the
galaxy disk.  In the interest of simplicity, here we use the term
`CGM' as a catchall to refer to bound gas in the outskirts of galaxy
halos, at a reasonable fraction $\sim(0.5$\,--\,$1)$ of the virial
radius.  Such gas is typically studied in absorption against
background quasars.  When that gas has a clear signature of outflow,
or is clearly unbound from the galaxy, we refer to it as a ``wind.''
And if that wind is seen in absorption against a quasar inside the
galaxy, we consider it a quasar outflow.  In this paper, ``HVCs''
refer to spatially resolved clouds in local group galaxies, typically
studied in 21\,cm emission.  These distinctions may be largely
incidental, and in many cases the same gas clouds could fall into
multiple categories, depending on how it is observed.

\subsection{Galaxy Halo Gas}
\label{subsec:results-galaxies}
\begin{table*}
\centering
% table.tex -- automatically generated by mktable.rb
%   edits to this file will be overwritten!
%
\begin{threeparttable}

% smoosh columns together to fit table on the page
\setlength{\tabcolsep}{4.275pt} % space between cols (6pt standard)

\caption{Observations of atomic gas in the CGM.  Though the CGM has no clear
  boundaries, in the interest of simplicity we attempt to show only
  bound gas in the outskirts of galaxy halos, at distances
  $\sim(0.5$\,--\,$1)\,r_{\text{vir}}$; we list more distant
  observations in the second group of rows, and discuss closer-in
  observations with HVCs in \S~\ref{subsec:results-hvcs}.  Most of
  these measurements come from photoionization modeling of absorption
  in sightlines to background quasars.  In a few cases, further
  constraints come from fluorescent $\text{Ly}_{\alpha}$ emission,
  from gravitationally-lensed background quasars which probe structure
  in the plane of the sky, or from collisionally-excited
  fine-structure lines; details are given in the notes below the
  table.  The first group of columns shows properties closely related
  to observable quantities; this includes the mass and redshift of the
  galaxy, its virial radius, and the total column of neutral hydrogen
  (which sets the depth of absorption lines).  The second group of
  columns shows properties derived from photoionization modeling;
  while there may be significant uncertainty in this modeling, in all
  cases the modeling indicates cold gas with a high density and tiny
  volume-filling fraction $f_{\text{V}}$.  The final group of columns
  lists predictions based on the shattering described in this paper;
  overall, we find excellent agreement with the observations, both in
  terms of the high density and small scale for the gas, and in terms
  of the large number of clouds along the line of sight, needed to
  explain broad line widths (see \S~\ref{subsec:results-lines},
  below).  The lone exception to this is \citet{Werk2014}; the low
  high column density $N$ and low volume density $n$ they report
  yields entirely different results.  \citet{Werk2014} are the only
  low-redshift observations we consider in this section, and are
  likely seeing a different phenomenon than the high-$z$ observations;
  we discuss this in
  section~\ref{sec:discussion}.\label{tab:obs-results}}

\begin{tabular}{LCCCCCCCCCC}
\toprule
%
% annotations
%
\multicolumn{4}{c}{Observationally Derived} &
\multicolumn{4}{c}{Photoionization Modeling} &
\multicolumn{3}{c}{Shattering Fragments}\\
\cmidrule(lr){1-4}
\cmidrule(lr){5-8}
\cmidrule(lr){9-11}
%
% heading
%
\multicolumn{1}{c}{$z$}
& M_{\text{halo}}\,(M_{\odot})
& r_{\text{vir}}\,(\text{kpc})
& N_{\text{HI}}\,(\text{cm}^{-2})
& N_{\text{H}}\,(\text{cm}^{-2})
& n_{\text{H}}\,(\text{cm}^{-3})
& f_{\text{V}}
& l_{\text{cold}}\,(\text{pc})
& n_{\text{cold}}\,(\text{cm}^{-3})
& \ell_{\text{cold}}\,(\text{pc})
& f_{\text{A}}
\\ \midrule
%
% entries
%
\multicolumn{11}{l}{\!\!\!\rule{0pt}{1ex}\textit{Bound gas near the virial radius:}} \\ \rule{0pt}{3ex}
%% Lau2015
%% 
2.0\tnote{a}  % z
& 10^{12.5}  % mhalo
& 75  % rvir
& 10^{18.0}  % NHI
& 10^{20.5}  % Nh
& 3  % nh
& 10^{-3.5}  % fv
& 34  % lcold
& 1.1  % ncold
& 10^{-1.0}  % Rcold
& 10^{3.5}  % Nclouds
\\
%% Hennawi2015
%% 
2.0\tnote{b}  % z
& 10^{14.5}  % mhalo
& 350  % rvir
& 10^{19.2}  % NHI
& 10^{20.4}  % Nh
& 2  % nh
& 10^{-4.0}  % fv
& 41  % lcold
& 24  % ncold
& 10^{-2.5}  % Rcold
& 10^{3.5}  % Nclouds
\\
%% Prochaska2009
%% 
2.4\tnote{c}  % z
& 10^{13.3}  % mhalo
& 111  % rvir
& 10^{18.6}  % NHI
& 10^{20.0}  % Nh
& 1.8  % nh
& 10^{-4.0}  % fv
& 18  % lcold
& 10  % ncold
& 10^{-2.0}  % Rcold
& 10^{3.0}  % Nclouds
\\
%% Lau2015
%% 
2.0\tnote{d}  % z
& 10^{12.5}  % mhalo
& 75  % rvir
& 10^{19.2}  % NHI
& 10^{19.5}  % Nh
& 100  % nh
& 10^{-6.0}  % fv
& 0.2  % lcold
& 1.1  % ncold
& 10^{-1.0}  % Rcold
& 316.2  % Nclouds
\\
%% Churchill2003
%% 
1.4\tnote{e}\,\tnote{\phantom{e}$\dagger$}  % z
& \multicolumn{2}{c}{\multirow{3}{*}{
\begin{varwidth}{0.75in}\centering
\textit{(galaxy not detected)}
\end{varwidth}}}  % mhalo
  % rvir
& 10^{19.1}  % NHI
& 10^{19.1}  % Nh
& 5  % nh
& \text{---}  % fv
& \scalebox{0.9}{
$\def\arraystretch{0.8}\left\{ \!\!\!\!
\begin{array}{c} 0.5^{(||)} \\ 135^{(\perp)} \end{array}
\right.$}  % lcold
& \text{---}  % ncold
& \text{---}  % Rcold
& 125.9  % Nclouds
\\
%% Rigby2002
%% 
0.8\tnote{f}\,\tnote{\phantom{f}$\dagger$}  % z
&    % mhalo
&    % rvir
& \text{---}  % NHI
& 10^{17.8}  % Nh
& 0.1  % nh
& \text{---}  % fv
& 2  % lcold
& \text{---}  % ncold
& \text{---}  % Rcold
& 6.3  % Nclouds
\\
%% Rauch1999
%% 
3.5\tnote{g}\,\tnote{\phantom{g}$\dagger$}  % z
&    % mhalo
&    % rvir
& 10^{16.0}  % NHI
& 10^{17.8}  % Nh
& 0.2  % nh
& \text{---}  % fv
& \scalebox{0.9}{
$\def\arraystretch{0.8}\left\{ \!\!\!\!
\begin{array}{c} 1.6^{(||)} \\ 26^{(\perp)} \end{array}
\right.$}  % lcold
& \text{---}  % ncold
& \text{---}  % Rcold
& 6.3  % Nclouds
\\
%% Stocke2013
%% 
0.04\tnote{h}\,\tnote{\phantom{h}\S}  % z
& 10^{12.4}  % mhalo
& 280  % rvir
& 10^{14.8}  % NHI
& 10^{17.0}  % Nh
& 10^{-3.5}  % nh
& 10^{-3.5}  % fv
& 80  % lcold
& 10^{-2.5}  % ncold
& 23.7  % Rcold
& 1.0  % Nclouds
\\
\midrule
\multicolumn{11}{l}{\!\!\!\rule{0pt}{1ex}\textit{Low-pressure systems, likely out in the IGM:}} \\ \rule{0pt}{3ex}
%% Crighton2013
%% 
2.4\tnote{i}\,\tnote{\phantom{i}$\dagger$}  % z
& 10^{11.5}  % mhalo
& 28  % rvir
& 10^{16.4}  % NHI
& 10^{19.0}  % Nh
& 10^{-2.0}  % nh
& 10^{-2.0}  % fv
& 324  % lcold
& \multicolumn{2}{c}{\multirow{3}{*}{
\begin{varwidth}{0.75in}\centering
\textit{(ambient pressure unknown)}
\end{varwidth}}}  % ncold
  % Rcold
& 100.0  % Nclouds
\\
%% Crighton2015
%% 
2.5\tnote{j}\,\tnote{\phantom{j}$\dagger$}  % z
& 10^{11.4}  % mhalo
& 24  % rvir
& 10^{15.2}  % NHI
& 10^{18.4}  % Nh
& 10^{-2.5}  % nh
& 10^{-2.0}  % fv
& 163  % lcold
&    % ncold
&    % Rcold
& 25.1  % Nclouds
\\
%% Schaye2007
%% 
2.3\tnote{k}\,\tnote{\phantom{k}$\dagger$}\,\tnote{\phantom{k$\dagger$}$\ddagger$}  % z
& \text{---}  % mhalo
& \text{---}  % rvir
& \text{---}  % NHI
& \text{---}  % Nh
& 10^{-3.5}  % nh
& \text{---}  % fv
& 100  % lcold
&    % ncold
&    % Rcold
& \text{---}  % Nclouds
\\
\midrule
\multicolumn{11}{l}{\!\!\!\rule{0pt}{1ex}\textit{Low-redshift, qualitatively distinct from shattering:}} \\ \rule{0pt}{3ex}
%% Werk2014
%% 
0.2\tnote{l}  % z
& 10^{12.0}  % mhalo
& 183  % rvir
& 10^{16.5}  % NHI
& 10^{20}  % Nh
& 10^{-4.0}  % nh
& 1.8  % fv
& \sim r_{\text{vir}}  % lcold
& 10^{-2.5}  % ncold
& 31.0  % Rcold
& \sim 1  % Nclouds
\\
\bottomrule
\end{tabular}

\begin{tablenotes}
\item[a] \citet{Lau2015} \dotfill{}H\,\textsc{i} absorption (median values from sample)\hspace*{0.125\textwidth}\linebreak[0]
\item[b] \citet{Hennawi2015} \hfill{}H\,\textsc{i} fluorescence + absorption\hspace*{0.125\textwidth}\linebreak[0]
\item[c] \citet{Prochaska2009} \hfill{}H\,\textsc{i} absorption, C\,\textsc{ii}\textsuperscript{*} \& Si\,\textsc{ii}\textsuperscript{*} (`C' component)\hspace*{0.125\textwidth}\linebreak[0]
\item[d] \citet{Lau2015} \hfill{}H\,\textsc{i} absorption, C\,\textsc{ii}\textsuperscript{*} (J1420+1603)\hspace*{0.125\textwidth}\linebreak[0]
\item[e] \citet{Churchill2003} \dotfill{}H\,\textsc{i} absorption; gravitationally lensed quasar\hspace*{0.125\textwidth}\linebreak[0]
\item[f] \citet{Rigby2002} \hfill{}Fe\,\textsc{ii}/Mg\,\textsc{ii}, C\,\textsc{iv}/Mg\,\textsc{ii} (systems 7, 13, \& 18)\hspace*{0.125\textwidth}\linebreak[0]
\item[g] \citet{Rauch1999} \hfill{}H\,\textsc{i}, O\,\textsc{i}/C\,\textsc{ii}, Si\,\textsc{ii}/Si\,\textsc{iv}; gravitationally lensed quasar\hspace*{0.125\textwidth}\linebreak[0]
\item[h] \citet{Stocke2013} \hfill{}absorption (PKS 1302-102)\hspace*{0.125\textwidth}\linebreak[0]
\item[i] \citet{Crighton2013} \dotfill{}H\,\textsc{i}, C\,\textsc{ii}, Si\,\textsc{ii}, Mg\,\textsc{ii} absorption\hspace*{0.125\textwidth}\linebreak[0]
\item[j] \citet{Crighton2015} \hfill{}absorption; multiple lines from Mg\,\textsc{ii}\,--\,C\,\textsc{iv}\hspace*{0.125\textwidth}\linebreak[0]
\item[k] \citet{Schaye2007} \hfill{}absorption\hspace*{0.125\textwidth}\linebreak[0]
\item[l] \citet{Werk2014} \hfill{}COS halos\hspace*{0.125\textwidth}\linebreak[0]
\vspace*{\baselineskip}
\item[$\dagger$] Indicates an overall absorption system that
resolves into individual components, each with approximately thermal
line widths.  In this case, the numbers $N$ and $n$ represent the
thinnest resolved component, rather than the system as a whole.
We see narrow lines only when the inferred number of clouds
$f_{\text{A}}$ along the line of sight (last column) is small.
\item[$\ddagger$] By design, \citet{Schaye2007} report only
conservative upper limits for the hydrogen column density.  In this
table, we quote only their conservative upper limit for $l_{\text{cold}}$.
\item[\S] \citet{Stocke2013} show a wide range of different
absorbing systems.  Here, we show only PKS 1302-102; this is representative
of absorbers close to the virial radius.  Of absorbers in the range
$(0.7$\,--\,$1.25)\,r_{\text{vir}}$, 13/14 show
${10^{17}}\,\text{cm}^{-2}\lesssim{}N_{\text{H}}\lesssim{10^{18.5}}\,\text{cm}^{-2}$.
\end{tablenotes}

\end{threeparttable}

\end{table*}
The gas in the outskirts of galaxy halos consists partly of cosmic gas
which accretes from the intergalactic medium and shocks to the
($\gtrsim10^6$\,K in local $\sim{L}_{*}$ galaxies) virial temperature
of the halo \citep{Birnboim2003}, and partly of gas recycled from the
galaxy disk in a galactic `fountain'
\citep[\eg][]{Shapiro1976,Oppenheimer2010}.  Simulations and analytic
arguments both suggest this halo gas persists as a long-lived,
virialized, approximately-hydrostatic atmosphere in the halos of
galaxies more massive than $\sim10^{12}{}M_{\odot}$
\citep[\eg][]{White1978,Birnboim2003,van2011,Fielding2016}.  This
reservoir of gas in galaxy halos seems to represent a significant
fraction of the universe's baryons
\citep{Maller2004,Stocke2013,Werk2014}, and may supply the fuel
necessary to explain continued star formation in galaxies
\citep[\eg][]{Bauermeister2010,Genzel2010}.

A number of observations suggest this halo gas plays an important role
in galaxy evolution.  The properties of halo gas correlate with galaxy
mass \citep{Prochaska2013,Prochaska2014}, with star formation activity
\citep{Tumlinson2011}, and possibly with redshift \citep{Werk2014},
suggesting the physics of halo gas somehow relates to these properties
of galaxies \citep{Sharma2012,Prochaska2014,Voit2015,Voit2015b}.
Though the causal relationships driving such correlations are not yet
understood, this is a rapidly evolving field of study, becoming
increasingly accessible to both simulations and observations as new
techniques are developed.

Since the virialized, $\sim10^{6}$\,K component of halo gas is too
faint to observe directly in emission,\footnote{Though this virialized
  CGM gas is invisible to current telescopes, there is little reason
  to doubt its existence: both analytic studies
  \citep[\eg][]{White1978,Birnboim2003,Maller2004} and cosmological
  simulations
  \citep[\eg][]{van2011,Fumagalli2014,Nelson2013,Fielding2016}
  consistently predict its presence, and stacked observations have
  directly detected virialized gas in galaxies only slightly more
  massive than the milky way \citep{Anderson2015,Planck2013}.
  Furthermore, circumstantial evidence from the confinement of gas
  clouds and stripping of satellite galaxies also points to a
  virialized CGM \citep[\eg][]{Maller2004,Fang2013,Stocke2013}.}
galaxy halo gas is instead studied in \textit{absorption} along
sightlines to more distant quasars; unfortunately, this technique is
sensitive only to atomic gas far cooler than the virial temperature
(comprising a small minority of the mass).  These absorption studies
indicate that a fraction of galaxy halo gas exists in a cool phase at
$\sim10^4$\,K; this is something which was not predicted by
cosmological simulations, and which remains a challenge to reproduce
quantitatively (\eg
\citealt{Fumagalli2014,Faucher2015,Fielding2016,Liang2016}; but see
\citealt{Faucher2016}).  In this section, we refer to \textit{all}
galaxy halo gas as the \textit{circumgalactic medium}, or CGM; this
includes shocked gas at the virial temperature, along with the cooler
phases probed in absorption sightlines.

One particularly surprising development in studies of the CGM has been
the frequent finding that the cool CGM in the outskirts of galaxy
halos comes in the form of dense ($\sim1\,\text{cm}^{-3}$), small
($\sim 1$\,--\,$100$\,pc) clouds distributed throughout the halo.
Though this cool CGM gas extends out to cosmological scales, as far
out as the virial radius of the halo, it is comparable in density to
the interstellar medium in the disk of the milky way galaxy!  This
unexpected result appears to be typical of the CGM, at least at the
redshifts $z\gtrsim2$ most accessible to quasar-absorption studies:
table~\ref{tab:obs-results} summarizes a number of such observations,
all of which arrive at this basic conclusion.  The first group of
columns in table~\ref{tab:obs-results} lists galaxy properties closely
related to observable quantities: this includes the mass\footnote{This
  mass is left blank in a few cases where the galaxy is not detected.
  Absorption is such a sensitive probe of neutral gas that in some
  cases the observations probe the diffuse gas in the outskirts of
  galaxy halos, even if the galaxy is too faint to be seen!} and
redshift of the galaxy halo, along with the column density of neutral
hydrogen.\footnote{Inferring $N_{\text{HI}}$ is sometimes complicated
  by the flatness of the curve of growth.  The potentially large
  uncertainty and degeneracy with \eg, turbulent broadening
  (particularly given the large amount of turbulent broadening we now
  advocate) should be kept in mind.}  The second group of columns in
table~\ref{tab:obs-results} lists results derived from photoionization
modeling, which yields the \textit{total} hydrogen column density
$N_{\text{H}}$, along with the volume density $n_{\text{H}}$, of the
cold gas.  The ratio of these densities
$l_{\text{cold}}\equiv{}N_{\text{H}}/n_{\text{H}}$ provides an
estimate of the total path length of cold gas along the line of sight.
In all cases, this path length $l_{\text{cold}}\sim(1$\,--\,$100)$\,pc
is \textit{far} smaller than the size of the galaxy, indicating a very
small characteristic size for cold clouds.  This is perhaps more
accurately interpreted as a tiny volume filling fraction of cold gas:
$f_{\text{V}}\sim{}N/(n\,R)\sim{}10^{-3}$\,--\,$10^{-4}$.

The observations in table~\ref{tab:obs-results} span some three orders
of magnitude in galaxy halo mass, along with a significant redshift
range $z\sim(0.04$\,--\,$3.5)$.  Thus, the unexpected finding that
galaxy halos are \textit{full} of tiny, dense clouds seems to be
generic.  This suggests that tying the cold gas to specific star
formation or feedback processes which depend sensitively on galaxy
mass or redshift may be misguided; extended, cold CGM gas seems to be
a nearly universal feature of galaxies.  Moreover,
table~\ref{tab:obs-results} shows that cool CGM gas is consistently
detected in quasar sightlines.  Often, these can be optically thin
with $N_{\text{HI}} < 10^{17}\,\text{cm}^{-3}$.  Specifically
quantifying optically thick covering fractions
\citep[\eg][]{Rudie2012,Prochaska2013} therefore may not probe the
physical origins of this gas.  The optically-thick covering fraction
is known to evolve rapidly with galaxy mass (cf.\ \citealt{Rudie2012}
and \citealt{Prochaska2013}; \citealt{Thompson2016}).
\citet{Faucher2016} successfully explain these high-column-density
absorbers in terms of feedback near the peak of star formation.  If
some cold gas is in fact always present in galaxy halos, then
radiative transfer effects may strengthen the process discussed in
\citet{Faucher2016}: if cold gas in halos is close to the self
shielding threshold, then a slight increase in the cold gas column
with halo mass could lead to a large apparent increase in the
optically thick covering fraction.

Absorbers start to self-shield once the neutral column
$N_{\text{HI}}>N_{\text{HI}}^{\text{LLS}}\sim\sigma_{\text{HI}}^{-1}\sim{}10^{17}\,{\text{cm}^{-2}}$.
Optically thin absorbers have a neutral fraction:
\begin{equation}
x_{\text{HI}}
= \frac{\alpha n_{\text{H}}}{\Gamma}
= 4 \times 10^{-3}
  \left( \frac{n}{10^{-2} \, {\text{cm}}^{-3}}  \right)
  \left( \frac{\Gamma}{10^{-12} \, {\text{s}^{-1}}} \right)^{-1}
\end{equation}
where the photo-ionization rate
$\Gamma\equiv{4\pi}\int_{\nu_{\text{th}}}^{\infty}(J/h\nu)\sigma(\nu)d\nu$.
Thus, absorbers with hydrogen column densities
$N_{\text{H}}>N_{\text{HI}}^{\text{LLS}}{}x_{\text{HI}}^{-1}=10^{20}\,{\text{cm}^{-3}}n_{-2}^{-1}\Gamma_{-12}\,{\text{cm}^{-2}}$
are capable of self-shielding.  Once absorbers exceed this critical
column density, they undergo a rapid transition to full neutrality
beyond the photosphere.  Thus, except for environments with extremely
weak radiation fields, an individual isolated cloudlet
$N_{\text{H}}\approx{}10^{17}\,{\text{cm}^{-2}}$ is optically thin and
highly ionized.  However, a cloud of droplets with
$N_{\text{H}}>N_{\text{H}}^{\text{SS}}$, or
$f_{\text{A}}>10^{3}n_{-2}^{-1}\Gamma_{-12}$ will start to
self-shield, rapidly approaching
$N_{\text{HI}}\rightarrow{}N_{\text{H}}$.  Note that for the purposes
of continuum radiative transfer, the properties of a cloud of droplets
is essentially identical to that of a monolithic slab of equivalent
column density.  However, this is not true for resonant line transfer,
since the cloud is able to sustain large shear and turbulent
velocities (see \S\ref{subsec:disc-future} for further discussion).

While there may be significant uncertainty in the photoionization
modeling (particularly in the assumed flux of ionizing photons), we
note that in many cases other constraints corroborate these results.
In some spectra, the C\,\textsc{ii}\textsuperscript{*} and
Si\,\textsc{ii}\textsuperscript{*} fine-structure lines, which are
thought to be collisionally excited, provide an independent estimate
of the electron number density within the cold gas
\citep{Prochaska2009,Lau2015}; this confirms the high volume densities
leading to such small inferred cloud sizes.  And in a few
extraordinary cases \citep[\eg][]{Rauch1999}, gravitationally lensed
quasars directly constrain the cloud sizes in the plane of the sky to
be less than $\sim30$\,pc; this is in excellent agreement with limits
along the line of sight derived from photoionization models.  Yet
another constraint comes from the ``slug nebula''
\citep{Cantalupo2014}, where the CGM is detected in fluorescent
emission: the $\text{Ly}_{\alpha}$ surface brightness ($\propto N\,n$)
and non-detection of He\,\textsc{ii} and C\,\textsc{iv} imply a large
volume density $n_{\text{H}}\gtrsim{3}\,\text{cm}^{-3}$, and thus a
column density $N_{\text{H}}\lesssim10^{20}\,\text{cm}^{-2}$.  This
corresponds to $l_{\text{cold}}\lesssim{20}$\,pc \citep{Arrigoni2015},
again in excellent agreement with the absorption measurements in
table~\ref{tab:obs-results}.  We emphasize that these
observationally-derived sizes are \textit{essentially all upper
  limits}: the quantity
$l_{\text{cold}}\sim{}N_{\text{H}}/n_{\text{H}}$ gives the integrated
amount of cold gas along the line of sight; this total column may come
in many small steps, however, if the cold gas is in fact made up of
much smaller clouds.  We argue below and in
section~\ref{subsec:results-lines} that this must be the case.

Currently, no explanation exists for such tiny cloud sizes in the CGM.
This result becomes significantly more bizarre, however, when one
considers the area covering fraction of the gas.  The tiny
volume-filling factors $f_{\text{V}}\lesssim{10^{-3}}$ in
table~\ref{tab:obs-results} suggest that sightlines through galaxies
should rarely intersect cold gas.  However, it is rare to find
sightlines \textit{without} absorption features: \cite{Prochaska2013}
find a large covering fraction ($\sim100\%$) of cold gas within the
virial radius for sufficiently low column density $N_{\text{HI}}$.
Moreover, in cases where the cool CGM is detected in fluorescent
emission \citep[\eg][]{Cantalupo2014,Hennawi2015,Borisova2016}, it
visibly fills the entire halo.  \textit{Despite such tiny absorption
  lengths along the line of sight (and correspondingly small total
  mass), this cool CGM gas somehow manages to cover the entire
  $\gtrsim100$\,kpc extent of the halo.}  In these galaxies,
therefore, the area-covering fraction of cold gas vastly exceeds its
volume-filling fraction $f_{\text{A}}\gg{}f_{\text{V}}$ , indicating
either that the absorption occurs in a thin shell
($\Delta{R}/R\lesssim{10^{-3}}$) surrounding the galaxy, or that the
absorption comes from a fog of tiny cloudlets distributed widely
throughout the halo, as might be produced by shattering.

Since a thin shell of gas would likely not fit the fluorescent
emission profiles in \citet{Cantalupo2014}, \citet{Hennawi2015}, and
\cite{Borisova2016}, these observations of galaxy halos are
qualitatively very suggestive of shattering: if the cold gas takes the
form of a fog of tiny, distributed cloudlets, a modest total mass (or
volume-filling fraction $f_{\text{V}}$) could naturally span the
entire galaxy.  As discussed in section~\ref{sec:intro}, the ratio of
the two fractions is
$f_{\text{A}}/f_{\text{V}}\sim{R_{\text{vir}}/\rcloudlet}\gtrsim10^6$.
Recall that we define $f_{\text{A}}$ as the mean number of cloudlets
along the line of sight; even if we require a large number of
cloudlets ($f_{\text{A}}\gg10^3$) to explain the smooth profiles and
broad line widths observed, shattering can easily accommodate the low
volume-filling fractions $f_{\text{V}}\ll1$ required by some
observations.  Whether this gas is accreted cold from the
intergalactic medium \citep[\eg][]{Keres2005}, forms in-situ via
thermal instability
\citep[\eg][]{Maller2004,McCourt2012,Sharma2012,Voit2015}, or is
launched from the galaxy disk in outflows
\citep[\eg][]{Faucher2016,Liang2016}, we expect the gas to fragment
into tiny cloudlets.  These cloudlets could then disperse throughout
the halo due to turbulence, producing the observed uniform covering
fractions.

While these indirect arguments are extremely suggestive, they do not
amount to direct observations of small scale structure.  The last
group of columns in table~\ref{tab:obs-results} lists the cloudlet
properties predicted by shattering.  The most direct comparison is to
see that the total (ionization-corrected) column density of cold gas
$N_{\text{H}}$ always exceeds the minimum scale
$\sim10^{17}\,\text{cm}^{-2}$ predicted by shattering.  Interestingly,
when the implied number of cloudlets along the line of sight
$f_{\text{A}}\sim{}N_{\text{H}}/(10^{17}\,\text{cm}^{-2})$ is large,
we typically find broad line widths, with velocities characteristic of
the virial velocity of the halo.  In the cases where $N_{\text{H}}$
dips below $\lesssim10^{18}\,\text{cm}^{-2}$, the lines resolve into
individual components, each with approximately thermal line widths (we
denote narrow lines with a $\dagger$ in the first column of
table~\ref{tab:obs-results}).  This transition from broad to narrow
lines around a column density of $\sim10^{17}\,\text{cm}^{-2}$ is
suggestive of shattering; we discuss this further in
section~\ref{subsec:results-lines}.

We furthermore show that the high volume densities and correspondingly
small spatial scales of the cold gas are suggestive of shattering,
though this requires assumptions about the ambient environment.  Since
we expect the cold clouds are pressure-confined, we define a typical
hydrostatic or ``virial'' pressure for the ambient halo gas as
$P_{\text{vir}}{}\sim{}f_{\text{b}}{}G{}M^2{}/{}R_{\text{vir}}^4$.  In
convenient units,
\begin{align}
  P_{\text{vir}} \sim (2000\,\text{cm}^{-3}\,\text{K})\,M_{12}^{2/3} h(z)^{8/3},\label{eq:pvir}
\end{align}
where $M_{12}\equiv{}M/10^{12}\,M_{\odot}$ is the total mass of the
galaxy, $h(z)\equiv{}H(z)/H_0$ is the Hubble parameter at redshift
$z$, and we have adopted the common definition for the virial radius
$R_{\text{vir}}\equiv{}R_{\text{200c}}$ as the radius within which the
mean density is 200 times the average energy density of the universe
(shown in the third column of table~\ref{tab:obs-results}).  The
baryon fraction $f_{\text{b}}$ is uncertain in galaxies, leading to
debate over the missing baryon problem; these constraints are even
weaker at the redshifts $z\gtrsim2$ relevant to quasar absorption
studies.  In the interest of simplicity, we adopt a cosmic baryon
fraction $f_{\text{b}}=\Omega_{\text{b}}/\Omega_{\text{m}}\sim{}0.17$,
which probably over-estimates the the pressure $P_{\text{vir}}$ in
equation~\ref{eq:pvir}.  In galaxy clusters, where the gas pressure
can be measured, $P_{\text{vir}}$ is a typical pressure within
$\sim{}0.1\,R_{\text{vir}}$, and the pressure falls steeply at large
radii \citep{Nagai2007,Arnaud2010}.  We note however that the pressure
profiles may be somewhat flatter, and lower in magnitude, in galaxies
(see, \eg, \citealt{Maller2004} or \citealt{Sharma2012}; but see
\citealt{Fielding2016}).  We adopt the pressure $P_0 \equiv
P_{\text{vir}}/100$ as a typical pressure between
$\sim{}R_{\text{vir}}/2$ and $\sim{}R_{\text{vir}}$ \citep[cf.][their
  figure 6]{Arnaud2010}.  Though this estimate for the pressure is
quite crude, our estimates for $n_{\text{cold}}$
($\equiv{P_{0}/10^4\,K}$) and $\ell_{\text{cold}}$
(equation~\ref{eq:rcloud}) may be easily scaled to more accurate
estimates ($\propto{}P$ and $\propto{}1/P$, respectively).  Given the
significant uncertainty in this gas pressure, we find remarkable
agreement with the densities inferred from photoionization models: the
high densities observed [$n\sim(1$\,--\,$10)\,\text{cm}^{-3}$] are
characteristic of pressure equilibrium in the halos of these massive,
high-redshift galaxies.  As expected, the cloudlet size \rcloudlet is
always smaller than the observationally-derived limit
$l_{\text{cold}}$; often by several orders of magnitude.  This implies
that each sightline intercepts a large number of cloudlets
($f_{\text{A}}$), consistent with the broad line widths typical of
these observations.

Taken together, these results suggest a model for the CGM.  When the
cold CGM is visible in fluorescent emission, the surface-brightness
profiles are strongly suggestive of distributed small cloudlets,
rather than a thin shell of gas.  The emission profiles in
\citet{Borisova2016} suggest a semi-universal mass density for the
cool CGM which scales as $\sim{}r^{-1.5}$; this is strikingly
consistent with predictions for the hydrostatic \textit{hot} gas from
some cosmological simulations \citep[\eg][]{Fielding2016}.  However,
the cold gas probed in these observations \textit{cannot} be in
hydrostatic equilibrium; balancing gravity with pressure at large
radii requires gas to be near the ($\sim10^6$\,K) virial temperature.
The observations in \citet{Borisova2016} therefore suggest a dense,
hydrostatic (but still invisible) CGM as is found in cosmological
simulations.  For reasons that remain unclear, but are possibly
related to galactic winds or thermal instability, a small fraction
[$\sim$(1\,--\,5)\%] of the CGM is cold; shattering would then imply
this cold gas takes the form of tiny cloudlets interspersed throughout
the hydrostatic CGM.  If the small cloudlets are dynamically coupled
to the hot gas (see section~\ref{subsec:results-winds}, below), this
could explain the hydrostatic-looking density profile for the cold
gas.

A strong dynamical coupling would also imply the cold gas is
long-lived (\ie, it remains suspended in the hot gas and does not fall
inward), explaining why it appears in 100\% of the galaxies targeted
by \citet{Borisova2016}, even out to the virial radius.  Indeed, this
potentially is relevant to why cold photoionized gas is seen in
absorption in both active and passive galaxies \citep{Werk2014} --
once formed, cold gas can be remarkably long-lived.  If correct, this
result is useful because it implies that the kinematic measurements in
\citet{Borisova2016} are in fact probing the dynamics of the
(invisible but dynamically dominant) hot CGM.  Such kinematics seem
qualitatively consistent with the cosmological simulations in
\citet{Fielding2016}, but we defer a detailed discussion to a later
study.

To summarize, recent studies of the CGM unexpectedly find cold gas in
a large fraction of sightlines through typical galaxies, at least at
the redshifts $z\gtrsim2$ most easily probed by observations.  In
cases where this gas can be imaged in fluorescent emission, it is
indeed seen to fill the galaxy halo.  Both absorption and emission
observations indicate a relatively small total mass of cold gas, which
is somehow present everywhere, and which follows a mass density
profile consistent with $\sim{}r^{-1.5}$.  These puzzling observations
are naturally explained if the cold gas comprises a fog of distributed
tiny cloudlets --- in fact, the absorption and emission observations
essentially \textit{require} this interpretation.  While any
individual observation listed in table~\ref{tab:obs-results} may have
significant uncertainty, the general conclusion that at least some
galaxies are filled with tiny clouds of cold, dense gas seems
inescapable.  More quantitatively, the lowest column-density
observations in table~\ref{tab:obs-results} show that the cloudlet
scale we identify is consistent with available observational limits.
While the origin of this cold gas is not known, the observations in
\citet{Hennawi2015}, \citet{Cantalupo2014}, and \citet{Borisova2016}
strongly suggest that the cloudlets are uniformly mixed with a
virialized, pressure-supported CGM as is predicted in observations.
We note that \citet{Stern2016} also interpret quasar-sightline
observations as indicative of a large number of small clouds.  This is
very different from the model proposed here, however, as they consider
a hierarchical model in which each absorption component corresponds to
a distinct cloud, and the different components are strongly out of
pressure balance.  In our picture, cold gas fragments entirely to
small scales and large column densities correspond to sightlines
intercepting multiple cloudlets.  We will make more detailed
comparisons to CGM observations in a future study.

\subsection{Turbulent Line Profiles}
\label{subsec:results-lines}
%
%% \todo{merge with following paragraph} Giant Ly$\alpha$ emission-line
%% nebulae from $z\sim 3-4$ quasar hosts all have extremely broad
%% line-widths ($\sim 500-2000 {\rm km \, s^{-1}}$;
%% \citet{Borisova2016}). Crucially, most line profiles are
%% singly-peaked, rather than doubly-peaked as characteristic for
%% resonantly trapped Ly$\alpha$ photons, which can be broadened by
%% radiative transfer (rather than kinematic) effects. Similarly, $z \sim
%% 2-3$ quasar hosts show metal line absorption (which is generally not
%% subject to radiative transfer effects) typically spanning $\sim 500 \,
%% {\rm km \, s^{-1}}$ \citep{Lau2015}. Interestingly, the metal lines
%% show significant substructure \citep{Prochaska2009,Lau2015}. This
%% could potentially arise if: (i) rather than a pervasive fog, the
%% cloudlets are grouped into distinct ‘clouds’; (ii) the velocity field
%% has significant bulk flows or a large coherence length; (iii) the host
%% quasar illuminates clouds anisotropically. Jointly modeling metal-line
%% and HI Ly$\alpha$ profiles could be fruitful and constraining.

The observations of the cold CGM in \citet{Borisova2016} show smooth,
broad emission lines.  While hydrogen $\text{Ly}_{\alpha}$ lines in
emission can potentially be broadened by radiative transfer effects,
optically-thin metal absorption lines sometimes show comparable
widths.  The width of these emission lines can vastly exceed the
$\sim10$\,km/s thermal width for gas at $10^4$\,K, indicating random
motions of $\sigma_{v}\sim1000$\,km/s in the cold gas.  If these
random motions are interpreted as turbulence within the bulk of the
cold gas, it corresponds to a Mach number $M\sim100$.  However, such
supersonic turbulence should quickly shock on a timescale
\begin{align*}
 t_{\text{diss}} < 10^6\,\,\text{year}
 \times \left(\frac{\ell_{\text{out}}}{\text{kpc}}\right),
\end{align*}
where $\ell_{\text{out}}$ is the outer scale of the turbulence (which
must be below the resolution limit of the observations in order to make
consistently broad line profiles).  Such shocks would heat the gas to
$\sim(10^6$\,--\,$10^7)$\,K, making it impossible to produce the
observed line emission.  This short dissipation timescale is
problematic given the relatively smooth and uniform nature of the
emission profiles: if the gas is destroyed on a timescale
$\lesssim10^6$\,year, it must somehow be continuously replenished
everywhere throughout the $\gtrsim100$\,kpc extent of the halo.  Since
this timescale is far below the cooling time of the hot gas, however,
it is not clear what this uniformly-distributed source of cold gas
could be, or what mechanism could provide the continued turbulent
stirring.  Moreover, since the observations in \citet{Borisova2016}
find broad lines in $\sim100\%$ of galaxies, the mechanism producing
them must be present in essentially all galaxies.

If the cold gas in the CGM represents a fog of tiny cloudlets, as
discussed above in section~\ref{subsec:results-galaxies}, the observed
column densities $N_{\text{H}}\lesssim10^{20}\,\text{cm}^{-2}$ imply
that sightlines intersect a large number ($\lesssim10^3$) of
cloudlets.  If these cloudlets have a velocity distribution
characteristic of the virial velocity of the halo, random motions
among cloudlets could easily explain the observed line widths.  As we
discuss in section~\ref{subsec:results-winds} below, if the droplets
are well dispersed and make up a small fraction of the CGM by mass, we
expect they are efficiently entrained in the ambient hot medium.  The
broad line widths should then reflect trans-sonic ($M\lesssim1$)
motions of the hot gas, which is typical for the turbulent motions in
the virialized gas found in cosmological simulations.  This picture
could be readily testable in the outer parts of the galaxies, where
cold gas column densities should drop: when the cold hydrogen column
falls below $N_{\text{H}}\lesssim10^{20}\,\text{cm}^{-2}$, a
high-resolution, high-signal to noise spectrum of the emission line
should resolve into individual sub-components with widths
$\lesssim20$km/s.  This could be tested by, \eg, looking for a
auto-correlation signal peaking at zero-lag within the line spectrum.
\citet{Arav1997} discuss the spectral resolution and signal-to-noise
necessary to perform such analysis.  If correct, this result is useful
because it implies that the kinematic measurements in
\citet{Borisova2016} are in fact probing the dynamics of the otherwise
invisible hot gas.  We discuss this possibility further in
section~\ref{subsec:disc-future}.  We note that suprathermal line
widths are also typical in galaxy winds
\citep[\eg][]{Martin2005,Schwartz2004}, which is also consistent with
shattering.  However, in winds it is also possible to model broad line
widths as resulting from a large-scale velocity shear over the scale
of the wind.

As we discuss in section~\ref{subsec:results-bals}, the atomic gas
near quasars also shows broad, smooth lines with widths corresponding
to random velocities of ($10^3$\,--\,$10^4$)\,km/s.  Interpreting
these random velocities as turbulence within the bulk of the cold gas
is also problematic, but in this case for a different reason.  Since
the cooling time for the dense gas near quasars is far shorter than
the dynamical time, this gas could easily radiate away any thermal
energy deposited by shocks.  However, this same mechanism should
rapidly dissipate the random motions, suggesting broad-line regions
are hard to maintain and should evolve on roughly the dynamical
timescale.  But broad lines are extremely characteristic of quasars
and AGN, and seen in a large fraction of observations (in some cases,
with an observational baseline of many dynamical times).  The broad
lines therefore cannot be a short-lived, transient phenomenon.  Again,
the problem may disappear if the cold gas takes the form of small,
distributed cloudlets: the broad line widths would then represent
virial motions of the confining hot gas, as traced by a population of
tiny, entrained cloudlets.  The predicted number of cloudlets within
BAL and BLR regions is too great to test this hypothesis directly by
resolving the lines into sub-components, however it may be possible in
more distant absorbers such as ``associated absorption lines,'' or
AALs; see section~\ref{subsec:results-bals} for more details.

In HVCs, \citet{Blagrave2016} show that the full, $\sim100$\,km/s
velocity spread of the gas splits into features with widths of
$\sim2$\,km/s, characteristic of the thermal width for the 100\,K gas
probed.  In both position and velocity space, they see structure all
the way down to the resolvable scale.  \citet{Richter2006} and
\citet{Braun2005} probe the smallest scales of HVCs yet identified,
with $N_{\text{HI}}\sim{}10^{18}\,{\text{cm}^{-2}}$ (and inferred
sizes of a few thousand AU).  Depending on the ionization fraction,
these observations may approach the scale of individual cloudlets.
And indeed the lines resolve into thermal widths, with resolved
Lorentzian profiles.

In similar observations, but at lower temperature, \citet{Hacar2011}
and \citet{Hacar2013} demonstrate in a careful analysis that molecular
filaments in the interstellar medium with strongly suprathermal line
widths resolve into $\sim10$s of smaller sub-filaments, each with
approximately thermal line widths.  The relative motions of filaments
sets the overall line width, and is consistent with subsonic motion of
each sub-filament through the confining hot medium.  While the
shattering discussed in this paper does not strictly apply at
temperatures far below $\sim10^4$\,K, these results are broadly
consistent with the notion that cold gas has a small characteristic
scale, and that broad lines can be made by superposing a large number
of narrow ones.  We discuss possible extensions of shattering to the
colder, 100\,K gas in star-forming regions in
section~\ref{sec:discussion}.

\subsection{Broad-Lines and Black Holes}
\label{subsec:results-bals}
\begin{table*}
\centering
\begin{threeparttable}
  \caption{The neutral gas around quasars comes in the form of small,
    dense, distributed cloudlets.  In (approximate) order of
    increasing distance from the black hole, we list example
    observations of the BLR, BAL, mini-BAL, FeLoBAL, and AAL regions
    around quasars (see \S~\ref{subsec:results-bals} for definitions
    of the acronyms; references given in the numbered notes below the
    table).  With the exception of the BLR, these are all seen in
    absorption.  Though this gas is cold and neutral, it moves with
    extreme velocity, sometimes a significant fraction of the speed of
    light $c$.  Turbulent broadening (estimated both from the width of
    the line, and also from radiative transfer effects influencing
    line ratios) indicate strongly supersonic random motions, in some
    cases with Mach numbers $\gtrsim10^3$ relative to the atomic gas.
    As argued in \citet{Arav1997} and in
    section~\ref{subsec:results-lines}, such broad suprathermal line
    widths suggest the emission comes from trans-sonic motion of cold
    cloudlets through the hot gas in a two-phase medium.  As in the
    case of the CGM (\S~\ref{subsec:results-galaxies}), the
    area-covering fraction of cold gas is $f_{\text{A}}\gtrsim1$, but
    photoionization modeling indicates high densities and hence a
    relatively small total amount of cold gas, implying volume-filling
    fractions as low as $f_{\text{V}}\sim10^{-6}$.  The last two
    columns in the table show the cloudlet size \rcloudlet computed
    from equation~\ref{eq:rcloud} (which can be compared to the
    observational upper limit $l_{\text{cold}}$) and the expected
    number of cloudlets along the line of sight $f_{\text{A}}$ (which
    can be compared to the observational lower limit
    $N_{\text{clouds}}^{\text{(region)}}$ based on line profiles).  In
    all cases, the small limit for the cloud size from photoionization
    modeling and the large number of clouds needed to produce smooth
    line profiles are both consistent with the shattering process
    described in section~\ref{sec:theory}.  \label{tab:bal-results}}

  \begin{tabular}{lR@{\,--\,}LR@{}LLCCCCCC}
\toprule
% pre-heading
\multirow{2}{*}{Object}
& \multicolumn{5}{c}{Line Shape}
& \multicolumn{4}{c}{Photoionization}
& \multicolumn{2}{c}{Shattering Fragments}
\\
\cmidrule(lr){2-6}
\cmidrule(lr){7-10}
\cmidrule(lr){11-12}
% heading
\phantom{---}
& \multicolumn{2}{c}{$v_{\text{outflow}}$}
& \multicolumn{2}{c}{\!\!$\sigma_{\text{turb}}/c_{\text{s}}$}
& N_{\text{clouds}}^{\text{(region)}}\tnote{$\dagger$}
& N\,(\text{cm}^{-2})
& n\,(\text{cm}^{-3})
& f_{\text{V}}
& l_{\text{cold}}\,\text{(pc)}
& \ell_{\text{cloudlet}}\,\text{(pc)}
& f_{\text{A}}\tnote{$\ddagger$}
\\ \midrule
\multicolumn{10}{l}{\rule{0pt}{2ex}\hspace*{-0.5em}\textit{Nearby absorbers, $\lesssim1$\,pc from the black hole:}} \\
\rule{0pt}{3ex}% small vertical space
BLR\tnote{1}
& (0.01 & 0.1)\,c
& \gtrsim & 10^3\tnote{a}
& >10^8
& 10^{23}
& 10^{10}
& 10^{-6}
& 10^{-6\phantom{.5}}
& 10^{-12}
& 10^{6\phantom{.5}}
\\
BAL\tnote{2}
& (0.1 & 0.3)\,c
& \gtrsim & 10^3
& >10^5
& 10^{24}
& 10^{11}
& 10^{-4}
& 10^{-6\phantom{.5}}
& 10^{-13}
& 10^{7\phantom{.5}}
\\
mini-BAL\tnote{3}
& (0.1 & 0.2)\,c
& \gtrsim & 10^3
& >10^5
& 10^{21}
& 10^{9\phantom{0}}
& 10^{-5}
& 10^{-7\phantom{.5}}
& 10^{-11}
& 10^{4\phantom{.5}}
\\ \\
\multicolumn{10}{l}{\hspace*{-0.5em}\textit{Distant absorbers, $\gtrsim1$\,--$100$\,\,kpc from the black hole:}} \\
\rule{0pt}{3ex}% small vertical space
FeLoBAL\tnote{4}
& (0.003 & 0.02)\,c
& \gtrsim & 10^2
& \text{---}
& 10^{21}
& 10^{4\phantom{0}}
& \multirow{3}{*}{\text{---}\tnote{b}}
& 10^{-1\phantom{.5}}
& 10^{-6\phantom{0}}
& 10^{5\phantom{.5}}
\\
AAL\tnote{5\,c}
& \multicolumn{2}{c}{$\sim 0.002\,c$}
& \gtrsim & 1
& >8\tnote{$\dagger\dagger$}
% & \multicolumn{2}{c}{---\tnote{b}}
& 10^{18}
& 10^{3\phantom{0}}
&
& 10^{-4.5}
& 10^{-5\phantom{0}}
& 10^{0.5}
\\
AAL\tnote{6}
& \multicolumn{2}{c}{$\sim 0.007\,c$}
& \sim & 1
& >1\tnote{$\dagger\dagger$}
% & \multicolumn{2}{c}{---\tnote{b}}
& \color{gray}{10^{18}\tnote{d}}
& 20
&
& \color{gray}{0.02\tnote{d}}
& 10^{-3\phantom{0}}
& 10^{0\phantom{.5}}
\\
\bottomrule
\end{tabular}

\begin{tablenotes}
\item[$\dagger$] The number $N_{\text{clouds}}^{\text{(region)}}$ here
  is derived from the smoothness of the line profile, assuming the
  line comes from adding together many lines with intrinsically
  thermal widths.  Since the quasar disk is not resolved, this
  represents the number of clouds \textit{in the entire region}, not
  along a geometric line of sight.  Also, it is strictly a lower
  limit.
\item[$\ddagger$] Here, the number $f_{\text{A}}$ is our estimate for
  the number of clouds \textit{along the line of sight}.  Since the
  quasar disk is not resolved, this can in principle be lower than the
  value $N_{\text{clouds}}^{\text{(region)}}$ derived from the line
  shape.  In practice, it seldom is.
\item[\!\!\!\!$\dagger\dagger$] In distant absorbers such as FeLoBALs
  and AALs, the observationally-derived limit on the number of clouds
  is more like the $f_{\text{A}}$ we calculate from shattering theory.
\vspace*{\baselineskip}
\item[1] \citet{Rees1987,Arav1997,Dietrich1999,Bottorff2000,Netzer2006}
\item[2] \citet{Elvis2000}
\item[3] \citet{Hamann2011,Hamann2013}
\item[4] \citet{Faucher2012}
\item[5] \citet{Finn2014}
\item[6] \citet{D2004} (UM\,680)
\vspace*{\baselineskip}
\item[a] In the BLR region, an additional constraint on
  $\sigma_{\text{turb}}$ comes from radiative transfer effects
  \citep{Bottorff2000}.
\item[b] We do not compute volume-filling fractions for FeLoBALs or
  AALs, since we have no constraints on the extent of the absorbing
  region.  These are probably not associated with the quasar, so the
  distance to the quasar is not likely representative of the size of
  the region.
\item[c] The observation in \citet{Finn2014} contains many different
  components with different sizes and densities.  Here, we simply
  quote the smallest size implied by the observation.
\item[d] \citet{D2004} give $N_{\text{HI}}\sim10^{15.2}$ and
  $U\sim10^{-2.3}$.  We (crudely) estimated the ionization correction
  of $\lesssim10^{-3}$ from figure~10 in \citet{Finn2014}, under the
  assumption that similar models for the ionizing photon spectrum
  apply to both systems.
\end{tablenotes}

\end{threeparttable}
\end{table*}
Here, we discuss broad atomic lines associated with black holes.  In
(approximate) order of increasing distance from the black hole, we
discuss emission lines from the broad-line region (BLR), followed by
absorption lines from broad-absorption line (BAL) region and slightly
more distant mini-BAL regions.  At yet larger distances, we discuss
absorbers known as FeLoBALs and AALs.

Quasars and AGNs often show broad emission lines.  These lines are
interpreted to come from bound gas quite close to the black hole, at a
distance $\sim0.1$\,pc from a typical quasar.  Amazingly, despite the
strong radiation field so close to the AGN (\eg, a hydrogen ionizing
photon flux
$\Phi_{\text{ion}}\sim10^{18.5}\text{cm}^{-2}\text{s}^{-1}$), and
despite its orbital motion with speeds approaching $\gtrsim0.1\,c$,
this gas is cold and neutral, with a temperature $\sim2\times10^4$\,K
\citep[\eg][]{Bottorff2000}.  Photoionization modeling indicates a
total column density $N_{\text{H}}\sim10^{23}\,\text{cm}^{-2}$ and
volume density $n\sim10^{10}\,\text{cm}^{-3}$ for the line-emitting
gas; the corresponding volume-filling fraction is
$f_{\text{V}}\sim10^{-6}$.  Broad-line regions thus appear to comprise
many small clouds which must be confined, either by ambient hot gas,
or else somehow by magnetic stresses (see, \eg, \citealt{Rees1987} or
\citealt{Netzer2006} for reviews of broad-line regions).  Despite
their broad widths, the emission lines appear perfectly smooth, even
in high signal-to-noise, high-resolution spectra; if they are composed
of individual clouds with thermal line widths, a minimum of
$10^6$\,--\,$10^8$ emitters must be present within the broad-line
region to explain the observed smooth line profiles
\citep{Arav1997,Dietrich1999}.

Qualitatively, a large number of confined, small clouds with short
cooling times seems consistent with the shattering process discussed
in this paper.  Shattering would produce cloudlets with a scale
$\rcloudlet\sim10^{-12}$\,pc, far smaller than the observational upper
limit $l_{\text{cold}}\sim{}N/n\sim10^{-7}$\,pc.  While it is
difficult to probe such small scales directly, this prediction that
$\rcloudlet\ll{}l_{\text{cold}}$ is consistent with the need for many
clouds ($\gtrsim10^8$) clouds in the broad-line region to explain the
smooth line profiles.  Radiative transfer effects are even more
suggestive of tiny clouds: the column density
$N_{\text{H}}\sim10^{23}\,\text{cm}^{-2}$ inferred from the ionization
fraction implies gas which is too optically thick at resonant line
frequencies to explain the observed spectra.  \citet{Bottorff2000}
show that turbulent broadening
$\sigma_{\text{turb}}\gtrsim{}10^3$\,km/s is needed to lower the
optical depth sufficiently to explain certain line ratios.  As
discussed above in section~\ref{subsec:results-lines}, such large
turbulent velocities seem impossible in cold dense gas.  If the cold
gas is a fog of small droplets confined in a virialized medium,
however, such broadening can naturally be explained by virial motions
of the cloudlets: even with intrinsically thermal line widths, virial
motions would imply a broadening $\sigma_{\text{turb}}\sim{v}$,
precisely as is observed.  Taken together, these observational
properties of BLRs -- small clouds, short cooling times, low
volume-filling fractions, suprathermal random motions, and unusual
radiative transfer -- all seem strongly suggestive of shattering.

In addition to the broad emission lines discussed above, approximately
10\% of quasars also show broad \textit{absorption} lines (BALs).
When they occur, such absorption lines come from gas with a volume
density $n\sim10^{11}\,\text{cm}^{-3}$, a total column $N_{\text{H}}
\sim10^{24}\,\text{cm}^{-2}$, and with velocities spanning a range
$\sim(0.1$\,--\,$0.3)\,c$.  \citet{Elvis2000} propose that these
clouds of cold neutral gas form in a thin, conical wind originating
from the quasar accretion disk at a distance only $\sim0.01$\,pc from
the black hole.  This implies a volume-filling fraction
$f_{\text{V}}\sim10^{-4}$ of cold gas in the absorbing region.  Even
stronger constraints come from mini-BALs, thought to come from
somewhat more distant gas with lower column density,
$N\sim10^{21}\,\text{cm}^{-2}$.  Photoionization modeling of mini-BAL
clouds indicates a volume-filling fraction
$f_{\text{V}}\lesssim10^{-5}$, while partial obscuration of the quasar
disk implies a typical area-covering fraction $f_{\text{A}}\sim0.15$
\citep{Hamann2013}.  As in the broad emission-line regions discussed
above, the low volume-filling fraction of the absorbing gas and the
smooth line profiles both suggest that the region consists of a large
number of small clouds, as might be produced by
shattering.\footnote{We note, however, that a large column density
  $N\gg{10^{17}}\text{cm}^{-2}$ seems inconsistent with a low covering
  fraction $f_{\text{A}}\lesssim{1}$ in our model.  In this case, the
  low covering fraction could represent holes in the large-scale
  distribution of cloudlets.}

Absorption lines originating at larger distances from the black hole
are also suggestive of shattering.  \citet{Faucher2012} discuss a
class of quasar absorption lines known as FeLoBALs, with lower
densities ($n\sim10^4\,\text{cm}^{-3}$), column densities
($N\sim10^{21}\,\text{cm}^{-2}$), and velocities ($v\sim10^3$\,km/s)
than the BALs discussed above.  These absorption lines are interpreted
to come from outflowing atomic gas at much larger distances
($\sim1$\,--\,$5$\,kpc) from the quasar.  As in the BALs, multiple
absorption components along the line of sight and smooth, suprathermal
line widths both indicate the presence of many small cloudlets.
\citet{Faucher2012} show that these absorbers can be understood as the
remnants of a dense molecular cloud which is overrun by the quasar
shock and shredded into small cloudlets entrained in the flow.  In
order for this fragmentation to work, the authors suggest the initial
cloud must be larger than
$v_{\text{shock}}{}t_{\text{cool}}\sim10\times\cstcool$,\footnote{In
  their paper, this is expressed as the requirement that
  $t_{\text{cool}}<t_{\text{crush}}$; for a density ratio of $10^3$
  and a Mach number of 1.5.  This is consistent with our requirement
  that $R\gg\cstcool$.  Though these two models differ slightly in
  detail, they yield similar predictions for the parameters relevant
  to FeLoBAL absorbers.} where
$v_{\text{shock}}\sim\text{few}\times{100}\,\text{km/s}$ is the speed
of the shock transmitted through the cloud.  We note that the density
ratio and shock speed in our simulations shown in
figure~\ref{fig:fragmentation} closely match some of the absorbers
discussed in this paper and find a similar minimum scale for
fragmenting cold gas.

At even larger distances, and correspondingly lower speeds,
\citet{Finn2014} report ``associated absorption lines'' (AALs) from a
quasar spanning a factor of $\sim30$ in density and temperature, seen
in lines ranging from H\,\textsc{i} to Ar\,\textsc{vii}.  In this
case, partial covering of the quasar disk indicates an upper limit on
cloud sizes $\ell\lesssim10^{-2.5}$\,pc.  Photoionization modeling
indicates upper limits as small as $\ell\lesssim10^{-4.5}$\,pc,
depending on the temperature component probed.  The lines are broad,
but resolve into eight distinct velocity components, each with
approximately $\sim$\,thermal line widths.  This observation is again
strongly suggestive of the shattering discussed in this paper.  While
in most cases, we compare only to observational upper limits, the tiny
size $\ell\lesssim10^{-4.5}$pc reported by \citet{Finn2014} is
actually within a factor $\sim10$ of the prediction by shattering,
comparable to or better than our expected theoretical uncertainty.
And the thermal line widths suggest this observation reaches a scale
containing a fairly small number of cloudlets.  This is the most
constraining observation we have found; it is extremely demanding to
explain, but appears consistent with shattering.

Table~\ref{tab:bal-results} summarizes these different observations of
atomic gas near quasars.  As in section~\ref{subsec:results-galaxies},
we compute a total absorbing length $l_{\text{cold}}\equiv{N/n}$ and a
volume-filling factor $f_{\text{V}}\equiv{l_{\text{cold}}/R}$ of cold
gas.  In all cases, the absorbing length $l_{\text{cold}}$ is much
larger than \rcloudlet, consistent with shattering.  Moreover, the
volume-filling factor is much smaller than the area-covering factor
$f_{\text{A}}\sim(0.1$\,--\,$1)$; as discussed above, this geometry is
highly suggestive of a fog-like collection of a large number of tiny
clouds.  The smoothness of the absorption lines also implies a large
number of independent clouds, shown as
$N_{\text{clouds}}^{(\text{region})}$ in the table.  We compute an
expected number of clouds along the line of sight
$f_{\text{A}}\equiv{l_{\text{cold}}}/\rcloudlet$ using equation 1; in
the case of distant absorbers such as FeLoBALs or AALs, this number
may be essentially directly compared with the observational limit
$N_{\text{clouds}}^{(\text{region})}$ derived from line shapes.  BLRs,
BALs, and mini-BALs are closer to the quasar, however, and the
observational limit $N_{\text{clouds}}^{(\text{region})}$ thus probes
the \textit{total} number of clouds in the system.  This can be
compared quantitatively with shattering theory only if we make
assumptions about the geometry of the system.  Nonetheless, the basic
requirement of a large number of distributed clouds is easily
accommodated by shattering.

Thus, atomic gas in quasar outflows appears to come in the form of
small, distributed cloudlets with a low volume-filling fraction.  This
conclusion holds over a wide range of density, velocity, and distance
from the quasar, and over at least a factor of $\sim30$ in
temperature.  Taken together, these observations are strongly
suggestive of shattering.

\subsection{High-Velocity Clouds}
\label{subsec:results-hvcs}
Galaxy surveys in 21\,cm emission have found that the milky way halo
is filled with large clouds of neutral gas.  These clouds are known as
either \textit{high-velocity clouds} (HVCs) or
\textit{intermediate-velocity clouds} (IVCs) depending on their motion
with respect to the local standard of rest.  While HVCs and IVCs may
have very different origins and dynamical histories, they appear to
show similar structure; for simplicity, we refer to both as
`HVCs.'\footnote{HVCs are traditionally observed in 21\,cm emission,
  which is only possible in the milky way and in M31.  However,
  Mg\,\textsc{ii} and Ca\,\textsc{ii} absorption studies likely probe
  similar gas in more distant galaxies;
  \citep[\eg][]{Ben2008,Richter2009,Richter2012}.}  HVCs show a wide
range of metallicity, indicating diverse origins including the
galactic fountain, stripping from the small Magellanic cloud, and
accretion of pristine intergalactic gas
\citep[\eg][]{Richter2001,Richter2006}.  Some fraction of HVCs might
represent accretion from the IGM \citep{Richter2009,Richter2012}, and
may provide roughly the amount of gas needed to fuel star formation in
low-redshift galaxies
\citep[\eg][]{Shull2009,Bauermeister2010,Genzel2010}.

Since HVCs are nearby and occupy a large ($\gtrsim5$°) region on the
sky, they are ideal for studying small-scale features of cold gas.
Though HVCs have $\sim$\,kpc spatial extents, they exhibit structure
down to scales of $\sim(10$\,--\,$100)$\,AU \citep{Richter2003}.
Draco is a nearby HVC which by chance lies along a relatively empty
sightline at high galactic latitude through the ISM.
\citet{Miville2016} show it has a wispy, clumpy structure down to the
0.05\,pc resolution limit of their observations.  They report that
50\% of the resolved clumps are less than two pixels across, while
90\% are less than three pixels across; since this is a projected
area, which likely consists of many overlapping cloudlets or
filaments, it seems likely this result is consistent with the
small-scale features being entirely unresolved.  The fact that large,
cold clouds such as the Draco nebula are made up of much smaller
cloudlets seems consistent with shattering.  However, we caution that
\citet{Miville2016} study dust emission at $\sim100$\,K; shattering as
described in this paper does not strictly apply to such cold gas.  We
discuss possible applications to gas below $10^4$\,K in
section~\ref{sec:discussion}.

HVCs are traditionally identified in 21\,cm emission, which imposes
the constraints $N_{\text{HI}}\gtrsim10^{19}\,\text{cm}^{-2}$ (in
order to be detected in all-sky surveys) and
$|v_{\text{l.o.s}}|\gtrsim150$\,km/s (in order to stand out from
neutral gas in the milky way disk).  As is the case in the CGM,
absorption studies of HVCs are more sensitive to small structures and
low column densities than emission studies.  \citet{Braun2005} and
\citet{Richter2005} study Ca\,\textsc{ii} absorption and 21\,cm
absorption toward bright quasars in the vicinity of HVCs, probing
lower column-density gas than can be seen in 21\,cm all-sky surveys.
They report tiny ($\sim{10^{-2}}$\,pc) clouds with column densities
$N_{\text{HI}}\sim(0.4$\,--\,$8)\times10^{18}\,\text{cm}^{-2}$,
pressures $\sim10^3\,\text{cm}^{-3}\,\text{K}$, and thermal line
widths, in excellent agreement with shattering.  This is in the limit
where each sightline passes through only a few cloudlets, as might be
expected near the periphery of HVCs \citep{Ben2008,Ben2009}.  As in
the high-redshift CGM, these observations find a high covering
fraction along with a tiny absorbing length of cold gas, suggesting a
population of tiny, distributed cloudlets.  Moreover, the narrow line
widths and multiple absorbing components along the line of sight
suggest dense, small, clumpy structures rather than diffuse, extended
clouds.  This is confirmed by the deep 21\,cm emission image in
\citet{Braun2005}.

\citet{Richter2009} present UV spectroscopy toward bright quasars and
identify a sample of even lower column density HVC absorbers.  The
two\footnote{We suspect a typo in their reported numbers for
  RXJ\,1230.0+0115 in table~4.  We therefore do not consider this
  sightline here.}  smallest clouds they identify have
ionization-corrected column densities
$N_{\text{H}}\sim10^{17.2}\,\text{cm}^{-2}$, consistent with single
clouds produced by shattering.  These sightlines also show narrow line
widths, $\lesssim10$\,km/s, as would be expected for emission from a
single cloudlet.  These absorption observations suggest that low
column-density clouds are extremely common, even in the most tenuous
parts of the milky way halo.  The lowest column densities probed are
consistent with predictions from shattering, over a fairly wide range
in ionization state (and thus in the observable column
$N_{\text{HI}}$).

As mentioned above, HVCs are classically defined to have column
densities $N_{\text{HI}}\sim10^{19}\,\text{cm}^{-2}$ and velocities
$|v_{\text{l.o.s}}|\gtrsim150$\,km/s.  \citet{Ben2008,Ben2009,Ben2012}
suggest that HVCs represent only the high-column-density tail of a
wide distribution of high-velocity gas.  By comparison with
high-redshift studies of the CGM, we furthermore suggest the milky way
CGM shows a wide range in velocity, with most of the gas falling far
below the threshold to be considered an HVC.  The low column density
$N_{\text{H}}\sim10^{17}\,\text{cm}^{-2}$ and area-covering fraction
$f_{\text{A}}\gtrsim20\%$ compared to the high-redshift CGM
observations discussed in section~\ref{subsec:results-galaxies} likely
reflect the velocity cut ($\gtrsim150$\,km/s) for material to be
considered an HVC.  This would indicate that only a small fraction of
the milky way's CGM lies above this threshold, as seems reasonable
(\citealt{Zheng2015}), and explains why the cold ($T<10^{6}$K) mass of
the Milky Way's CGM appears 1\,--\,2 orders of magnitude less than
other $L_{*}$ spirals \citep{Stocke2013,Werk2014}, where this velocity
cut is not applied.

\subsection{Entrainment in Multiphase Galactic Winds}
\label{subsec:results-winds}
Galactic winds are a common feature of rapidly star-forming galaxies.
These winds play an important role in galaxy formation: not only do
they influence both the baryon budget and the star formation rate in
galaxies, but galactic winds also control the chemical evolution of
the intergalactic medium.  One striking feature of galactic winds is
the presence of multiphase gas spanning a wide range of density and
temperature: molecular, atomic, and x-ray emitting gas are all
detected, often moving at similar velocity (see section~1 of
\citealt{Zhang2015} for a summary of the available observations).

Such co-moving multiphase gas is difficult to understand
theoretically, because the acceleration timescale for drag forces to
make different gas phases co-move:
\begin{subequations}
\begin{align}
  t_{\text{acc}} \sim
  \left(\frac{\rho_{\text{cloud}}}{\rho_{\text{hot}}}\right)
  \frac{R_{\text{cloud}}}{v_{\text{rel}}},\label{eq:tacc}
\end{align}
is much longer than the timescale to destroy the cooler phase:
\begin{align}
  t_{\text{crush}} \sim
  \left(\frac{\rho_{\text{cloud}}}{\rho_{\text{hot}}}\right)^{1/2}
  \frac{R_{\text{cloud}}}{v_{\text{rel}}},\label{eq:tcrush}
\end{align}
\end{subequations}
by either shock-heating (i.\,e.\ ``crushing''), or by hydro
instabilities such as the Kelvin-Helmholtz or Rayleigh-Taylor
instabilities \citep[\eg][]{Klein1994,Mac1994}.  The ratio between
these two timescales,
$\sim(\rho_{\text{cloud}}/\rho_{\text{hot}})^{1/2}$, can be very large
in strongly multi-phase gas as observed in galaxy winds, suggesting we
should a priori not expect multi-phase gas to co-move (see
\citealt{Zhang2015} and \citealt{Bruggen2016} for a detailed analysis
of this problem).

Some studies have identified mechanisms to delay the disruption of gas
clouds, possibly enabling entrainment over a range of parameters
\citep[\eg][]{Cooper2009,McCourt2015}.  However, the entrainment
problem is significantly exacerbated by the observation that cold gas
is not only accelerated before it is destroyed, but is accelerated to
very high velocity (>1,000\,km/s in some cases) even at small
distances, close to the wind launching region.  These observations
suggest that that the acceleration timescale is in fact much shorter
than the crushing timescale: $t_{\text{acc}}{\ll}t_{\text{crush}}$.

One possible explanation is that the cold gas arises from thermal
instability in the hot wind, and thus could be born comoving with the
hot gas \citep[\eg][]{Martin2015,Thompson2016}, although this
possibility has not yet been verified in numerical simulations.
Regardless of its origin, however, shattering substantially modifies
the dynamics of cold gas clouds and may imply near-instantaneous
entrainment.  As an initially monolithic cloud travels through a hot
medium, it shatters into tiny cloudlets with a size
$\rcloudlet\sim\cstcool\sim(0.1\,\text{pc})/n$
(equation~\ref{eq:rcloud}).  The simulations in
figures~\ref{fig:fragmentation} and~\ref{fig:entrainment} suggest this
fragmentation enables rapid entrainment of cold gas.  While the system
of cloudlets as a whole may move rapidly, the interstitial hot gas
between the droplets shows little random motion with respect to the
cloud center of mass.  This implies little velocity shear on the scale
of the cloudlets to shred them and mix the gas phases.  However, these
results must be confirmed in a more detailed, 3D study, probably
including magnetic fields.

From a macroscopic perspective, the ``entrainment in trouble'' problem
comes from taking the ratio of equations~\ref{eq:tacc}
and~\ref{eq:tcrush}, which are appropriate for single clouds embedded
in an otherwise undisturbed flow.  Shattering the cloud into tiny
cloudlets vastly decreases the column density through the cloud
material, thereby increasing the surface area to mass ratio, like the
unfurling of a parachute.  This not only helps explain the puzzling
QSO sightline observations, in which a small total amount of cold gas
manages to cover a vast region of space, but also dramatically
amplifies the drag force on the cold gas, closely coupling the
dynamics of the hot and cold phases.

Implicit in this discussion is the assumption that cold gas mass is
conserved, \ie, that little or no mixing takes place.  Testing this
hypothesis will be require significantly more detailed work.  However,
we note that, in all of the observations we mentioned, cold gas
appears to be stable, ubiquitous, and long-lived, even on small
scales.  We do not find any compelling observational evidence for
mixing between the gas phases, or for conductive evaporation of cold
gas.  Radiative cooling is already known to strongly suppress the
small-scale mixing due to the Kelvin Helmholtz instability
\citep{Mellema2002,Cooper2009}, and this effect becomes progressively
stronger with increasing numerical resolution, ultimately resulting in
the shattering discussed in this paper.  We note that uncontrolled
numeric instabilities likely over-estimate mixing in our simulations,
perhaps significantly (Appendix~\ref{sec:sim-problems} and
\citealt{Lecoanet2016}).  We suspect that once the cooling length
\cstcool is resolved and numeric instabilities are controlled,
simulations will find that mixing is strongly inhibited.

\section{Discussion}
\label{sec:discussion}
As we discuss in Appendix~\ref{sec:sim-problems}, hydro simulations of
multiphase gas may be prone to a number of poorly-understood errors.
More direct constraints may therefore come from laboratory
experiments.  \citet{Suzuki2015} study the bow shock generated by the
collision of two counter-streaming supersonic plasma jets, meant to
mimic bow shocks due to jets in young stellar objects.  While the
plasma cooling function in these experiments is quite different from
astrophysical plasmas, it nonetheless contains a peak, and the authors
find behavior consistent with shattering: the initially smooth bow
shock fragments via thermal instability into small scale clumps of
order $\sim\cstcool$.  Such laboratory experiments are a potential way
of calibrating and corroborating numerical calculations of strongly
multiphase gas.

In the remainder of this section, we discuss theoretical uncertainties
in our work, which must be addressed in a future study
(section~\ref{subsec:disc-uncertain}), along with promising directions
for future research (section~\ref{subsec:disc-future}).  We discuss
numerical uncertainties in Appendix~\ref{sec:sim-problems}.

\subsection{Theoretical Uncertainties}
\label{subsec:disc-uncertain}
In this paper we have made a number of significant assumptions and
idealizations, whose effects should be explored in future work.  We
list several below.

\textit{\textbf{Applicability of hydrodynamics:}}
The mean free path of charged particles due to Coulomb collisions is
roughly independent of their mass.  For the conditions relevant to the
CGM, we find a collisional mean free path
$\lambda_{\text{mfp}}^{\text{(hot)}}\sim{3}\,\text{pc}$, which is much
larger than the expected size of cloudlets we consider.  We can
quantify the collisionality via the Knudsen number:
\begin{align}
  \sigma \equiv \frac{\lambda_{\text{mfp}}^{\text{(hot)}}}{\rcloudlet} 
  \sim 300 \times \left(\frac{T}{10^6\,\text{K}}\right)^2,
  \label{eqn:sat_parameter}
\end{align}
where we have assumed thermal pressure balance between the hot and
cold phases.  This large disparity calls into question where
hydrodynamics is at all applicable at the cloudlet
scale.\footnote{Note that the fluid approximation is certainly valid
  in the cold gas, where $\sigma{}\ll{}1$.  It is only invalid in the
  tenuous hot gas on cloudlet scales.}  Since particles cannot
effectively exchange momentum on such small scales, one might expect
that pressure confinement is impossible and that the cold cloudlets
would simply disperse.\footnote{If pressure confinement were not an
  issue, this large mean free path implies a very low Reynolds number
  for droplet dynamics.}

One possible conclusion is that cloud clouds only shatter down to
scales of order the mean free path.  However, even a weak magnetic
field significantly influences particle motion in dilute plasmas and
the mean free path in equation~\ref{eqn:sat_parameter} is not likely
representative if magnetic fields drape around the cloudlets.  It may
be that a fully kinetic calculation is necessary, but in this paper,
we have simply assumed that standard hydrodynamics/MHD still applies.
This assumption must ultimately be tested by PIC simulations; however,
several arguments suggest the possible validity of a fluid approach:
(i) Magnetic fields.  In the cross-field direction, the characteristic
scale is the gyro-radius $r_{\text{L}}\ll\rcloudlet$.  Swept up
magnetic fields draped across a cloudlet enable a contact
discontinuity to be maintained, as observed in galaxy cluster ``cold
fronts'' \citep{Markevitch2007}.  (ii) Plasma instabilities.  Halo gas
at cloudlet scales is similar to the solar wind: a $\beta{}\sim{}1$,
weakly collisional plasma, where plasma instabilities such as the
firehose and mirror instability, which feed off pressure anisotropies,
provide an effective scattering mechanism and can reduce the particle
mean free paths to orders of magnitude below their Coulomb value
\citep{Sharma2006,Bale2009,Kunz2014}.  The collisionless shocks
observed in galaxy clusters provide another example of regions where
density discontinuities are maintained by collisionless processes.
(iii) Galaxy clusters.  The mean free path in galaxy clusters is
$\lambda_{\text{mfp}}\sim20\,\text{kpc}$, which is comparable to the
size of cluster cores.  Yet clusters are routinely modeled with MHD,
producing results which closely match x-ray observations.  We also
note that cold gas filaments are sometimes observed in cluster cores
with scales far below the Coulomb mean free path.  (iv) Finally,
perhaps the best argument is the observational evidence, both direct
and indirect, in very different environments, for small scale
structure in cold gas.  These observations may in turn represent an
opportunity to learn and constrain plasma processes on these
collisionless scales.

\textit{\textbf{Thermal Conduction:}}
Thermal conduction erases small scale temperature structure and, since
the hot gas is collisionless on the expected scale of the cloudlets,
we might expect they are subject to a very large conductive heat flux.
\citet{Cowie1977} identify three regimes based on the saturation
parameter
$\sigma_{0}^{\prime}\equiv{}1.84\times\lambda_{\text{mfp}}^{\text{(hot)}}/\rcloudlet$:
(i) classical diffusive conduction, with $\sigma_{0}^{\prime}<1$; (ii)
saturated conduction, with $ 1<\sigma_{0}^{\prime}<100$, in which the
heat flux saturates at $Q \sim 5 P c_{\text{s}}$ since energy cannot
be transported at a rate significantly larger than the electron
thermal velocity; and (iii) a two-fluid regime with
$\sigma_{0}^{\prime}>100$, in which hot gas particles freely penetrate
the cloud \citep{Balbus1982}.  From equation \ref{eqn:sat_parameter},
we are either in regimes (ii) or (more often) (iii).  In regime (ii),
the heating timescale
\begin{subequations}
\begin{align}
  t_{\text{heat}} 
  \sim \frac{P V}{Q \times 4 \pi R^{2}} 
  \sim \frac{1}{15} \left(\frac{c_{\text{s}}^{\text{(cold)}}}{c_{\text{s}}^{\text{(hot)}}}\right) t_{\text{cool}}
  \ll t_{\text{cool}},
\end{align}
which implies rapid evaporation.  In regime (iii), hot electrons
freely permeate the cold gas (leaving aside thorny questions of
confinement and plasma instabilities).  The Coulomb heating rate is
$\Gamma=(5.5\times10^{-14}\,\text{erg/s/cm}^{-3})\,n_{\text{hot}}\,T_{\text{hot}}^{-1/2}$
\citep{Schunk1971}.  If we compare this to the cooing rate, we obtain:
\begin{align}
  t_{\text{heat}} = \left(10^{-2}\,t_{\text{cool}}\right)
  \times \left(\frac{T_{\text{hot}}}{10^7\,\text{K}}\right)^{3/2}
  \ll t_{\text{cool}},
\end{align}
\end{subequations}
where we have assumed pressure balance.  This also implies rapid
evaporation.  Thus, in principle whenever hot gas directly abutts cold
gas without a thermal conduction front, the cloudlets are evaporated.

3D simulations of cold clouds in galactic winds with classical,
diffusive thermal conduction also conclude that either thermal
conduction must be suppressed, or the cold clouds seen in outflows do
not originate from the galaxy \citep{Bruggen2016}.  In particular, low
column density clouds
$N_{\text{H}}<(10^{18}\,\text{cm}^{-2})\,(T/10^7\,\text{K})^{2}$
evaporate and disrupt rapidly.

In this paper, we assume that thermal conduction is suppressed, either
by magnetic fields or plasma instabilities.  We believe this is
well-motivated.  Conduction is essentially observationally
unconstrained; to our knowledge, there is no \textit{direct} evidence
for thermal conduction in the ISM.  The only environment where Spitzer
heat fluxes have been directly measured is in the solar wind, which is
defined by a large-scale, ordered magnetic field \citep{Bale2013}.
Even in the hot intra-cluster medium (ICM), where conduction might be
expected to be extremely effective, there is no compelling
observational evidence for conduction (\eg, broad conductive
interfaces, though note \citealt{Sparks2009}), and the literature
contains a number of assumptions ranging from complete suppression to
fully Spitzer values.  Interestingly, simulations of magnetic draping
along cold fronts suggest that significant suppression of conduction
along field lines is still required to explain the observed sharpness
of fronts \citep{ZuHone2013}, as magnetic insulation always has some
imperfections.  The required reduction could potentially come from
small-scale plasma instabilities which scatter particles and reduce
the mean free path \citep{Komarov2016}.

With regard to our specific applications to the CGM, it is worth
noting that even if we abandon our droplet model and adopt the
conventional monolithic absorber model in which a single cloud of size
$l_{\text{cold}}\sim{}N/n$ accounts for the observed absorption,
conduction \textit{still} must be suppressed in order to explain these
observational upper limits: (i) in Table~\ref{tab:obs-results}, for
the $z\sim 2$ systems \textit{a\,--\,d}, for instance,
$l_{\text{cold}}\sim20\,\text{pc}{}\ll{}\lambda_{\text{mfp}}^{\text{(hot)}}\sim(250\text{pc})\,T_{7}^{2}n_{-3}^{-1}$,
while the Field length
$\lambda_{\text{F}}^{\text{(hot)}}\sim(80\,\text{kpc})T_{7}^{2}n_{-3}^{-1/2}$.
If there is no conductive interface, and we adopt the saturated
conduction rate, the cloud will be destroyed on a hot gas
sound-crossing time:
$t_{\text{heat}}\sim{R/c_{\text{s}}^{\text{(hot)}}}\sim2\times10^{4}\,\text{year}$.
(ii) Even in stable or condensing scenarios, where radiative losses
can outstrip conductive heat supply, thermal contact must be
sufficiently poor that conduction only supplies a small fraction of
radiated energy.  Otherwise, the cold gas would rapidly cool the hot
gas, leading to a massive cooling flow.  Given that
$L\propto\int{}n^{2}\Lambda(T)\,dV{}\propto{}M{}n{}\Lambda(T)$, the
radiative cooling rate in cold gas can be significantly higher than in
the hot gas:
\begin{align}
  \frac{L_{\text{cold}}}{L_{\text{hot}}} 
  &\sim \frac{n_{\text{cold}} M_{\text{cold}} \Lambda(T_{\text{cold}})}{n_{\text{hot}} M_{\text{hot}} \Lambda(T_{\text{hot}})} \nonumber \\
  &\sim 100 \left( \frac{\delta}{10^{3}} \right) \left( \frac{f_{\text{M}}}{10^{-2}} \right) \left( \frac{\Lambda(T_{\text{cold}})/\Lambda(T_{\text{hot}})}{10} \right)
\end{align}
where
$\delta\equiv{}n_{\text{cold}}/n_{\text{hot}}\sim{}T_{\text{hot}}/T_{\text{cold}}$
is the density contrast, and $f_{\text{M}}$ is the cold gas mass
fraction.  This results in a cooling time for the hot gas of:
\begin{align}
  t_{\text{cool}}^{\text{(hot)}} 
  &\sim \frac{M_{\text{hot}} k T_{\text{hot}}}{n_{\text{cold}} M_{\text{cold}} \Lambda(T_{\text{cold}})} 
  = \frac{\delta}{f_{\text{M}}} t_{\text{cool}}^{\text{(cold)}} \nonumber \\
  &\sim \left(5 \times 10^{7}\,\text{year}\right)\times\left( \frac{\delta}{10^{3}} \right) \left( \frac{f_{\text{M}}}{10^{-2}} \right)^{-1}
\end{align}
which would cause a complete collapse of the hot hydrostatic
atmosphere.  This implies that we require $t_{\text{conduct}} \gg
t_{\text{cool}}$ in the cold gas and strongly argues against scenarios
where cold clouds are continuously evaporated and reformed: they are
too efficient as coolants of the hot gas.  Since the
$\text{Ly}_{\alpha}$ luminosity is comparable to the X-ray luminosity
of low-redshift clusters, this is similar to the cooling flow problem
there, which is generally resolved by heat input into the hot phase
(\eg, by AGN).  Having cold clouds as a coolant is especially
pernicious, however, since it produces a cooling flow problem in
regions which otherwise have a long cooling time (\eg, the outskirts
of galaxy clusters) and where heating is not expected to be effective.
Since we directly observe the cold gas in absorption, we interpret
this as strong evidence that thermal conduction is suppressed, at
least at the interface between cold and hot gas.

\textit{\textbf{Effect of Magnetic Fields:}}
As already discussed, magnetic fields affect diffusion transport
coefficients such as conduction and viscosity.  The effects of
magnetic pressure and tension can also alter the dynamics of
shattering.  In preliminary work, we have found that thermal
instability in a stratified medium is \textit{enhanced} by magnetic
tension, since tension suppresses the internal gravity waves which
quench instability (Ji et al.\ 2016, in preparation).  In principle,
magnetic pressure could inhibit shattering, if the cold phase is
magnetically rather than thermally supported.  In an extreme case,
where the plasma $\beta \ll 1$, isochoric cooling without shattering
might be feasible.  In practice, we expect magnetic fields to
introduce anisotropy into the shattering process.  Note that in the
ISM, the magnetic field strength appears to be independent of
temperature or phase for $n\leq300\,\text{cm}^{-3}$
\citep{Crutcher2010}, which is potentially explicable by reconnection
diffusion \citep{Lazarian2014}.  Non-thermal pressure support appears
to be important primarily in much denser gas, which is
self-gravitating.

\textit{\textbf{Photoionization:}}
We have assumed cooling functions appropriate for gas in collisional
ionization (coronal) equilibrium.  However, ISM/CGM gas is illuminated
by both the local galactic radiation field and the metagalactic
UV/X-ray background.  Both photo-heating and photoionization of
important atomic coolants can significantly change cooling
efficiencies as a function of temperature
\citep{Wiersma2009,Cantalupo2010,Gnedin2012,Kannan2014}.  For typical
radiation fields, with gas at low density
($n\sim10^{-3}\,\text{cm}^{-3}$) and optically thin to H/He
photoionizing radiation, the effect is modest at $T>10^{5}$\,K (when
metal line cooling prevails), but leads to a drastic reduction of
cooling below $ T \sim 10^{5}$K and, depending on the hardness and
strength of the radiation field, an elevated equilibrium temperature
$10^{4}\,\text{K}<T<10^{5}\,\text{K}$.  Thus, it is unlikely to be
important in high-redshift systems (\eg, items \textit{a\,--\,e} in
Table~\ref{tab:obs-results}), which are much denser
($n\sim1\,\text{cm}^{-3}$) and self-shielding
($N_{\text{HI}}>10^{17}\,\text{cm}^{-2}$), with inferred neutral
fractions from photoionization modeling
$x_{\text{HI}}\sim0.1$\,--\,$1$.  However, these effects are
potentially important in low-redshift systems \citep{Stocke2013,
  Werk2014}, which have much lower densities and are optically thin to
H/He ionizing radiation.  Naively, if we assume shattering halts at
$T\sim10^{5}$\,K, then from the red curves in
figure~\ref{fig:cooling-curve}, there would only be an order of
magnitude increase in $N_{\text{cloudlet}}$.  However, the dynamics of
shattering may change, given a reduced temperature contrast between
ambient and cool gas.  These considerations may also be important for
the abundance of O\,\textsc{vi} absorbers (since the collisional
ionization abundance of O\,\textsc{vi} peaks at
$T\sim2\times10^{5}$\,K).

\textit{\textbf{Relevance to low-z halos:}}
Both magnetic fields and photoionization are potentially relevant to
the qualitatively distinct features inferred in low-redshift CGM by
\citet{Werk2014}.  In particular, \citet{Werk2014} report very low
densities for the cold gas in low-redshift galaxies (comparable to the
expected density for the hot gas), along with high column densities of
cold gas, $N_{\text{H}}\sim{10^{20}}\,\text{cm}^{-2}$.  The large
volume filling factor implied by this low density is more consistent
with a single, monolithic absorber than with a fog of cloudlets.
However, these inferences are more model-dependent than high-$z$
observations:\footnote{Or BAL observations, where extremely high
  densities imply low ionization parameters and high neutral
  fractions.} unlike the latter, the low-redshift absorbers are
optically thin and highly photoionized, with a large (and uncertain)
photoionization correction ($x_{\text{HI}}\sim10^{-3.5}$, compared to
$x_{\text{HI}}\sim 0.1$\,--\,$1$ at high redshift).  This leads to
sensitivity to the assumed ionizing spectrum, as well as the
assumption of a single temperature and density for all absorption
lines.  Models which allow for a range of densities obtain
qualitatively different results \citep{Stern2016}.  Moreover, assuming
pure photoionization with all gas at $T=10^{4}$\,K ignores higher
temperature collisionally ionized gas.  Cosmological simulations which
can account for this would report artificially low gas densities if
pure photoionization is assumed (A.\ Kravtsov, private communication).
Nonetheless, if these low reported densities and large cloud sizes
continue to be found, this potentially points toward different cooling
or shattering dynamics in the CGM of low-redshift galaxies,
potentially due to non-thermal pressure support (such as Wiener et
al. 2016, in preparation) or photoionization effects.

Another low-redshift observation which seems inconsistent with our
model is the study of absorption sightlines through the Virgo cluster
in \citet{Yoon2012}.  While these authors find a large covering
fraction of neutral hydrogen, the column densities
($N_{\text{HI}}\sim10^{13}\,\text{cm}^{-2}$) are very low compared to
the column $N_{\text{cloudlet}}\sim10^{17}\,\text{cm}^{-2}$ we expect
through a single cloudlet.

\textit{\textbf{Collisions and Coagulation:}}
While we have identified a characteristic scale for cold gas, not all
the cold gas necessarily shatters down to this scale.  For instance,
the Jeans mass implies a characteristic scale for fragmentation, but
collisions and coagulation allow for a broad (power-law) range of
masses which cuts off at the Jeans mass \citep{Bonnell2003}.  The
coagulation process also depends on non-thermal components such as
magnetic fields, turbulence and details of turbulent dissipation and
mixing between hot and cold phases.  These are prone to artificial
coagulation if under-resolved (see the third panel in
figure~\ref{fig:cloud-plot} and discussion in text) and demand
detailed 3D simulations, which are beyond the scope of this paper.

\textit{\textbf{Cooling below 10\textsuperscript{4}\,K:}}
In this paper, we have only considered a two-phase medium, with
$T\sim10^{4}$\,K gas and a significantly hotter phase.  We have not
explored whether there could be shattering in even colder gas, which
cools via atomic fine structure lines, molecular, or dust cooling.  We
conjecture (but do not show) that pressure confined gas shatters down
to the \textit{minimum} in the parameter \cstcool.  Since this minimum
generally lies at $T\sim10^{4}$\,K, this would imply that gas cooling
to much lower temperatures would not shatter any further but remain at
the scale \rcloudlet we have identified.  In other words, below
$T\sim10^{4}$\,K, clouds cool isobarically and always remain in sonic
contact with their surroundings.  Note, however, that cold molecular
gas tends to be self-gravitating and highly supersonically turbulent,
so the governing dynamics change dramatically.

\subsection{Implications and Directions for Future Work}
\label{subsec:disc-future}
Since multi-phase gas is ubiquitous in astrophysics, and we expect
shattering to occur whenever the cold phase is pressure confined, the
work presented here may have broad applications.  We list a few
possibilities here.

\textit{\textbf{Turbulence in Galaxy and Cluster Halos:}}
Even weak turbulence in the hot hydrostatic atmospheres of galaxy
clusters can have an outsized impact on their evolution: turbulence
advects metals and entropy, affects the topology and (via the
turbulent dynamo) strength of magnetic fields, and thus influences the
diffusion coefficients such as thermal conduction and viscosity.  At
larger amplitudes, turbulence might be an important source of pressure
support, influencing both the evolution of galaxy clusters and their
utility for constraining dark energy.  Thus, directly measuring
turbulent line broadening in clusters was an important goal for
Hitomi, which unfortunately in its brief lifetime was able to make
only a single measurement, finding a turbulent energy density
$U_{\text{turb}}\sim 4\%\,U_{\text{therm}}$ in the core of the Perseus
cluster \citep{Hitomi2016}.

If shattering enables the rapid entrainment of cold gas as we suggest
in sections~\ref{subsec:sims} and~\ref{subsec:results-galaxies}, cold
gas may directly reflect the kinematics of its surrounding hot gas.
Indeed, this is our interpretation of the broad hydrogen
$\text{Ly}_{\alpha}$ line widths seen at high redshift
\citep{Hennawi2015,Borisova2016}.  This would imply that Keck
spectroscopy of cold gas in clusters could measure ICM turbulence with
much higher signal-to-noise and spectral resolution than was possible
with Hitomi.\footnote{Hitomi instrumental resolution was comparable to
  the expected turbulent broadening ($\sim\,150\,\text{km/s}$),
  whereas optical/UV spectroscopy has an order of magnitude higher
  resolution.}  Moreover, Hitomi only had the surface brightness
sensitivity to observe a small handful of nearby clusters, largely in
their cores.  By contrast, even if the $\text{Ly}_{\alpha}$ surface
brightness is too faint to image in nearby clusters, absorption-line
spectroscopy toward background quasars could still be performed for a
large number of systems.  Unlike emission-line observations, this
sensitive probe extends to high redshift, to lower mass systems, and
out to the virial radius (as the \citealt{Hennawi2015} observations
make clear).  At high spectral resolution, the details of the line
shape can yield rich information about turbulence, such as its
volume-filling factor and the importance of bulk motion
\citep{Shang2012}.  If cold CGM gas such as that found in
\cite{Hennawi2015} in fact persists in low-redshift clusters, then
targeted spectroscopy of background quasars behind galaxy clusters
could yield rich dividends.

\textit{\textbf{UV Escape Fractions from Low-Mass Galaxies:}}
Shattering vastly increases surface area to volume for neutral gas,
implying that the area covering fraction $f_{\text{A}}$ may be large,
even when the mass fraction of cold gas is negligible.  For a fixed
column density of neutral gas, this effectively increases the optical
depth and suggests that the escape fraction of ionizing photons
$f_{\text{esc}}$ could be very small due to a fog of obscuring cold
gas which envelops the entire halo.  The escape fraction
$f_{\text{esc}}$ is perhaps the most uncertain and hotly debated
parameter in models of reionization, with few available empirical
constraints \citep[\eg,][]{Siana2010}.  In our model, the escape
fraction is likely to \textit{increase} in lower mass halos (which are
critical for reionization) for three reasons: (i) below
$\sim{}10^{11}\,M_{\odot}$, galaxy halos may lose the hydrostatic
atmospheres in which droplets can remain suspended
\citep{Birnboim2003,Keres2005,Fielding2016}; (ii) even if halo gas
persists in halos less massive than $\sim{}10^{11}\,M_{\odot}$ (\eg,
due to heating processes or galactic winds), the reduced temperature
contrast between a lower virial temperature and the $\sim10^{4}$\,K
cold phase may inhibit shattering; and (iii) the reduced column
density of gas in smaller halos could also mean that cold gas is no
longer self-shielding to the external UV background.\footnote{There
  appears to be tentative evidence for such a sharp transition in the
  covering fraction of optically thick gas around quasar halos
  \citep{Prochaska2013} ($f_{\text{cov}}\sim65\%$) compared to that
  around Lyman break galaxies \citep{Rudie2012}
  ($f_{\text{cov}}\sim{}30\%$) at $z\sim{}2$\,--\,$2.5$, over a factor
  of only $\sim{}3$ in typical halo mass.}  Observationally, comparing
the emissivity inferred from the $\text{Ly}_{\alpha}$ forest and UV
luminosity functions implies that the escape fraction must increase
with redshift \citep{Kuhlen2012,Becker2013}, which could be linked to
tentative evidence that $f_{\text{esc}}$ increases at lower UV
luminosity \citep{Dijkstra2016}.

This fog of cold gas also has crucial effects on radiative transfer of
resonant line photons, most notably $\text{Ly}_{\alpha}$.  Each of the
cloudlets is still optically thick to $\text{Ly}_{\alpha}$
($N_{\text{HI}}>10^{14}\,\text{cm}^{-2}$ for typical ionization
corrections, particularly once shadowing and mutual shielding from the
radiation field is taken into account).  Radiative transfer in a
multi-phase medium, with scattering off numerous optically thick
clouds \citep{Neufeld1991,Hansen2006} has distinct and unique
properties.  The line profile, escape fraction in the presence of
dust, etc., all change considerably.  Moreover, a fog of cloudlets has
orders of magnitude more scattering clouds than have been considered
in theoretical studies to date.  As previously discussed, due to their
entrainment within the hot gas, these cloudlets may sustain large
relative velocities.  In this scenario, the mean free path of
$\text{Ly}_{\alpha}$ photons could be comparable to the correlation
length of the velocity field, invalidating the common assumption of
micro-turbulent broadening (\ie, by adopting a turbulent
$b$-parameter).  Detailed comparison of $\text{Ly}_{\alpha}$ radiative
transfer calculations to observations could sharpen and test our
models for cold gas.  An even more extreme environment where all these
effects would be taking place is the broad-line region around quasars
(\S~\ref{subsec:results-bals}).

\textit{\textbf{Cosmological Simulations:}}
Our results have important consequences for cosmological simulations.
While zoom-in simulations achieve high ($\sim\text{pc}$) resolution in
star forming regions, the effective Lagrangian nature of AMR or SPH
codes implies comparatively poor resolution ($\sim\text{kpc}$) in
galaxy halos.  It is currently infeasible to resolve shattering in
cosmological simulations, and this failure may result in artificial
mixing and destruction of cold gas.\footnote{Interestingly, the poor
  mixing seen in ``old'' SPH or low resolution codes may in fact be a
  more accurate representation of the two gas phases.}  Instead, we
advocate that cold gas should be modeled in a sub-grid fashion similar
to how stars are modeled, with source and sink terms, and with a drag
force coupling it to the hot phase.  We plan to develop and test such
an implementation in the future.

\textit{\textbf{Small Coherence Scale in Low-Column Absorbers:}}
A prediction of our model is that the coherence scale of low column
density absorbers ($N_{\text{H}}<10^{18}\,\text{cm}^{-2}$) should be
undetectably small for absorption line measurements along multiple
sightlines to lensed quasars.  Alternatively, in the Milky Way, the
very small size of a low column density cloudlet could potentially be
constrained by proper motion: the absorption signature against a
background quasar could disappear on $\sim$\,decade long timescales.
This could also be tested using time variability of absorption lines
in quasar spectra, both from local absorbers
\citep[\eg][]{Hamann1995}, and from absorption by intervening galaxies
\citep[\eg][]{Hacker2013}.

\section{Summary}
\label{sec:summary}
As discussed in section~\ref{sec:intro}, several observational
properties seem to be typical of cold ($10^4$\,K) gas in astrophysics,
while at the same time very difficult to explain.  These include broad
absorption and emission lines, large area-covering fractions relative
to the volume-filling fraction, and co-moving hot and cold phases; in
most cases, \textit{none} of these observations are predicted by
current theoretical models of cold gas.  In this paper, we suggest
that such apparent inconsistencies stem from an implicit assumption
that cold gas in astrophysics comes in the form of contiguous,
monolithic ``clouds.''  We suggest that many observations are
naturally explained if cold gas instead takes the form of tiny
cloudlets, distributed sparsely throughout space in a manner analogous
to a mist or a fog.  In this sense, we suggest that astrophysical
``clouds'' have a similar form to actual, meteorological,
\textit{clouds}.

In section~\ref{subsec:hand-wavy} \citep[cf.][]{Voit1990}, we describe
a hydrodynamic process which quickly fragments cold gas to tiny scales
and may produce a fog-like collection of small, distributed cloudlets
as implied by observations.  This process, which we call
``shattering,'' is closely analogous to fragmentation by the Jeans
instability, but with contraction driven by cooling and compression by
ambient pressure instead of by self-gravity.  We expect cold gas to
fragment down to the lengthscale $\cstcool\,{\sim}\,0.1\,\text{pc}/n$,
where $n$ is the volume density in $\text{cm}^{-3}$, such that
individual cloudlets have a characteristic column density
$N_{\text{cloudlet}}\sim{10^{17}}\,\text{cm}^{-2}$, relatively
independent of their environment.  Interestingly, this column density
implies that, under a wide range of conditions, individual cloudlets
are (depending on ionization corrections) either optically thin or at
most marginally optically thick to continuum radiation, but optically
thick to resonant lines.  We motivate the scale \cstcool in
section~\ref{subsec:hand-wavy}, we demonstrate fragmentation to this
characteristic scale in simulations in section~\ref{subsec:sims} and
in figure~\ref{fig:fragmentation}, we compare our results to recent
laboratory experiments in section~\ref{sec:discussion}, and we discuss
connections to previous work in Appendix~\ref{sec:sim-problems}.  Our
simulations and calculations nonetheless contain significant
uncertainty; we discuss the most relevant shortcomings in
Appendix~\ref{sec:sim-problems} and in
section~\ref{subsec:disc-uncertain}.  We therefore rely heavily on
observational data for interpreting and corroborating our model.

We compare our results to astrophysical observations of cold gas in
section~\ref{sec:results}, finding evidence for shattering in the
circumgalactic medium in galaxy halos
(\S~\ref{subsec:results-galaxies}), in the broad line-widths
frequently observed (\S~\ref{subsec:results-lines}), in quasar spectra
(\S~\ref{subsec:results-bals}), in HVCs
(\S~\ref{subsec:results-hvcs}), and in galactic winds
(\S~\ref{subsec:results-winds}).  In these cases, the small
characteristic lengthscale implied by shattering may explain
observational features which are otherwise very difficult to
understand.  Our comparison with observations is necessarily
incomplete, but suggests how shattering may be tested quantitatively
in the future with more detailed studies.

In this paper, we are intentionally agnostic about the \textit{origin}
of cold gas, and thus cannot specify the amount of cold gas present in
any particular environment; we simply show that shattering is
inevitable once it forms.  In galactic halos, perhaps the two most
likely sources of gas are thermal instability and galactic winds.
Another source of cold gas may be cold streams accreting from the
cosmic web.  An analysis of the adiabatic case indicates they may be
unstable to the Kelvin-Helmholtz instability \citep{Mandelker2016}.
These streams would likely shatter in the presence of radiative
cooling, perhaps explaining the apparent prevalence of cold gas even
in the extreme outskirts of galaxy halos.

\appendix
\section{Simulation Setup}
\label{sec:sim-method}
\begin{figure*}
  \centering

  \hspace*{\fill}$R_0\sim{100}\,\cstcool$\hspace*{\fill}%
  \hspace*{0.05\linewidth}%
  \hspace*{\fill}$R_0\sim{30}\,\cstcool$\hspace*{\fill}

  \hspace*{\fill}$t\sim{7}\,t_{\text{crush}}$\hspace*{\fill}%
  \hspace*{0.05\linewidth}%
  \hspace*{\fill}$t\sim{7}\,t_{\text{crush}}$\hspace*{\fill}

  \vspace*{\baselineskip}

  \includegraphics[width=0.475\linewidth]{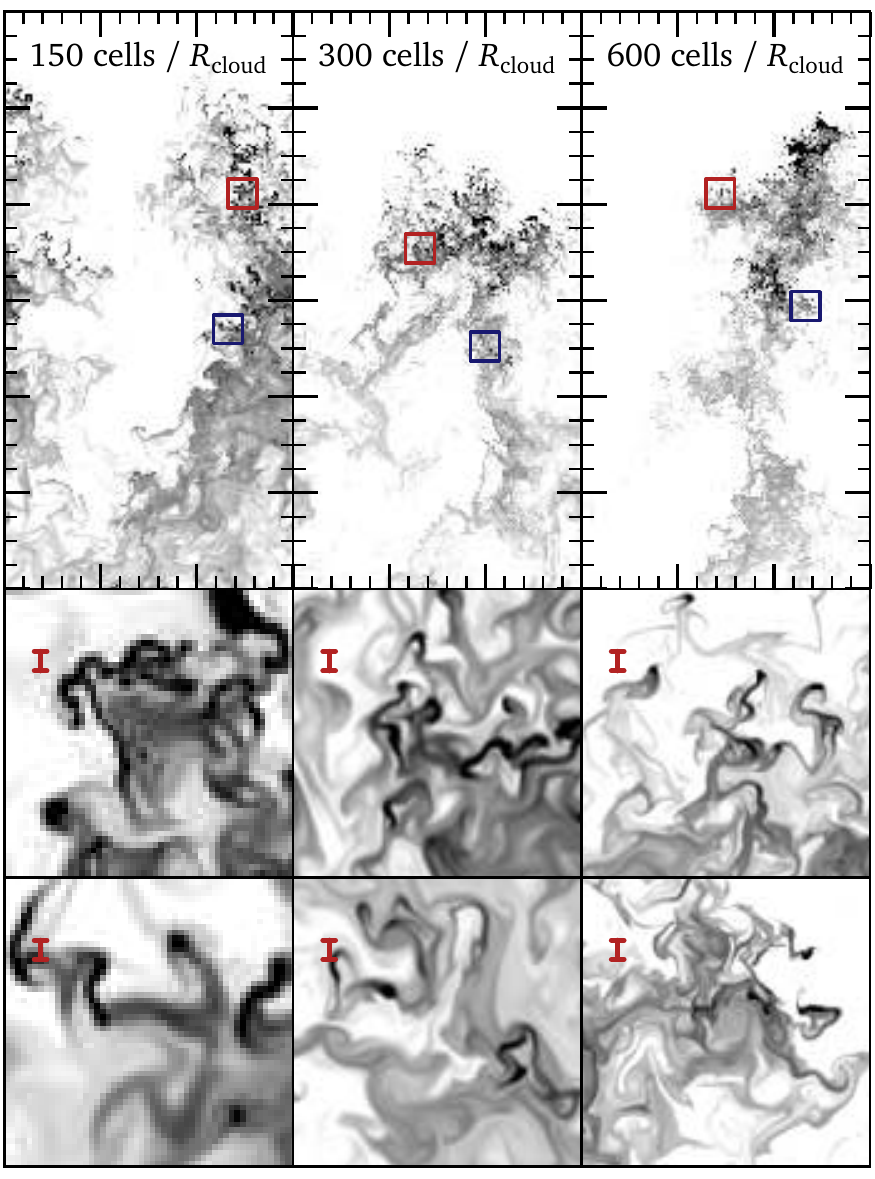}%
  \hspace*{\fill}%
  \includegraphics[width=0.475\linewidth]{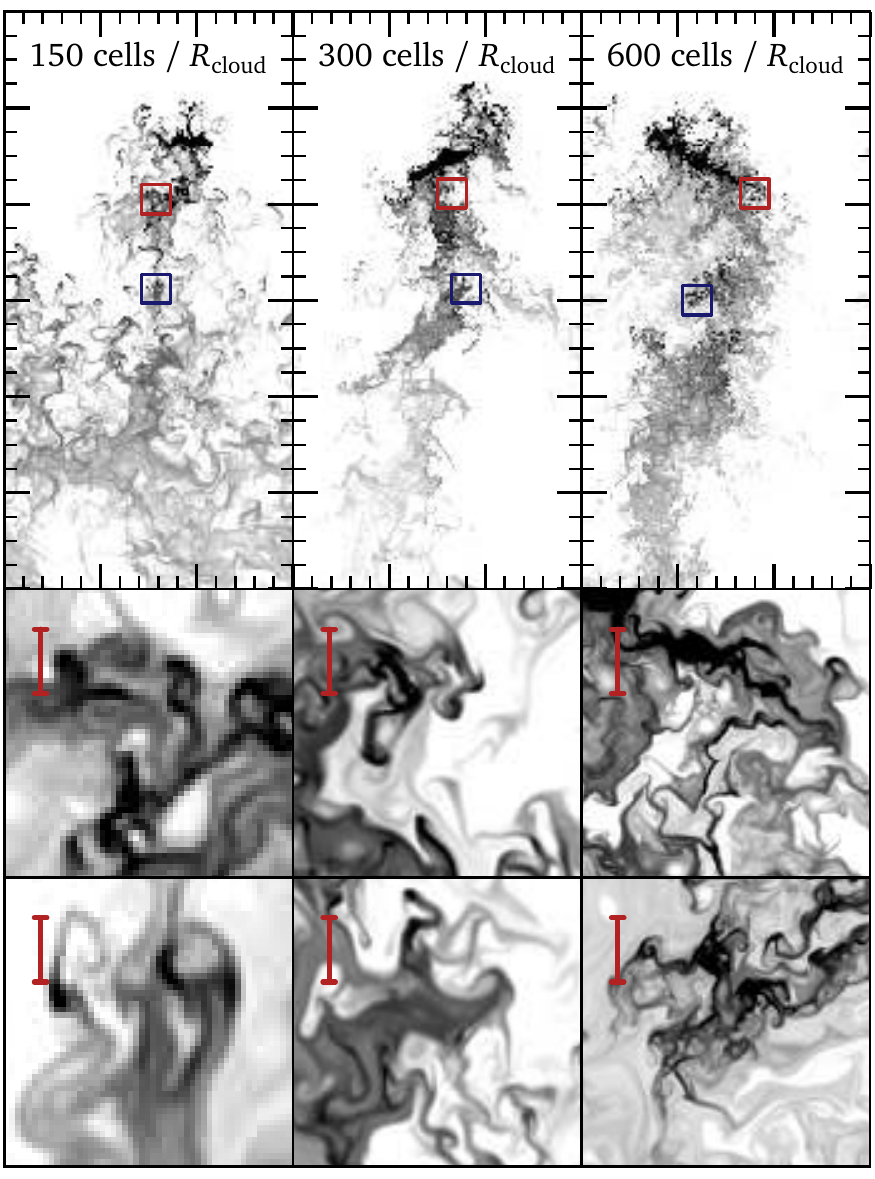}

  \caption{Resolution studies for ``cloud crushing'' simulations shown
    in figure~\ref{fig:fragmentation}.  The \textit{left} plot shows a
    simulation with an initial cloud size $R_0\sim{100}\,\cstcool$
    (second from the right in figure~\ref{fig:fragmentation}), and the
    \textit{right} plot shows a simulation with an initial cloud size
    $R_0\sim{30}\,\cstcool$ (not shown in
    figure~\ref{fig:fragmentation}).  We show simulations with
    resolutions of 150, 300, and 600 cells per cloud radius; we show
    simulations with a resolution of 300 cells/$R_{\text{cloud}}$ in
    the rest of our figures.  As in figure~\ref{fig:fragmentation},
    the squares (\textit{red, top; blue, bottom}) mark regions show in
    the insets below; zooming in reveals that the apparently wispy
    regions in the wakes behind the clouds are in fact filled with
    small dense knots of cold gas.  The simulations do not converge,
    are are not expected to since they contain unresolved gradients
    and do not include explicit dissipation; however, the knots of
    cold gas appear to track the lengthscale \cstcool, marked in the
    insets with a \textit{red} error bar.\label{fig:res-1e2}}
\end{figure*}
\begin{figure}
  \centering
  \includegraphics[width=\linewidth]{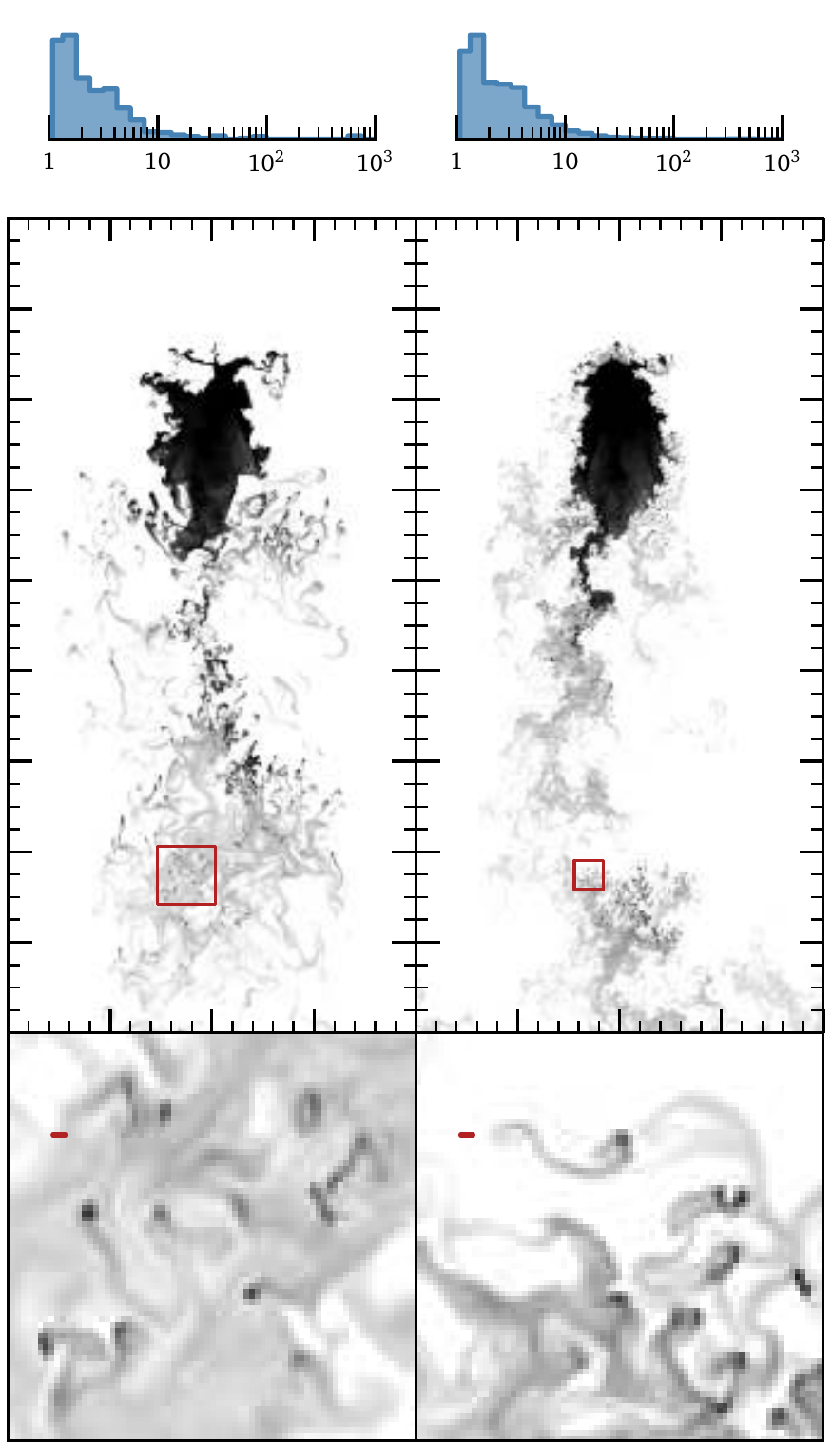}
  \caption{Clump formation in unresolved simulations with an initial
    cloud size $R_0={10^4}\,\cstcool$.  Mixed, $\sim10^{5}$\,K gas in
    the wake of the cloud is thermally unstable and rapidly cools to
    form small clumps of cold gas.  These clumps form at roughly the
    grid scale of the simulation: higher resolution simulations
    produce more, and increasingly smaller, clumps.  The clump size
    histograms shown at the top are essentially identical when put in
    units of the cell size; increasing the resolution by a factor of 4
    (from 150\,cells per cloud radius to 600) simply decreased all of
    the clump size by the same factor.  In both simulations, the clump
    sizes are therefore consistent with zero; they are totally
    unresolved. \label{fig:unresolved-convergence}}
\end{figure}
\begin{figure*}
  \centering
  \includegraphics[width=\linewidth]{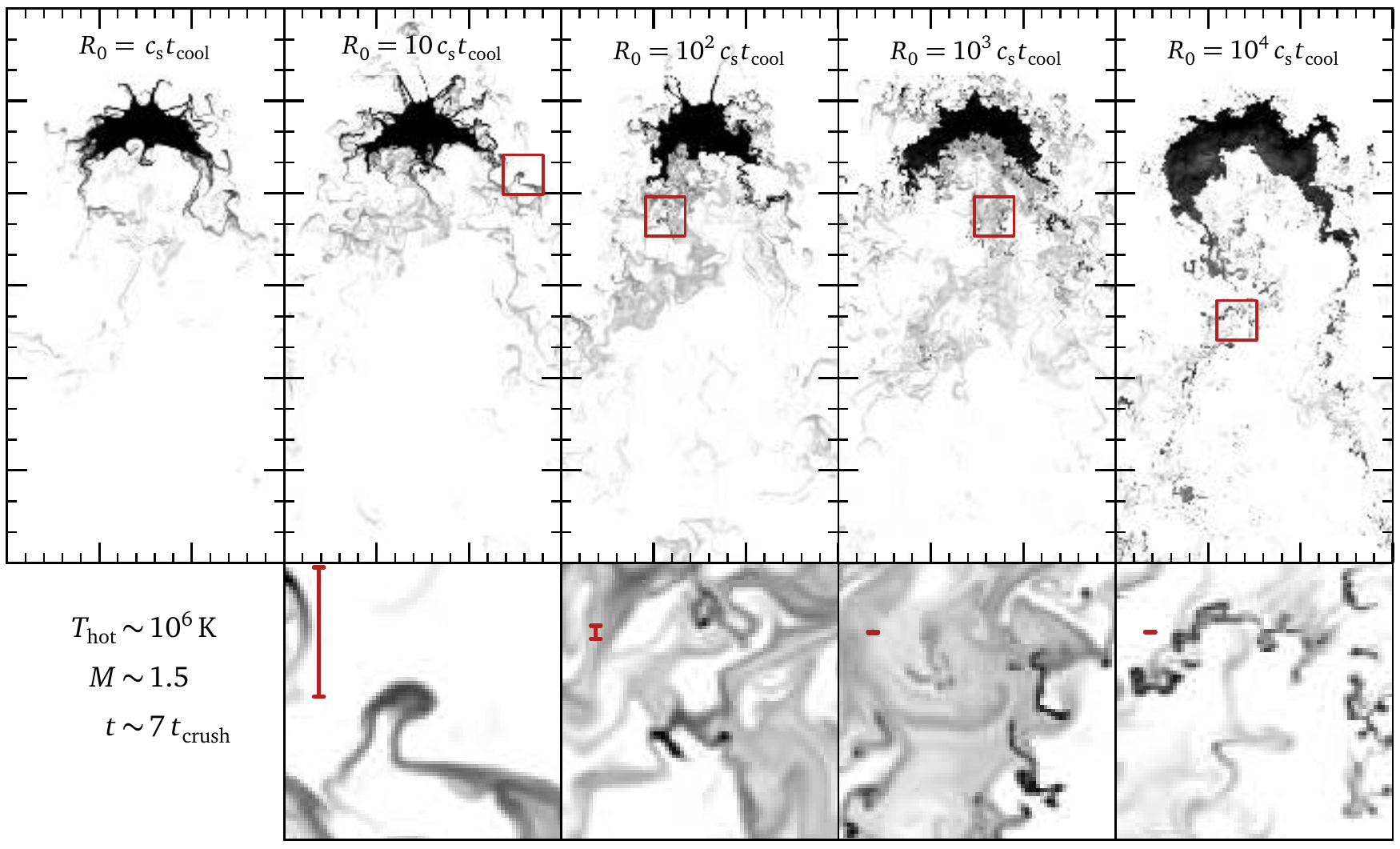}

  \vspace*{\baselineskip}

  \includegraphics[width=\linewidth]{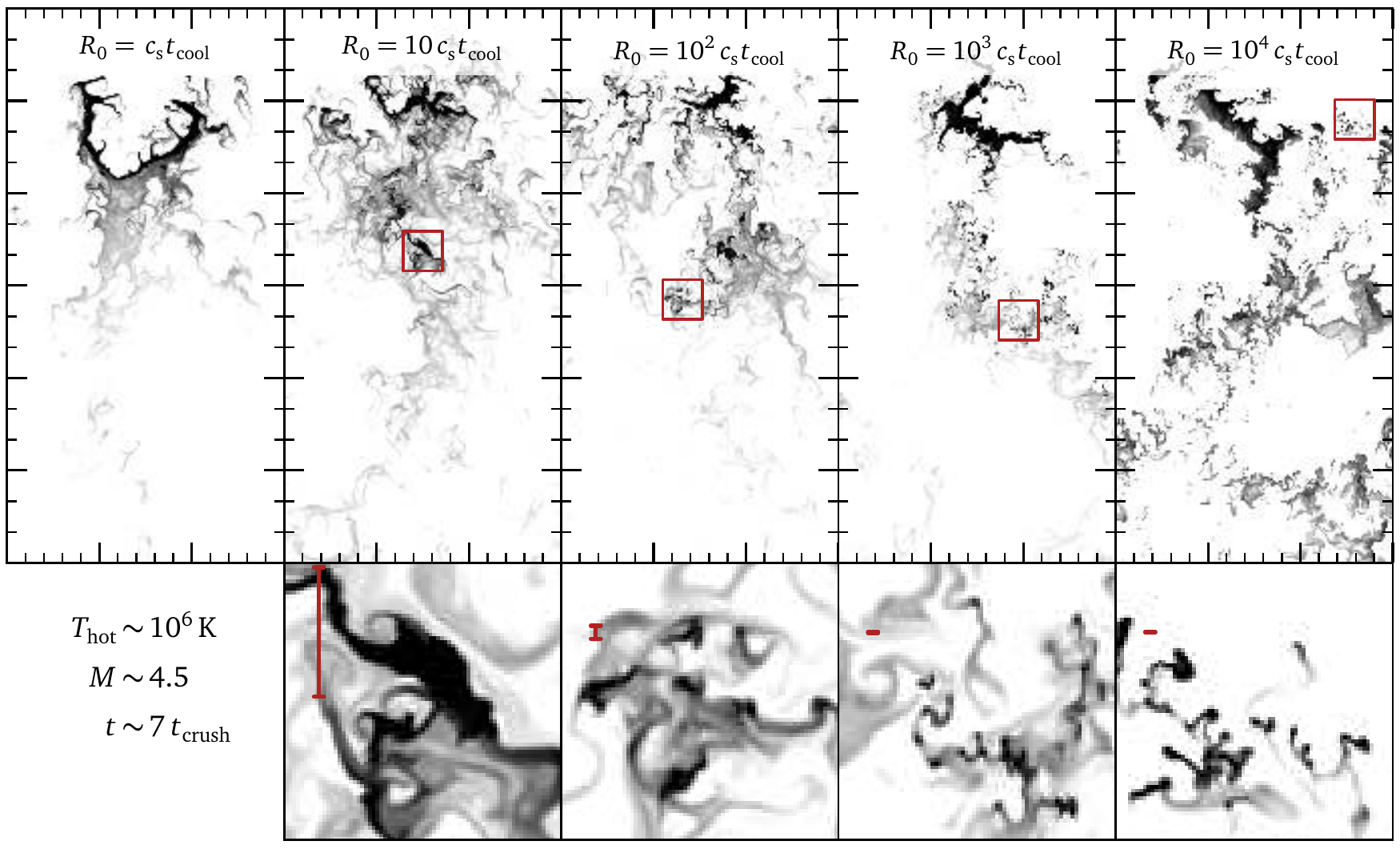}

  \caption{Same as figure~\ref{fig:fragmentation} but for a $10^6$\,K
    halo.  Though the disruption of the initial cloud proceeds very
    differently here due to the lower density ration, the clump size
    distributions are virtually identical to those in
    figure~\ref{fig:fragmentation}.  This strongly suggests that it is
    the cooling rate and properties of the cold gas which determines
    the cloud size, rather than the properties of the hot gas as
    suggested by \citet{Burkert2000} and \citet{Audit2005}.  The lower
    plot shows a repeat simulation with Mach number $\sim{}4.5$ (\ie,
    with the same velocity as in figure~\ref{fig:fragmentation}).  In
    this case, increasing the Mach number by a factor $\sim{}3$
    increases the cloudlet size by a factor $\sim{}2$, suggesting that
    heating does play a small role in determining the cloudlet
    size\label{fig:fragmentation-1e6}}
\end{figure*}
This section details our computational method.  In all of our
simulations, we integrate the usual equations of ideal hydrodynamics
using the conservative code Athena \citep{Stone2008, Gardiner2008}.
In order to ensure that we resolve the dynamics in the wakes of
disrupting clouds, we use a fixed Cartesian grid with equal resolution
throughout the domain.  Simulations of multiphase gas contain very
high Mach numbers and sharp density gradients; since this combination
makes conservative codes prone to crashes, we use the ``van Leer''
integrator implemented in Athena, with second-order reconstruction in
primitive variables.  This choice is more diffusive, and hence more
robust, than the Athena's default integrator.  We have checked that
shattering still happens with the CTU integrator with 3rd order
reconstruction; if anything, we suspect our choice overestimates
mixing and the destruction of cold cloudlets.

We note that every simulation in this paper assumes ideal
hydrodynamics and is two-dimensional; we discuss these significant
limitations in Appendix~\ref{sec:sim-problems} and in
section~\ref{subsec:disc-uncertain}.

By construction, in many of our simulations the cooling time of the
gas becomes very short compared to the dynamical time.  This is
expensive to simulate using methods which add cooling as a source term
to the integrator, as these methods limit the simulation time-step to
a small fraction of the cooling time.  Instead, we implemented the
``exact'' cooling algorithm described in \citet{Townsend2009}, which
rewrites the operator-split energy equation as
\begin{align}
  t_{\text{cool}} \frac{d T}{d t} = - T_0  \frac{\Lambda(T)}{\Lambda(T_0)},\label{eq:townsend-cool}
\end{align}
where $T_0$ is the temperature at the start of a time-step.
\citet{Townsend2009} solves equation~\ref{eq:townsend-cool} by
separating variables.  This integral has a closed-form solution for
piecewise power-law cooling curves $\Lambda(T)$ such as our fit to the
\citet{Sutherland1993} curve shown in figure~\ref{fig:cooling-curve},
enabling an efficient and accurate representation of cooling, even for
very short cooling times.

Since it is already computationally expensive to resolve the scale
\cstcool, we apply a temperature floor at $10^4$\,K to prevent
cloudlets from cooling further and contracting to yet smaller scales.
Cooling below $10^4$\,K must be addressed in a future study.

\subsection{Thermal Instability Simulations}
The simulations are shown in figure~\ref{fig:cloud-plot}.  We begin
with a domain in which $\rho\sim{1}$ and $T\sim{1}$.  We chose our
unit system such that this corresponds to diffuse gas with a density
$n\sim{10^{-4}}\,\text{cm}^{-3}$ and $T\sim{10^7}$\,K.  We use a high
temperature such that this gas has a long cooling time and we do not
need to include heating or feedback processes to maintain an ambient
hot phase.

We generate a Gaussian-random scalar field $\delta$ which has a power
spectrum $\propto{k^{-1}}$ over a range of scales from $k\sim{2\pi}/L$
to $k\sim{40\times}{2\pi}/L$, where $L$ is the smallest size of the
domain.  We normalize the scalar field $\delta$ to have an RMS of
unity, then apply the density perturbation:
\begin{subequations}
\begin{align}
  \rho \to \rho \times e^{2+\delta},
\end{align}
followed by the correction
\begin{align}
  \rho = \begin{dcases}
    \rho, & \rho \leq 10 \\
    10 + \sqrt{\rho-10}, & \rho > 10,
\end{dcases}
\end{align}
\end{subequations}
which effectively flattens the high-density regions, preventing mass
from concentrating in a small fraction of the domain.  We perturb
density but not pressure, so that the temperature $T\propto{1/\rho}$.
This arbitrary perturbation generates the initial condition shown in
figure~\ref{fig:cloud-plot}: roughly half of the domain is at
$\sim{10^6}$\,K, with a density of $\rho\sim{10}$.  While this setup
may seem contrived, it ensures that (i) we do not seed the simulation
with tiny-scale structure such as shattering might produce, and (ii)
that the perturbation occupies a large volume, making it easier to
resolve.  We do not expect our results to be very sensitive to this
initial perturbation.

We run our simulations for $\sim{1}$ cooling time in the $10^6$\,K
gas, and scale the domain to $L\sim{10^5}\times\cstcool$
(\textit{top}), $L\sim{10^3}\times\cstcool$ (\textit{middle}), and
$L\sim{10}\times\cstcool$ (\textit{bottom}).  Since we run for a fixed
cooling time, simulations with small domain sizes are much more
expensive due to the short sound-crossing time.  We therefore use a
resolution of $(4096\times{2048})$, $(512\times{256})$, and
$(256\times{128})$ for the simulations.  The different behaviour is
not simply due to the differing resolutions, however; we do see
shattering even in lower-resolution simulations of big clouds.

\subsection{Cloud Crushing Simulations}
These simulations are shown in figures~\ref{fig:fragmentation},
\ref{fig:res-1e2}, \ref{fig:unresolved-convergence},
and~\ref{fig:fragmentation-1e6}.

This setup is essentially identical to the one used in
\citet{McCourt2015}, except that we do not use magnetic fields and we
run our simulations in 2D.  Following \citet{Shin2008}, we use a
passively advected scalar to keep track of the cloud and we boost the
simulation domain after every time-step to keep the cloud from leaving
the boundaries.  This technique both decreases the size of the
computational grid needed to track the cloud material, and reduces the
truncation errors by minimizing the relative velocity between the
cloud and the computational grid.

In all of our simulations except those in the bottom panel of
figure~\ref{fig:fragmentation-1e6}, we impose a steady wind with
$v_x{=}1.5\,c_{\text{s}}$ at the upstream boundary, where
$c_{\text{s}}$ is the sound speed in the external, confining medium.
This velocity seems appropriate for dense clouds free-falling through
hot virialized gas, as might be found in the CGM or in galactic winds.
Figure~\ref{fig:fragmentation-1e6} shows that our results are not
sensitive to the particular choice, however.  We use an outflow
(zero-gradient) boundary condition downstream, and periodic boundary
conditions in the directions orthogonal to the flow.  We choose
periodic boundary conditions for their simplicity and numerical
stability.

These simulations have domain sizes of
$(6.666\times26.666)\,R_{\text{cloud}}$; \ie, with an aspect ratio of
4:1.  We show only the upper half of the domain in these figures.
Unless otherwise specified, these simulations have a resolution of
$(300\,\text{cells})/R_{\text{cloud}}$, or $(2048\times8192)$\,cells
across the domain.

\subsection{Entrainment Simulations}
The simulation shown in figure~\ref{fig:entrainment} uses a setup
essentially identical to that in figure~\ref{fig:fragmentation},
however the background is at $10^6$\,K instead of $10^7$\,K and the
cold gas is distributed in a large number of small chunks rather than
in a single, monolithic entity.  We use a resolution of
$(4096\times16384)$\,cells across the domain, such that each cloudlet
is resolved with $(64)^2$\,cells.  Each cloudlet has a size
$\sim10\times\cstcool$.  Since this simulation runs for a timescale
longer than the cooling time in the ambient hot gas, we add a crude
form of ``feedback'' heating in order to maintain a well-defined hot
phase: after every call of \citet{Townsend2009}'s integrator, we take
the total thermal energy lost to cooling and add it back into the
domain at a constant rate per unit volume.

\section{Numerical Uncertainties and Comparison to Previous Work}
\label{sec:sim-problems}
The dynamics of multiphase gas is unfortunately difficult to simulate
reliably: both SPH-type codes and Godunov-type codes have severe
problems modeling the contact discontinuities separating the hot and
cold phases \citep[\eg][]{Read2010,Lecoanet2016}.  Numeric errors can
either enhance or suppress mixing between hot and cold gas, and very
little experimental data is available to confirm numerical results in
the regimes relevant to astrophysical multiphase gas (with
$\delta\rho/\rho\gg\{\delta{P}/P,1\}$ and $\text{Re}\gg{}1$).

\citet{Lecoanet2016} show that Godunov-type grid codes are susceptible
to an artificial instability which likely over-estimates cloud
destruction and mixing in our simulations; we have made no attempt to
control such errors in this paper, however, as it would be
prohibitively expensive to do so (even in 2D; shattering is likely not
possible in 1D).  The possibility of uncontrolled artificial
instabilities represents a significant limitation of our simulations
which we are currently unable to address.  Nonetheless, our basic
result that large clouds with short cooling times are prone to
fragmentation appears to be robust, even if the details of the cloud
disruption in figure~\ref{fig:fragmentation} are not.  While the shear
instabilities destroying the clouds in figure~\ref{fig:fragmentation}
(and especially the numerical errors associated with these
instabilities; \citealt{Lecoanet2016}) depend \textit{strongly} on the
density ratio between hot and cold gas, we find a lengthscale for cold
gas which is independent of this density ratio
(cf.\ figures~\ref{fig:fragmentation}
and~\ref{fig:fragmentation-1e6}).  Numeric errors also increase
substantially with increasing resolution for the setup shown in
figure~\ref{fig:fragmentation} \citep{Lecoanet2016}, however we find
similar cloudlet sizes even at different resolutions (see, \eg,
figure~\ref{fig:res-1e2}).  Moreover, while numeric errors should
generically grow at the grid scale, where the timescale is the
fastest, we find shattering takes place at a physical lengthscale,
$\sim\cstcool$, and timescale, $\sim{}t_{\text{cool}}$, which can be
resolved if the domain size is made sufficiently small
(figures~\ref{fig:cloud-plot}, \ref{fig:fragmentation},
and~\ref{fig:res-1e2}); this behavior is expected for a physical
process, but not for any known artificial instability due to errors in
the method.  We furthermore note that we use the van Leer integrator
with second-order reconstruction in the simulations shown in this
paper (see Appendix~\ref{sec:sim-method}); this is more diffusive than
the default integrator implemented in Athena and therefore less
susceptible to numeric instabilities.  We obtain similar results using
Athena's default CTU integrator with third-order reconstruction.

The above remarks suggest that the shattering we highlight in
figures~\ref{fig:cloud-plot}, \ref{fig:fragmentation},
and~\ref{fig:fragmentation-1e6} is not simply the result of an
artificial numeric instability, even though such instabilities may be
present in our simulations.  We also find results suggestive of
shattering in a number of previously published simulations; these
simulations use a variety of codes and setups, and would presumably
exhibit different numerical problems.  The 2D hydrodynamic simulations
in \citet{Mellema2002} and \citet{Koyama2002}, and 3D hydro
simulations in \citet{Cooper2009} and \citet{Schneider2016}, who all
study some variant of the cloud crushing problem, show qualitatively
similar fragmentation driven by cooling.  This behavior is also
consistent with \citet{Audit2005} and \citet{Hennebelle2007}, who also
find in high-resolution 2D simulations that clouds fragment at the
grid scale.  (These authors furthermore show that a stable two-phase
medium in approximate pressure equilibrium, as we appeal to in
section~\ref{subsec:hand-wavy}, makes sense even in a strongly
turbulent flow.)  Though \citet{Kwak2011} and \citet{Gritton2014} do
not study fragmentation directly in their simulations of HVCs, it
appears from their plots that larger initial cloud sizes are more
prone to breaking apart, which is again consistent with our findings
here.

Since most previous studies have focused on macroscopically large
clouds, with $R_0\gg{}\cstcool$, we expect the cloudlets are
essentially always unresolved in currently published studies.  More
work is therefore needed to confirm the cloudlet size of
$\sim{}\cstcool$ as we find here, and to identify its dependence on
parameter choices and on other processes.  The impact of magnetic
fields on the formation and on the morphologies of these cloudlets
will be especially important to study.

The references above suggest that there is precedent for shattering in
the astrophysical literature.  But if shattering is a general
hydrodynamic process, why don't all simulations with cooling show
fragmentation to the grid scale?  We suspect this is because
shattering is particularly sensitive to numerical resolution.  In
figure~\ref{fig:unresolved-convergence}, the blue histograms above
each column show clump size distributions\footnote{We define
  ``clumps'' as contiguous blocks of cells with a density $\geq$ half
  the density of the initial cloud.  In these 2D simulations, the we
  quote the clump size as the square-root of the area, measured in
  units of grid cells.} in terms of the cell size of the simulation
grid.  The distribution in the leftmost panel represents unresolved
clumps in which fragmentation is determined by grid-scale effects; we
confirmed this by re-running this simulation with $4\times$ higher
resolution, obtaining the exact same distribution in units of the cell
size.  \textit{We note that, surprisingly, even clumps with sizes of
  $\sim$10 cells (100 cells in area) are entirely unresolved}; though
these clumps seem much larger than the grid scale, we find the size is
proportional to the grid cell size when we compare simulations with
different resolutions.  The fragmentation of clumps up to
$\sim10$\,cells in size is therefore entirely determined by grid-scale
effects.  Thus, we expect that only fragments with a size $\gtrsim$20
cells are capable of splitting --- though this exact cutoff depends on
the numeric method used, we expect it is typical of the astrophysical
grid codes in current use.  In our simulations, we find a resolution
of $\gtrsim150$ cells per cloud radius is necessary to see the
repeated fragmentation leading to shattering in 2D.  (However this
requirement seems to be somewhat lower in 3D; O'Leary et al., in
prep.)  Since cloud-crushing simulations typically have resolutions of
32\,--\,64 cells per cloud radius
\citep[\eg][]{McCourt2015,Scannapieco2015}, this may not be sufficient
to see the shattering we discuss here.  In recent years, however, a
few 3D simulations have been published approaching the required
resolution
\citep[\eg]{Scannapieco2015,Schneider2016,Banda2016,Bruggen2016}.

\citet{Audit2005} and \citet{Burkert2000} present very high resolution
simulations with cooling, which should in principle exhibit
shattering.  The high resolution 2D simulations in \citet{Audit2005}
do show rapid fragmentation to the grid scale, representing an upper
limit since they were not designed to resolve the scales discussed
here.  The 1D simulations in \citet{Burkert2000} study only purely
single-mode perturbations.  These single-mode simulations enable an
impressive and thorough analysis in \citet{Burkert2000}, but may
impose too much symmetry to fragment in the manner discussed here; it
is also unclear whether fragmentation is possible in 1D.  Moreover,
the simulations in \citet{Burkert2000} run only for $\sim1$ cooling
time, and so do not follow the full non-linear evolution.  It is
likely that shattering happens on a somewhat longer timescale (though
still short compared to the dynamical time).

Finally, we note that the scale $\sim\cstcool$ is only accurate to
within an order of magnitude.  It scales (crudely) as $\sim{}T^{3}$
above $10^4$\,K, so heating the clouds only slightly above the
temperature at which \cstcool is minimized could make them
substantially larger.  When we increase the Mach number of the cloud
by a factor of 3, we find cloudlets which are larger by a factor of
$\lesssim{}2$, suggesting that heating does in fact play a secondary
role in determining the final cloud size (see
figure~\ref{fig:fragmentation-1e6}).  Though this is an interesting
possibility, we unfortunately cannot confidently pursue it here.  As
discussed in section~\ref{subsec:disc-uncertain}, these simulations
omit a number of potentially important physical processes.  We will
consider these in future work.

\section*{Acknowledgments}
\noindent{} We thank Andrey Kravtsov and Eliot Quataert for insightful
conversations and suggestions, and for detailed comments on the
manuscript.  We are grateful to Crystal Martin and Mark Dijkstra for
interesting conversations, which shaped the early development of this
work.  We are also grateful to Anna Barnacka, Cara Battersby, Joe
Hennawi, Chris McKee, Eve Ostriker, Jim Stone, Drummond Fielding, and
James Guillochon for interesting and helpful conversations.  Drummond
Fielding generously helped run a parameter study of the simulations
shown in figure~\ref{fig:cloud-plot}.  MM and SPO were supported by
NASA grants NNX15AK81G and HST-AR-14307.001-A.  MM was also partially
supported by NASA grant HST-HF2-51376.001-A, under NASA contract
NAS5-26555.  SPO thanks KITP for hospitality; this research was
supported in part by the National Science Foundation under Grant
No.\ NSF PHY11-25915.  RMO acknowledges the support provided by NSF
grant AST-1313021.  A-MM was partially supported by a TAC fellowship
at UC Berkeley's Theoretical Astrophysics Center.  We ran our
simulations on the Stampede supercomputer, under XSEDE grants
TG-AST140047 and TG-AST140083.  XSEDE is supported by National Science
Foundation grant number ACI-1053575.  The authors acknowledge the
Texas Advanced Computing Center (TACC) at The University of Texas at
Austin for providing resources that have contributed to the research
results reported within this paper.  We made all of our plots using
the open-source software Tioga, and this research made use of NASA's
ADS system.

\small
\bibliographystyle{mn2e}
\bibliography{clumps}

\begin{thebibliography}{135}
\expandafter\ifx\csname natexlab\endcsname\relax\def\natexlab#1{#1}\fi

\bibitem[{{Anderson}, {Churazov} \& {Bregman}(2015){Anderson}, {Churazov}, \&
  {Bregman}}]{Anderson2015}
{Anderson} M.~E., {Churazov} E., {Bregman} J.~N., 2015, \mnras, 452, 3905

\bibitem[{{Arav} {et~al}\mbox{.}(1997){Arav}, {Barlow}, {Laor}, \&
  {Blandford}}]{Arav1997}
{Arav} N. {et~al.}, 1997, \mnras, 288, 1015

\bibitem[{{Arnaud} {et~al}\mbox{.}(2010){Arnaud}, {Pratt}, {Piffaretti},
  {B{\"o}hringer}, {Croston}, \& {Pointecouteau}}]{Arnaud2010}
{Arnaud} M. {et~al.}, 2010, \aap, 517, A92

\bibitem[{{Arrigoni Battaia} {et~al}\mbox{.}(2015){Arrigoni Battaia},
  {Hennawi}, {Prochaska}, \& {Cantalupo}}]{Arrigoni2015}
{Arrigoni Battaia} F. {et~al.}, 2015, \apj, 809, 163

\bibitem[{{Audit} \& {Hennebelle}(2005)}]{Audit2005}
{Audit} E., {Hennebelle} P., 2005, \aap, 433, 1

\bibitem[{{Balbus} \& {McKee}(1982)}]{Balbus1982}
{Balbus} S.~A., {McKee} C.~F., 1982, \apj, 252, 529

\bibitem[{{Bale} {et~al}\mbox{.}(2009){Bale}, {Kasper}, {Howes}, {Quataert},
  {Salem}, \& {Sundkvist}}]{Bale2009}
{Bale} S.~D. {et~al.}, 2009, Physical Review Letters, 103, 211101

\bibitem[{{Bale} {et~al}\mbox{.}(2013){Bale}, {Pulupa}, {Salem}, {Chen}, \&
  {Quataert}}]{Bale2013}
{Bale} S.~D. {et~al.}, 2013, \apjl, 769, L22

\bibitem[{{Banda-Barrag{\'a}n} {et~al}\mbox{.}(2016){Banda-Barrag{\'a}n},
  {Parkin}, {Federrath}, {Crocker}, \& {Bicknell}}]{Banda2016}
{Banda-Barrag{\'a}n} W.~E. {et~al.}, 2016, \mnras, 455, 1309

\bibitem[{{Bauermeister}, {Blitz} \& {Ma}(2010){Bauermeister}, {Blitz}, \&
  {Ma}}]{Bauermeister2010}
{Bauermeister} A., {Blitz} L., {Ma} C.-P., 2010, \apj, 717, 323

\bibitem[{{Becker} \& {Bolton}(2013)}]{Becker2013}
{Becker} G.~D., {Bolton} J.~S., 2013, \mnras, 436, 1023

\bibitem[{{Ben Bekhti} {et~al}\mbox{.}(2008){Ben Bekhti}, {Richter},
  {Westmeier}, \& {Murphy}}]{Ben2008}
{Ben Bekhti} N. {et~al.}, 2008, \aap, 487, 583

\bibitem[{{Ben Bekhti} {et~al}\mbox{.}(2009){Ben Bekhti}, {Richter}, {Winkel},
  {Kenn}, \& {Westmeier}}]{Ben2009}
{Ben Bekhti} N. {et~al.}, 2009, \aap, 503, 483

\bibitem[{{Ben Bekhti} {et~al}\mbox{.}(2012){Ben Bekhti}, {Winkel}, {Richter},
  {Kerp}, {Klein}, \& {Murphy}}]{Ben2012}
{Ben Bekhti} N. {et~al.}, 2012, \aap, 542, A110

\bibitem[{{Birnboim} \& {Dekel}(2003)}]{Birnboim2003}
{Birnboim} Y., {Dekel} A., 2003, \mnras, 345, 349

\bibitem[{{Blagrave} {et~al}\mbox{.}(2016){Blagrave}, {Martin}, {Joncas},
  {Kothes}, {Stil}, {Miville-Deschenes}, {Lockman}, \& {Taylor}}]{Blagrave2016}
{Blagrave} K. {et~al.}, 2016, ArXiv e-prints

\bibitem[{{Bonnell}, {Bate} \& {Vine}(2003){Bonnell}, {Bate}, \&
  {Vine}}]{Bonnell2003}
{Bonnell} I.~A., {Bate} M.~R., {Vine} S.~G., 2003, \mnras, 343, 413

\bibitem[{{Borisova} {et~al}\mbox{.}(2016){Borisova}, {Cantalupo}, {Lilly},
  {Marino}, {Gallego}, {Bacon}, {Blaizot}, {Bouch{\'e}}, {Brinchmann},
  {Carollo}, {Caruana}, {Finley}, {Herenz}, {Richard}, {Schaye}, {Straka},
  {Turner}, {Urrutia}, {Verhamme}, \& {Wisotzki}}]{Borisova2016}
{Borisova} E. {et~al.}, 2016, ArXiv e-prints

\bibitem[{{Bottorff} {et~al}\mbox{.}(2000){Bottorff}, {Ferland}, {Baldwin}, \&
  {Korista}}]{Bottorff2000}
{Bottorff} M. {et~al.}, 2000, \apj, 542, 644

\bibitem[{{Braun} \& {Kanekar}(2005)}]{Braun2005}
{Braun} R., {Kanekar} N., 2005, \aap, 436, L53

\bibitem[{{Br{\"u}ggen} \& {Scannapieco}(2016)}]{Bruggen2016}
{Br{\"u}ggen} M., {Scannapieco} E., 2016, \apj, 822, 31

\bibitem[{{Burkert} \& {Lin}(2000)}]{Burkert2000}
{Burkert} A., {Lin} D.~N.~C., 2000, \apj, 537, 270

\bibitem[{{Cantalupo}(2010)}]{Cantalupo2010}
{Cantalupo} S., 2010, \mnras, 403, L16

\bibitem[{{Cantalupo} {et~al}\mbox{.}(2014){Cantalupo}, {Arrigoni-Battaia},
  {Prochaska}, {Hennawi}, \& {Madau}}]{Cantalupo2014}
{Cantalupo} S. {et~al.}, 2014, \nat, 506, 63

\bibitem[{{Churchill} {et~al}\mbox{.}(2003){Churchill}, {Mellon}, {Charlton},
  \& {Vogt}}]{Churchill2003}
{Churchill} C.~W. {et~al.}, 2003, \apj, 593, 203

\bibitem[{{Cooper} {et~al}\mbox{.}(2009){Cooper}, {Bicknell}, {Sutherland}, \&
  {Bland-Hawthorn}}]{Cooper2009}
{Cooper} J.~L. {et~al.}, 2009, \apj, 703, 330

\bibitem[{{Cowie} \& {McKee}(1977)}]{Cowie1977}
{Cowie} L.~L., {McKee} C.~F., 1977, \apj, 211, 135

\bibitem[{{Crighton}, {Hennawi} \& {Prochaska}(2013){Crighton}, {Hennawi}, \&
  {Prochaska}}]{Crighton2013}
{Crighton} N.~H.~M., {Hennawi} J.~F., {Prochaska} J.~X., 2013, \apjl, 776, L18

\bibitem[{{Crighton} {et~al}\mbox{.}(2015){Crighton}, {Hennawi}, {Simcoe},
  {Cooksey}, {Murphy}, {Fumagalli}, {Prochaska}, \& {Shanks}}]{Crighton2015}
{Crighton} N.~H.~M. {et~al.}, 2015, \mnras, 446, 18

\bibitem[{{Crutcher} {et~al}\mbox{.}(2010){Crutcher}, {Wandelt}, {Heiles},
  {Falgarone}, \& {Troland}}]{Crutcher2010}
{Crutcher} R.~M. {et~al.}, 2010, \apj, 725, 466

\bibitem[{{Dietrich} {et~al}\mbox{.}(1999){Dietrich}, {Wagner}, {Courvoisier},
  {Bock}, \& {North}}]{Dietrich1999}
{Dietrich} M. {et~al.}, 1999, \aap, 351, 31

\bibitem[{{Dijkstra}, {Gronke} \& {Venkatesan}(2016){Dijkstra}, {Gronke}, \&
  {Venkatesan}}]{Dijkstra2016}
{Dijkstra} M., {Gronke} M., {Venkatesan} A., 2016, \apj, 828, 71

\bibitem[{{D'Odorico} {et~al}\mbox{.}(2004){D'Odorico}, {Cristiani}, {Romano},
  {Granato}, \& {Danese}}]{D2004}
{D'Odorico} V. {et~al.}, 2004, \mnras, 351, 976

\bibitem[{{Elvis}(2000)}]{Elvis2000}
{Elvis} M., 2000, \apj, 545, 63

\bibitem[{{Fang}, {Bullock} \& {Boylan-Kolchin}(2013){Fang}, {Bullock}, \&
  {Boylan-Kolchin}}]{Fang2013}
{Fang} T., {Bullock} J., {Boylan-Kolchin} M., 2013, \apj, 762, 20

\bibitem[{{Faucher-Giguere} {et~al}\mbox{.}(2016){Faucher-Giguere}, {Feldmann},
  {Quataert}, {Keres}, {Hopkins}, \& {Murray}}]{Faucher2016}
{Faucher-Giguere} C.-A. {et~al.}, 2016, ArXiv e-prints

\bibitem[{{Faucher-Gigu{\`e}re} {et~al}\mbox{.}(2015){Faucher-Gigu{\`e}re},
  {Hopkins}, {Kere{\v s}}, {Muratov}, {Quataert}, \& {Murray}}]{Faucher2015}
{Faucher-Gigu{\`e}re} C.-A. {et~al.}, 2015, \mnras, 449, 987

\bibitem[{{Faucher-Gigu{\`e}re}, {Quataert} \&
  {Murray}(2012){Faucher-Gigu{\`e}re}, {Quataert}, \& {Murray}}]{Faucher2012}
{Faucher-Gigu{\`e}re} C.-A., {Quataert} E., {Murray} N., 2012, \mnras, 420,
  1347

\bibitem[{{Field}(1965)}]{Field1965}
{Field} G.~B., 1965, \apj, 142, 531

\bibitem[{{Field}, {Goldsmith} \& {Habing}(1969){Field}, {Goldsmith}, \&
  {Habing}}]{Field1969}
{Field} G.~B., {Goldsmith} D.~W., {Habing} H.~J., 1969, \apjl, 155, L149

\bibitem[{{Fielding} {et~al}\mbox{.}(2016){Fielding}, {Quataert}, {McCourt}, \&
  {Thompson}}]{Fielding2016}
{Fielding} D.~B. {et~al.}, 2016, {The Impact of Star Formation Feedback on the
  Circumgalactic Medium}, {in prep.}

\bibitem[{{Finn} {et~al}\mbox{.}(2014){Finn}, {Morris}, {Crighton}, {Hamann},
  {Done}, {Theuns}, {Fumagalli}, {Tejos}, \& {Worseck}}]{Finn2014}
{Finn} C.~W. {et~al.}, 2014, \mnras, 440, 3317

\bibitem[{{Fumagalli} {et~al}\mbox{.}(2014){Fumagalli}, {Hennawi}, {Prochaska},
  {Kasen}, {Dekel}, {Ceverino}, \& {Primack}}]{Fumagalli2014}
{Fumagalli} M. {et~al.}, 2014, \apj, 780, 74

\bibitem[{{Gardiner} \& {Stone}(2008)}]{Gardiner2008}
{Gardiner} T.~A., {Stone} J.~M., 2008, Journal of Computational Physics, 227,
  4123

\bibitem[{{Genzel} {et~al}\mbox{.}(2010){Genzel}, {Tacconi}, {Gracia-Carpio},
  {Sternberg}, {Cooper}, {Shapiro}, {Bolatto}, {Bouch{\'e}}, {Bournaud},
  {Burkert}, {Combes}, {Comerford}, {Cox}, {Davis}, {Schreiber},
  {Garcia-Burillo}, {Lutz}, {Naab}, {Neri}, {Omont}, {Shapley}, \&
  {Weiner}}]{Genzel2010}
{Genzel} R. {et~al.}, 2010, \mnras, 407, 2091

\bibitem[{{Gnedin} \& {Hollon}(2012)}]{Gnedin2012}
{Gnedin} N.~Y., {Hollon} N., 2012, \apjs, 202, 13

\bibitem[{{Gritton}, {Shelton} \& {Kwak}(2014){Gritton}, {Shelton}, \&
  {Kwak}}]{Gritton2014}
{Gritton} J.~A., {Shelton} R.~L., {Kwak} K., 2014, \apj, 795, 99

\bibitem[{{Hacar} \& {Tafalla}(2011)}]{Hacar2011}
{Hacar} A., {Tafalla} M., 2011, \aap, 533, A34

\bibitem[{{Hacar} {et~al}\mbox{.}(2013){Hacar}, {Tafalla}, {Kauffmann}, \&
  {Kov{\'a}cs}}]{Hacar2013}
{Hacar} A. {et~al.}, 2013, \aap, 554, A55

\bibitem[{{Hacker} {et~al}\mbox{.}(2013){Hacker}, {Brunner}, {Lundgren}, \&
  {York}}]{Hacker2013}
{Hacker} T.~L. {et~al.}, 2013, \mnras, 434, 163

\bibitem[{{Hamann} {et~al}\mbox{.}(1995){Hamann}, {Barlow}, {Beaver},
  {Burbidge}, {Cohen}, {Junkkarinen}, \& {Lyons}}]{Hamann1995}
{Hamann} F. {et~al.}, 1995, \apj, 443, 606

\bibitem[{{Hamann} {et~al}\mbox{.}(2013){Hamann}, {Chartas}, {McGraw},
  {Rodriguez Hidalgo}, {Shields}, {Capellupo}, {Charlton}, \&
  {Eracleous}}]{Hamann2013}
{Hamann} F. {et~al.}, 2013, \mnras, 435, 133

\bibitem[{{Hamann} {et~al}\mbox{.}(2011){Hamann}, {Kanekar}, {Prochaska},
  {Murphy}, {Ellison}, {Malec}, {Milutinovic}, \& {Ubachs}}]{Hamann2011}
{Hamann} F. {et~al.}, 2011, \mnras, 410, 1957

\bibitem[{{Hansen} \& {Oh}(2006)}]{Hansen2006}
{Hansen} M., {Oh} S.~P., 2006, \mnras, 367, 979

\bibitem[{{Hennawi} {et~al}\mbox{.}(2015){Hennawi}, {Prochaska}, {Cantalupo},
  \& {Arrigoni-Battaia}}]{Hennawi2015}
{Hennawi} J.~F. {et~al.}, 2015, Science, 348, 779

\bibitem[{{Hennebelle} \& {Audit}(2007)}]{Hennebelle2007}
{Hennebelle} P., {Audit} E., 2007, \aap, 465, 431

\bibitem[{{Hennebelle} \& {P{\'e}rault}(1999)}]{Hennebelle1999}
{Hennebelle} P., {P{\'e}rault} M., 1999, \aap, 351, 309

\bibitem[{{Hitomi Collaboration} {et~al}\mbox{.}(2016){Hitomi Collaboration},
  {Aharonian}, {Akamatsu}, {Akimoto}, {Allen}, {Anabuki}, {Angelini}, {Arnaud},
  {Audard}, {Awaki}, {Axelsson}, {Bamba}, {Bautz}, {Blandford}, {Brenneman},
  {Brown}, {Bulbul}, {Cackett}, {Chernyakova}, {Chiao}, {Coppi}, {Costantini},
  {de Plaa}, {den Herder}, {Done}, {Dotani}, {Ebisawa}, {Eckart}, {Enoto},
  {Ezoe}, {Fabian}, {Ferrigno}, {Foster}, {Fujimoto}, {Fukazawa}, {Furuzawa},
  {Galeazzi}, {Gallo}, {Gandhi}, {Giustini}, {Goldwurm}, {Gu}, {Guainazzi},
  {Haba}, {Hagino}, {Hamaguchi}, {Harrus}, {Hatsukade}, {Hayashi}, {Hayashi},
  {Hayashida}, {Hiraga}, {Hornschemeier}, {Hoshino}, {Hughes}, {Iizuka},
  {Inoue}, {Inoue}, {Ishibashi}, {Ishida}, {Ishikawa}, {Ishisaki}, {Itoh},
  {Iyomoto}, {Kaastra}, {Kallman}, {Kamae}, {Kara}, {Kataoka}, {Katsuda},
  {Katsuta}, {Kawaharada}, {Kawai}, {Kelley}, {Khangulyan}, {Kilbourne},
  {King}, {Kitaguchi}, {Kitamoto}, {Kitayama}, {Kohmura}, {Kokubun}, {Koyama},
  {Koyama}, {Kretschmar}, {Krimm}, {Kubota}, {Kunieda}, {Laurent}, {Lebrun},
  {Lee}, {Leutenegger}, {Limousin}, {Loewenstein}, {Long}, {Lumb}, {Madejski},
  {Maeda}, {Maier}, {Makishima}, {Markevitch}, {Matsumoto}, {Matsushita},
  {McCammon}, {McNamara}, {Mehdipour}, {Miller}, {Miller}, {Mineshige},
  {Mitsuda}, {Mitsuishi}, {Miyazawa}, {Mizuno}, {Mori}, {Mori}, {Moseley},
  {Mukai}, {Murakami}, {Murakami}, {Mushotzky}, {Nagino}, {Nakagawa},
  {Nakajima}, {Nakamori}, {Nakano}, {Nakashima}, {Nakazawa}, {Nobukawa},
  {Noda}, {Nomachi}, {O'Dell}, {Odaka}, {Ohashi}, {Ohno}, {Okajima}, {Ota},
  {Ozaki}, {Paerels}, {Paltani}, {Parmar}, {Petre}, {Pinto}, {Pohl}, {Porter},
  {Pottschmidt}, {Ramsey}, {Reynolds}, {Russell}, {Safi-Harb}, {Saito},
  {Sakai}, {Sameshima}, {Sato}, {Sato}, {Sato}, {Sawada}, {Schartel},
  {Serlemitsos}, {Seta}, {Shidatsu}, {Simionescu}, {Smith}, {Soong}, {Stawarz},
  {Sugawara}, {Sugita}, {Szymkowiak}, {Tajima}, {Takahashi}, {Takahashi},
  {Takeda}, {Takei}, {Tamagawa}, {Tamura}, {Tamura}, {Tanaka}, {Tanaka},
  {Tanaka}, {Tashiro}, {Tawara}, {Terada}, {Terashima}, {Tombesi}, {Tomida},
  {Tsuboi}, {Tsujimoto}, {Tsunemi}, {Tsuru}, {Uchida}, {Uchiyama}, {Uchiyama},
  {Ueda}, {Ueda}, {Ueno}, {Uno}, {Urry}, {Ursino}, {de Vries}, {Watanabe},
  {Werner}, {Wik}, {Wilkins}, {Williams}, {Yamada}, {Yamaguchi}, {Yamaoka},
  {Yamasaki}, {Yamauchi}, {Yamauchi}, {Yaqoob}, {Yatsu}, {Yonetoku}, {Yoshida},
  {Yuasa}, {Zhuravleva}, \& {Zoghbi}}]{Hitomi2016}
{Hitomi Collaboration} {et~al.}, 2016, \nat, 535, 117

\bibitem[{{Jeans}(1901)}]{Jeans1901}
{Jeans} J.~H., 1901, Proceedings of the Royal Society of London Series I, 68,
  454

\bibitem[{{Kannan} {et~al}\mbox{.}(2014){Kannan}, {Stinson}, {Macci{\`o}},
  {Hennawi}, {Woods}, {Wadsley}, {Shen}, {Robitaille}, {Cantalupo}, {Quinn}, \&
  {Christensen}}]{Kannan2014}
{Kannan} R. {et~al.}, 2014, \mnras, 437, 2882

\bibitem[{{Kere{\v s}} {et~al}\mbox{.}(2005){Kere{\v s}}, {Katz}, {Weinberg},
  \& {Dav{\'e}}}]{Keres2005}
{Kere{\v s}} D. {et~al.}, 2005, \mnras, 363, 2

\bibitem[{{Klein}, {McKee} \& {Colella}(1994){Klein}, {McKee}, \&
  {Colella}}]{Klein1994}
{Klein} R.~I., {McKee} C.~F., {Colella} P., 1994, \apj, 420, 213

\bibitem[{{Komarov} {et~al}\mbox{.}(2016){Komarov}, {Churazov}, {Kunz}, \&
  {Schekochihin}}]{Komarov2016}
{Komarov} S.~V. {et~al.}, 2016, \mnras, 460, 467

\bibitem[{{Koyama} \& {Inutsuka}(2002)}]{Koyama2002}
{Koyama} H., {Inutsuka} S.-i., 2002, \apjl, 564, L97

\bibitem[{{Kuhlen} \& {Faucher-Gigu{\`e}re}(2012)}]{Kuhlen2012}
{Kuhlen} M., {Faucher-Gigu{\`e}re} C.-A., 2012, \mnras, 423, 862

\bibitem[{{Kunz}, {Schekochihin} \& {Stone}(2014){Kunz}, {Schekochihin}, \&
  {Stone}}]{Kunz2014}
{Kunz} M.~W., {Schekochihin} A.~A., {Stone} J.~M., 2014, Physical Review
  Letters, 112, 205003

\bibitem[{{Kwak}, {Henley} \& {Shelton}(2011){Kwak}, {Henley}, \&
  {Shelton}}]{Kwak2011}
{Kwak} K., {Henley} D.~B., {Shelton} R.~L., 2011, \apj, 739, 30

\bibitem[{{Lau}, {Prochaska} \& {Hennawi}(2015){Lau}, {Prochaska}, \&
  {Hennawi}}]{Lau2015}
{Lau} M.~W., {Prochaska} J.~X., {Hennawi} J.~F., 2015, ArXiv e-prints

\bibitem[{{Lazarian}(2014)}]{Lazarian2014}
{Lazarian} A., 2014, \ssr, 181, 1

\bibitem[{{Lecoanet} {et~al}\mbox{.}(2016){Lecoanet}, {McCourt}, {Quataert},
  {Burns}, {Vasil}, {Oishi}, {Brown}, {Stone}, \& {O'Leary}}]{Lecoanet2016}
{Lecoanet} D. {et~al.}, 2016, \mnras, 455, 4274

\bibitem[{{Liang}, {Kravtsov} \& {Agertz}(2016){Liang}, {Kravtsov}, \&
  {Agertz}}]{Liang2016}
{Liang} C.~J., {Kravtsov} A.~V., {Agertz} O., 2016, \mnras, 458, 1164

\bibitem[{{Low} \& {Lynden-Bell}(1976)}]{Low1976}
{Low} C., {Lynden-Bell} D., 1976, \mnras, 176, 367

\bibitem[{{Mac Low} {et~al}\mbox{.}(1994){Mac Low}, {McKee}, {Klein}, {Stone},
  \& {Norman}}]{Mac1994}
{Mac Low} M.-M. {et~al.}, 1994, \apj, 433, 757

\bibitem[{{Maller} \& {Bullock}(2004)}]{Maller2004}
{Maller} A.~H., {Bullock} J.~S., 2004, \mnras, 355, 694

\bibitem[{{Mandelker} {et~al}\mbox{.}(2016){Mandelker}, {Padnos}, {Dekel},
  {Birnboim}, {Burkert}, {Krumholz}, \& {Steinberg}}]{Mandelker2016}
{Mandelker} N. {et~al.}, 2016, \mnras

\bibitem[{{Markevitch} \& {Vikhlinin}(2007)}]{Markevitch2007}
{Markevitch} M., {Vikhlinin} A., 2007, \physrep, 443, 1

\bibitem[{{Martin}(2005)}]{Martin2005}
{Martin} C.~L., 2005, \apj, 621, 227

\bibitem[{{Martin} {et~al}\mbox{.}(2015){Martin}, {Dijkstra}, {Henry}, {Soto},
  {Danforth}, \& {Wong}}]{Martin2015}
{Martin} C.~L. {et~al.}, 2015, \apj, 803, 6

\bibitem[{{McCourt} {et~al}\mbox{.}(2015){McCourt}, {O'Leary}, {Madigan}, \&
  {Quataert}}]{McCourt2015}
{McCourt} M. {et~al.}, 2015, \mnras, 449, 2

\bibitem[{{McCourt} {et~al}\mbox{.}(2012){McCourt}, {Sharma}, {Quataert}, \&
  {Parrish}}]{McCourt2012}
{McCourt} M. {et~al.}, 2012, \mnras, 419, 3319

\bibitem[{{Mellema}, {Kurk} \& {R{\"o}ttgering}(2002){Mellema}, {Kurk}, \&
  {R{\"o}ttgering}}]{Mellema2002}
{Mellema} G., {Kurk} J.~D., {R{\"o}ttgering} H.~J.~A., 2002, \aap, 395, L13

\bibitem[{{Miville-Desch{\^e}nes} {et~al}\mbox{.}(2016){Miville-Desch{\^e}nes},
  {Salom{\'e}}, {Martin}, {Joncas}, {Blagrave}, {Dassas}, {Abergel}, {Beelen},
  {Boulanger}, {Lagache}, {Lockman}, \& {Marshall}}]{Miville2016}
{Miville-Desch{\^e}nes} M.-A. {et~al.}, 2016, ArXiv e-prints

\bibitem[{{Nagai}, {Vikhlinin} \& {Kravtsov}(2007){Nagai}, {Vikhlinin}, \&
  {Kravtsov}}]{Nagai2007}
{Nagai} D., {Vikhlinin} A., {Kravtsov} A.~V., 2007, \apj, 655, 98

\bibitem[{{Nelson} {et~al}\mbox{.}(2013){Nelson}, {Vogelsberger}, {Genel},
  {Sijacki}, {Kere{\v s}}, {Springel}, \& {Hernquist}}]{Nelson2013}
{Nelson} D. {et~al.}, 2013, \mnras, 429, 3353

\bibitem[{{Netzer}(2006)}]{Netzer2006}
{Netzer} H., 2006, in Lecture Notes in Physics, Berlin Springer Verlag, Vol.
  693, Physics of Active Galactic Nuclei at all Scales, {Alloin} D., ed., p.~1

\bibitem[{{Neufeld}(1991)}]{Neufeld1991}
{Neufeld} D.~A., 1991, \apjl, 370, L85

\bibitem[{{Oppenheimer} {et~al}\mbox{.}(2010){Oppenheimer}, {Dav{\'e}},
  {Kere{\v s}}, {Fardal}, {Katz}, {Kollmeier}, \& {Weinberg}}]{Oppenheimer2010}
{Oppenheimer} B.~D. {et~al.}, 2010, \mnras, 406, 2325

\bibitem[{{Planck Collaboration} {et~al}\mbox{.}(2013){Planck Collaboration},
  {Ade}, {Aghanim}, {Arnaud}, {Ashdown}, {Atrio-Barandela}, {Aumont},
  {Baccigalupi}, {Balbi}, {Banday}, \& et~al.}]{Planck2013}
{Planck Collaboration} {et~al.}, 2013, \aap, 557, A52

\bibitem[{{Prochaska} \& {Hennawi}(2009)}]{Prochaska2009}
{Prochaska} J.~X., {Hennawi} J.~F., 2009, \apj, 690, 1558

\bibitem[{{Prochaska}, {Hennawi} \& {Simcoe}(2013){Prochaska}, {Hennawi}, \&
  {Simcoe}}]{Prochaska2013}
{Prochaska} J.~X., {Hennawi} J.~F., {Simcoe} R.~A., 2013, \apjl, 762, L19

\bibitem[{{Prochaska}, {Lau} \& {Hennawi}(2014){Prochaska}, {Lau}, \&
  {Hennawi}}]{Prochaska2014}
{Prochaska} J.~X., {Lau} M.~W., {Hennawi} J.~F., 2014, \apj, 796, 140

\bibitem[{{Rauch}, {Sargent} \& {Barlow}(1999){Rauch}, {Sargent}, \&
  {Barlow}}]{Rauch1999}
{Rauch} M., {Sargent} W.~L.~W., {Barlow} T.~A., 1999, \apj, 515, 500

\bibitem[{{Read}, {Hayfield} \& {Agertz}(2010){Read}, {Hayfield}, \&
  {Agertz}}]{Read2010}
{Read} J.~I., {Hayfield} T., {Agertz} O., 2010, \mnras, 405, 1513

\bibitem[{{Rees}(1987)}]{Rees1987}
{Rees} M.~J., 1987, \mnras, 228, 47P

\bibitem[{{Richter}(2006)}]{Richter2006}
{Richter} P., 2006, in Reviews in Modern Astronomy, Vol.~19, Reviews in Modern
  Astronomy, {Roeser} S., ed., p.~31

\bibitem[{{Richter}(2012)}]{Richter2012}
{Richter} P., 2012, \apj, 750, 165

\bibitem[{{Richter} {et~al}\mbox{.}(2009){Richter}, {Charlton}, {Fangano},
  {Bekhti}, \& {Masiero}}]{Richter2009}
{Richter} P. {et~al.}, 2009, \apj, 695, 1631

\bibitem[{{Richter}, {Sembach} \& {Howk}(2003){Richter}, {Sembach}, \&
  {Howk}}]{Richter2003}
{Richter} P., {Sembach} K.~R., {Howk} J.~C., 2003, \aap, 405, 1013

\bibitem[{{Richter} {et~al}\mbox{.}(2001){Richter}, {Sembach}, {Wakker},
  {Savage}, {Tripp}, {Murphy}, {Kalberla}, \& {Jenkins}}]{Richter2001}
{Richter} P. {et~al.}, 2001, \apj, 559, 318

\bibitem[{{Richter}, {Westmeier} \& {Br{\"u}ns}(2005){Richter}, {Westmeier}, \&
  {Br{\"u}ns}}]{Richter2005}
{Richter} P., {Westmeier} T., {Br{\"u}ns} C., 2005, \aap, 442, L49

\bibitem[{{Rigby}, {Charlton} \& {Churchill}(2002){Rigby}, {Charlton}, \&
  {Churchill}}]{Rigby2002}
{Rigby} J.~R., {Charlton} J.~C., {Churchill} C.~W., 2002, \apj, 565, 743

\bibitem[{{Rudie} {et~al}\mbox{.}(2012){Rudie}, {Steidel}, {Trainor}, {Rakic},
  {Bogosavljevi{\'c}}, {Pettini}, {Reddy}, {Shapley}, {Erb}, \&
  {Law}}]{Rudie2012}
{Rudie} G.~C. {et~al.}, 2012, \apj, 750, 67

\bibitem[{{Scannapieco} \& {Br{\"u}ggen}(2015)}]{Scannapieco2015}
{Scannapieco} E., {Br{\"u}ggen} M., 2015, \apj, 805, 158

\bibitem[{{Schaye}, {Carswell} \& {Kim}(2007){Schaye}, {Carswell}, \&
  {Kim}}]{Schaye2007}
{Schaye} J., {Carswell} R.~F., {Kim} T.-S., 2007, \mnras, 379, 1169

\bibitem[{{Schneider} \& {Robertson}(2016)}]{Schneider2016}
{Schneider} E.~E., {Robertson} B.~E., 2016, ArXiv e-prints

\bibitem[{{Schunk} \& {Hays}(1971)}]{Schunk1971}
{Schunk} R.~W., {Hays} P.~B., 1971, \planss, 19, 113

\bibitem[{{Schwartz} \& {Martin}(2004)}]{Schwartz2004}
{Schwartz} C.~M., {Martin} C.~L., 2004, \apj, 610, 201

\bibitem[{{Shang} \& {Oh}(2012)}]{Shang2012}
{Shang} C., {Oh} S.~P., 2012, \mnras, 426, 3435

\bibitem[{{Shapiro} \& {Field}(1976)}]{Shapiro1976}
{Shapiro} P.~R., {Field} G.~B., 1976, \apj, 205, 762

\bibitem[{{Sharma} {et~al}\mbox{.}(2006){Sharma}, {Hammett}, {Quataert}, \&
  {Stone}}]{Sharma2006}
{Sharma} P. {et~al.}, 2006, \apj, 637, 952

\bibitem[{{Sharma} {et~al}\mbox{.}(2012){Sharma}, {McCourt}, {Parrish}, \&
  {Quataert}}]{Sharma2012}
{Sharma} P. {et~al.}, 2012, \mnras, 427, 1219

\bibitem[{{Shin}, {Stone} \& {Snyder}(2008){Shin}, {Stone}, \&
  {Snyder}}]{Shin2008}
{Shin} M.-S., {Stone} J.~M., {Snyder} G.~F., 2008, \apj, 680, 336

\bibitem[{{Shull} {et~al}\mbox{.}(2009){Shull}, {Jones}, {Danforth}, \&
  {Collins}}]{Shull2009}
{Shull} J.~M. {et~al.}, 2009, \apj, 699, 754

\bibitem[{{Siana} {et~al}\mbox{.}(2010){Siana}, {Teplitz}, {Ferguson}, {Brown},
  {Giavalisco}, {Dickinson}, {Chary}, {de Mello}, {Conselice}, {Bridge},
  {Gardner}, {Colbert}, \& {Scarlata}}]{Siana2010}
{Siana} B. {et~al.}, 2010, \apj, 723, 241

\bibitem[{{Sparks} {et~al}\mbox{.}(2009){Sparks}, {Pringle}, {Donahue},
  {Carswell}, {Voit}, {Cracraft}, \& {Martin}}]{Sparks2009}
{Sparks} W.~B. {et~al.}, 2009, \apjl, 704, L20

\bibitem[{{Stern} {et~al}\mbox{.}(2016){Stern}, {Hennawi}, {Prochaska}, \&
  {Werk}}]{Stern2016}
{Stern} J. {et~al.}, 2016, ArXiv e-prints

\bibitem[{{Stocke} {et~al}\mbox{.}(2013){Stocke}, {Keeney}, {Danforth},
  {Shull}, {Froning}, {Green}, {Penton}, \& {Savage}}]{Stocke2013}
{Stocke} J.~T. {et~al.}, 2013, \apj, 763, 148

\bibitem[{{Stone} {et~al}\mbox{.}(2008){Stone}, {Gardiner}, {Teuben}, {Hawley},
  \& {Simon}}]{Stone2008}
{Stone} J.~M. {et~al.}, 2008, \apjs, 178, 137

\bibitem[{{Sutherland} \& {Dopita}(1993)}]{Sutherland1993}
{Sutherland} R.~S., {Dopita} M.~A., 1993, \apjs, 88, 253

\bibitem[{{Suzuki-Vidal} {et~al}\mbox{.}(2015){Suzuki-Vidal}, {Lebedev},
  {Ciardi}, {Pickworth}, {Rodriguez}, {Gil}, {Espinosa}, {Hartigan},
  {Swadling}, {Skidmore}, {Hall}, {Bennett}, {Bland}, {Burdiak}, {de Grouchy},
  {Music}, {Suttle}, {Hansen}, \& {Frank}}]{Suzuki2015}
{Suzuki-Vidal} F. {et~al.}, 2015, \apj, 815, 96

\bibitem[{{Thompson} {et~al}\mbox{.}(2016){Thompson}, {Quataert}, {Zhang}, \&
  {Weinberg}}]{Thompson2016}
{Thompson} T.~A. {et~al.}, 2016, \mnras, 455, 1830

\bibitem[{{Townsend}(2009)}]{Townsend2009}
{Townsend} R.~H.~D., 2009, \apjs, 181, 391

\bibitem[{{Tumlinson} {et~al}\mbox{.}(2011){Tumlinson}, {Thom}, {Werk},
  {Prochaska}, {Tripp}, {Weinberg}, {Peeples}, {O'Meara}, {Oppenheimer},
  {Meiring}, {Katz}, {Dav{\'e}}, {Ford}, \& {Sembach}}]{Tumlinson2011}
{Tumlinson} J. {et~al.}, 2011, Science, 334, 948

\bibitem[{{van de Voort} {et~al}\mbox{.}(2011){van de Voort}, {Schaye},
  {Booth}, {Haas}, \& {Dalla Vecchia}}]{van2011}
{van de Voort} F. {et~al.}, 2011, \mnras, 414, 2458

\bibitem[{{Voit} {et~al}\mbox{.}(2015{\natexlab{a}}){Voit}, {Bryan}, {O'Shea},
  \& {Donahue}}]{Voit2015}
{Voit} G.~M. {et~al.}, 2015{\natexlab{a}}, \apjl, 808, L30

\bibitem[{{Voit} \& {Donahue}(1990)}]{Voit1990}
{Voit} G.~M., {Donahue} M., 1990, \apjl, 360, L15

\bibitem[{{Voit} {et~al}\mbox{.}(2015{\natexlab{b}}){Voit}, {Donahue}, {Bryan},
  \& {McDonald}}]{Voit2015b}
{Voit} G.~M. {et~al.}, 2015{\natexlab{b}}, \nat, 519, 203

\bibitem[{{Wakker} \& {van Woerden}(1997)}]{Wakker1997}
{Wakker} B.~P., {van Woerden} H., 1997, \araa, 35, 217

\bibitem[{{Werk} {et~al}\mbox{.}(2014){Werk}, {Prochaska}, {Tumlinson},
  {Peeples}, {Tripp}, {Fox}, {Lehner}, {Thom}, {O'Meara}, {Ford}, {Bordoloi},
  {Katz}, {Tejos}, {Oppenheimer}, {Dav{\'e}}, \& {Weinberg}}]{Werk2014}
{Werk} J.~K. {et~al.}, 2014, \apj, 792, 8

\bibitem[{{White} \& {Rees}(1978)}]{White1978}
{White} S.~D.~M., {Rees} M.~J., 1978, \mnras, 183, 341

\bibitem[{{Wiersma}, {Schaye} \& {Smith}(2009){Wiersma}, {Schaye}, \&
  {Smith}}]{Wiersma2009}
{Wiersma} R.~P.~C., {Schaye} J., {Smith} B.~D., 2009, \mnras, 393, 99

\bibitem[{{Yoon} {et~al}\mbox{.}(2012){Yoon}, {Putman}, {Thom}, {Chen}, \&
  {Bryan}}]{Yoon2012}
{Yoon} J.~H. {et~al.}, 2012, \apj, 754, 84

\bibitem[{{Zhang} {et~al}\mbox{.}(2015){Zhang}, {Thompson}, {Quataert}, \&
  {Murray}}]{Zhang2015}
{Zhang} D. {et~al.}, 2015, ArXiv e-prints

\bibitem[{{Zheng} {et~al}\mbox{.}(2015){Zheng}, {Putman}, {Peek}, \&
  {Joung}}]{Zheng2015}
{Zheng} Y. {et~al.}, 2015, \apj, 807, 103

\bibitem[{{ZuHone} {et~al}\mbox{.}(2013){ZuHone}, {Markevitch}, {Ruszkowski},
  \& {Lee}}]{ZuHone2013}
{ZuHone} J.~A. {et~al.}, 2013, \apj, 762, 69

\end{thebibliography}

\end{document}